\documentclass[final]{IEEEtran}
\usepackage{mathtools}
\usepackage{stfloats}
\usepackage{graphicx}
\usepackage{amsmath}
\usepackage{amssymb}
\usepackage{epsf}
\usepackage{enumitem}
\usepackage{latexsym}
\usepackage{cite}
\usepackage{caption}
\usepackage{subcaption}
\usepackage{color}
\usepackage{psfrag}
\usepackage{float}
\usepackage{amsfonts}
\usepackage[normalem]{ulem}
\usepackage{soul}

\newcommand{\Tr}{\operatorname{Tr}}

\newtheorem{remark}{Remark}
\newtheorem{theorem}{Theorem}

\newtheorem{corollary}{Corollary}
\newtheorem{example}{Example}

\usepackage{authblk}
\usepackage{comment}
\usepackage{lipsum}
\usepackage{algorithm}
\usepackage{algpseudocode}

\usepackage{array}  % For \newcolumntype and \setlength{\extrarowheight}
\usepackage{booktabs} % For \toprule, \midrule, and \bottomrule
\algrenewcommand\algorithmicindent{1.0em}%
\DeclareMathOperator*{\argmax}{arg\,max}
\def\BibTeX{{\rm B\kern-.05em{\sc i\kern-.025em b}\kern-.08em
    T\kern-.1667em\lower.7ex\hbox{E}\kern-.125emX}}
\allowdisplaybreaks
\usepackage{mathtools}

\usepackage{derivative}
\makeatletter
\renewcommand*\env@matrix[1][*\c@MaxMatrixCols c]{
  \hskip -\arraycolsep
  \let\@ifnextchar\new@ifnextchar
  \array{#1}}
\makeatother

\usepackage{mathrsfs}
\usepackage[switch, pagewise]{lineno}  % Use the "pagewise" option
\usepackage{xcolor} % This package is required for coloring text
\title{Matrix Pencil-Based DoA Estimation for Hybrid Receivers in 
Snapshot-Limited Scenarios}
\author{Mona Mostafa, Ramy H. Gohary, Amr El-Keyi,  and Yahia A. Eldemerdash Ahmed
\thanks{The first two authors are with the Department of Systems and Computer Engineering, Carleton University, Ottawa, ON K1S5B6, Canada. 
The last two authors are with 
Ericsson Canada Inc., Mississauga, ON, Canada.
}}

\begin{document}

\maketitle

\begin{abstract}\label{sec:abstract}
%The objective of
The goal of this paper is to estimate the directions of arrival~(DoAs) for %of the signals impinging on the antenna arrays of 
 hybrid analog/digital~(HAD) receivers when the number of snapshots is too small for %available. This limitation makes 
 statistical averaging to be reliable. 
%  In such cases, reliable DoA estimation requires %, thereby calling for 
% the structure embedded in the received signals to be exploited.
This goal is achieved in fully-digital receivers  by employing the matrix pencil method~(MPM). % to exploit the structure of the received signal. 
%using the matrix pencil method~(MPM). 
Unfortunately,  the MPM cannot be directly applied in HAD receivers %architecture prevents the MPM from directly applied 
%is unwieldy 
because of the entanglement induced by the underlying analog combiners on the output signals.
%the analog combiners in those receivers result in tangling the output signals.
%at the output of the HAD receiver, 
%rendering direct application of the MPM rather unwieldy.  
Furthermore, these analog combiners project the received signal onto a low-dimensional space, jeopardizing the reception of signals arriving from particular DoA ranges.
%corresponding to specific spatial sectors, 
%causing significant attenuation for DoA ranges corresponding to the null space of those hyperplanes. 
To circumvent these difficulties, 
we propose two approaches to enable the MPM to extract the DoAs in HAD receivers.
%expose the structure of the output signal of the analog combiner. 
%\blu{% first draft: The first approach avoids severe signal attenuation by using an exhaustive set of analog combiners along with post-detection side information on the transmitted signal vector to disentangle the output of the HAD receiver. Our second approach eliminates contingency on side information by using a finer set of analog combiners, thereby trading off side information for snapshots. In our final approach, we dispense with both side information and exhaustive sets of analog combiners characteristic of the first two approaches. 
The two approaches avoid severe attenuation induced by low-dimensional projection %of the received signals 
by % onto low-dimensional spaces by 
cycling over an exhaustive set of analog combiners, collectively spanning the entire space. The first approach can be applied to both fully-connected~(FC) and partially-connected~(PC) HADs and relies on the availability of periodic, potentially unknown, signals to disentangle the output of the HAD receiver. %, \blu{either fully-connected~(FC) or partially-connected~(PC). 
%\org{Our second approach eliminates contingency on periodic signals by using a larger set of analog combiners, each spanning a narrow spatial sector.} 
The second approach applies to PC-HADs only, and eliminates contingency on periodic signals by exploiting  the underlying block diagonal structure. % of  PC-HADs. % using the same set of analog combiners as in the first approach.} %wherein each analog combiner is constructed using one column of the .}
% narrower than its counterpart in the first approach. %That is,  trading periodic signals for snapshots. 
% The third approach dispenses with both periodic signals and the exhaustive set of analog combiners. % characteristic of the first two approaches. 
%  %To do so,  we introduce 
% This approach relies on
% a novel class of analog combiners. %, that are neither PC nor FC. 
% Unlike their predecessors, the new combiners act as spatial all-pass filters 
% guaranteeing that no signal suffers from severe attenuation, irrespective of its DoA. 
%As such, these combiners function as spatial all-pass filters requiring neither exhaustive sets of analog combiners nor periodic signals. 
%which ensures that, irrespective of the DoA, no signal suffers from severe attenuation. 
The superiority of the proposed approaches %\blu{over their existing counterparts} 
is demonstrated via numerical simulations
and comparisons with the %corresponding 
Cram\'{e}r-Rao lower bound.
\end{abstract}
%\begin{IEEEkeywords}
%DoA estimation, hybrid analog/digital, Matrix pencil, CRLB.
%\end{IEEEkeywords}

% INTRODUCTION: 2 - 3 columns
\section{Introduction}\label{sec:introduction}
% Move 1: Establish a research territory 
% Why is your research important?
% What is known about the topic?
% What are your hypotheses?
% What are your objectives?

%https://typeset.io/?mibextid=Zxz2cZ&fbclid=IwAR1pXfVBjIX_Rdy_KnqUKsMjqD-zJRl-TGNCntRG2qtFxMTVLV-z0tVFCFg

% Explain why DoA estimation is important. 
% The 1st paragraph should be compressed

% reduce applications 
% \blu{Direction of arrival (DoA) estimation is integral to wireless communications, both military and civilian~\cite{tuncer2009classical,heath2016overview}. 
% %~\cite{tuncer2009classical,  heath2016overview}
% DoA estimation is necessary for secure data transmission, aircraft, and maritime tracking in radar systems, in addition to the Internet of Things~(IoT) applications~\cite{kaur2015energy} and unmanned aerial vehicles~(UAVs) communications~\cite{zeng2017energy}.
% In wireless sensor networks~(WSNs), DoA estimation enhances environmental monitoring and industrial automation~\cite{fei2016survey}. For 5G and beyond, millimeter-wave~(mm-Wave) massive multiple-input multiple-output~(MIMO) systems use DoA estimation to enhance data rates~\cite{wang2015energy}.} 
Direction of arrival~(DoA) estimation is essential for both military and civilian wireless communications~\cite{tuncer2009classical, heath2016overview}, including secure data transmission,  aircraft, and maritime tracking, Internet of Things~(IoT) applications~\cite{kaur2015energy,fei2016survey}, unmanned aerial vehicles~(UAVs) communications~\cite{zeng2017energy}, and emerging, 5G+, cellular systems~\cite{wang2015energy}.
% In wireless sensor networks~(WSNs), DoA estimation improves environmental monitoring and industrial automation~\cite{fei2016survey}. 
%In wireless sensor networks~(WSNs), DoA estimation can be used to improve environmental monitoring~\cite{fei2016survey}.
%5G communications, %including millimeter-wave~(mm-Wave) massive multiple-input multiple-output~(MIMO) ones, 
%DoA estimation is crucial for the base-station to determine transmission directions~\cite{wang2015energy}. 
DoA estimation techniques have traditionally relied on collecting a large number of snapshots. However, %collecting a large number of snapshots becomes practically infeasible 
when the wireless channel undergoes rapid changes, % is highly dynamic, %particularly in vehicular and UAV communications. In such cases, 
collecting a large number of snapshots becomes practically infeasible, especially under stringent processing time constraints~\cite{liu2020doa}. This calls for techniques that 
%Hence, it is desirable to develop techniques that
enable accurate DoA estimation using a small number of snapshots to be developed.

In addition to having access to a large number of snapshots, accurate DoA estimation requires the receiver to be equipped with a large number of antennas, to improve its 
%In particular, 
%increasing the number of antennas enhances 
%the 
spatial resolution, i.e., 
its ability to distinguish between neighbouring DoAs~\cite{heath2016overview}.
Having a large number of antenna elements comes at a heavy price: increased hardware cost and complexity,  and power consumption of the radio frequency~(RF) chains needed for down-conversion and digitization~\cite{zhang2021direction,zhang2005variable, shi2011iteratively}.
%large-scale antenna arrays~\cite{zhang2021direction} suffer from two main challenges, namely, hardware complexity and power consumption. 
These challenges are exacerbated in classical fully-digital~(FD) receivers, wherein each antenna is connected to a dedicated RF chain.
%, each comprising an RF low-noise amplifier, an analog-to-digital converter~(ADC), a mixer, and a low-pass filter~\cite{zhang2005variable, shi2011iteratively}. %shi2011iteratively, ,sun2020principal
%To elaborate,   
%For example, considering fully digital~(FD) baseband processing in conjunction with a large-scale antenna array at the base station~(BS) is infeasible due to various practical considerations such as hardware complexity, wherein a dedicated radio frequency~(RF) chain is required per antenna element, which consist of analog to digital converters~(ADCs), mixers and power amplifier~\cite{zhang2005variable}, \cite{venkateswaran2010analog}, which will impose high
%power consumption, and cost~\cite{sun2020principal}. %~\cite{telatar1999capacity, dahrouj2010coordinated, shi2011iteratively}.
One approach to overcome this limitation is to replace FD receivers with their analog counterparts~\cite{doan2004design, roh2014millimeter}. 
%These receivers perform a discretized search over all possible DoAs by controlling the direction of reception using a network of analog phase shifters. 
%Unfortunately, this approach limits the ability of the receiver to estimate multiple sources simultaneously~\cite{zhang2005variable, han2015large}.
%The main functional difference between FD and fully analog receivers is that the former controls the phases and amplitudes of the received signals, whereas the latter controls the phases only. 
%this approach leads to decreased estimation accuracy and limits 
Unfortunately, the accuracy of the estimates generated by fully analog receivers is significantly less than that generated by their FD counterparts.
%imits the estimation accuracy and  
%the ability of the receiver to estimate multiple sources simultaneously. T
This is because FD receivers process the amplitudes and phases of the received  signals, whereas fully analog receivers process their phases only~\cite{zhang2005variable, han2015large}.
A versatile class of architectures that enable trading hardware-related difficulties for estimation accuracy 
%to FD and fully analog receivers 
is that comprising hybrid analog/digital~(HAD) receivers~\cite{han2015large,alkhateeb2014channel,el2014spatially,sohrabi2016hybrid}. %and achieve comparable performance to that of FD systems while maintaining reasonable hardware complexity and power consumption~\cite{alkhateeb2014channel,el2014spatially,sohrabi2016hybrid}. %It balances the demand for sufficient beamforming gains to overcome the high propagation loss in the mmWave communication systems and the need to reduce the hardware cost and power consumption~\cite{lin2019hybrid}.

% Write about HAD structure: PC and FC
% https://wirelesspi.com/what-is-the-difference-between-analog-digital-and-hybrid-beamforming/
%In hybrid receivers, the antenna processing unit is divided into a large-scale analog beamformer implemented using analog phase shifters and a small-scale digital beamformer with a limited number of RF chains~\cite{el2014spatially}. The analog beamformer handles down-conversion and signal filtering, while the digital beamformer manages signal processing and parameter estimation~\cite{du2018hybrid}. Compared to analog beamforming, hybrid architectures offer increased flexibility in hybrid beamformer design and enhance energy efficiency through adaptive configurations while achieving a balance between estimation accuracy and complexity~\cite{sun2020principal}. 

In HAD receivers, the antenna processing unit is divided into an analog combiner, implemented using analog phase shifters, and a digital combiner connected to a relatively small number of RF chains~\cite{el2014spatially}. By choosing the number of antennas connected to each RF chain, % and the number of antennas connected to each RF chain,
HAD receivers offer a trade-off between estimation accuracy and hardware complexity~\cite{du2018hybrid}. %~\cite{sun2020principal,du2018hybrid}
%In particular, having an RF chain per antenna results in an FD receiver which offers the m
%by merging analog phase shifters and reducing the number of RF chains~\cite{sun2020principal}. 
%The analog beamformer is responsible for RF signal processing, while the digital beamformer manages both signal processing and parameter estimation tasks~\cite{du2018hybrid}. 
% Among the HAD architectures,  HAD can be configured 
%Hybrid architectures are either fully-connected (FC)~\cite{yu2016alternating} or partially-connected~(PC)~\cite{ mendez2016hybrid}. 
Among existing HAD receivers, are the fully-connected~(FC)~\cite{yu2016alternating} and  partially-connected~(PC)~\cite{mendez2016hybrid,gao2016energy} ones.
In FC-HAD, each RF chain is connected to all the antennas through phase shifters,  whereas in PC-HAD, % counterpart, %each RF chain 
only a subset of antennas is connected to each RF chain. 
%Hence, the PC-HAD architecture features a reduced number of phase shifters and avoids power splitters and combiners required in the FC-HAD architecture~\cite{gao2016energy}. 
%%%%%%%%%%%%%%%%%%%%%%%%%%%%%%%%%%%%%%%%%%%%%%%%%%%%%%%%%%%%%%%%
 %The reduction in complexity offered by the PC HAD architecture comes at a price: a reduction in array gain, resulting in a degradation in the DoA estimation accuracy.

%, partially-connected (PC)~\cite{nguyen2019unequally, mendez2016hybrid}. %, and switches-based (SE)~\cite{mendez2016hybrid}.
%partially-connected (PC)~\cite{nguyen2019unequally}, and switches-based (SE)~\cite{mendez2016hybrid}
%In FC structures, each RF chain connects to all antenna elements through phase shifters. In contrast, in PC structures, each RF chain to a subset of antennas, reducing the need for phase shifters and avoiding power splitters and combiners~\cite{guo2019energy}. 
%However, PC structures may experience performance loss due to reduced array gains~\cite{gao2016energy}. %In addition, SE structures deploy switches instead of \red{PS} in analog processing stages, reducing cost, complexity, and power consumption. 

% Literature review on DoA estimation algorithms and work that was done with HBF
%2. Introduce and review items of previous research in the area.
% traditional techniques
% mention the techniques and criticize each of them and involve the techniques that use the classical techniques in the HAD receivers. How the classical has been adopted in the hybrid mention Refs.

Several DoA estimation techniques have been developed for classical FD receivers, e.g.,~\cite{krim1996two, capon2005maximum, stoica1989music, kumaresan1983estimating, schmidt1986multiple,barabell1983improving,roy1989esprit, yilmazer2006utilization, sarkar1995using, gaber2014study, hua1990matrix,yilmazer2008multiple}. 
Among these techniques are the Bartlett and the minimum variance distortionless response~(MVDR) ones~\cite{krim1996two,capon2005maximum}, which involve a linear search over %by steering the antenna array in 
all possible DoAs. % and measuring power. 
Another class of techniques relies on  received signal statistics. For instance, the multiple signal classification~(MUSIC) and the minimum norm~(NM) techniques rely on estimating the covariance matrix and subsequently restricting the linear search to the subspace dominantly occupied by the received signal~\cite{stoica1989music, kumaresan1983estimating, schmidt1986multiple}. 
Related techniques that incur reduced  %To reduce the %The linear search inherent in the aforementioned techniques renders the 
computational complexity %of these techniques rather impractical. Techniques that circumvent this difficulty  
include root-MUSIC~\cite{barabell1983improving} and the estimation of signal parameters via rotational invariance technique~(ESPRIT)~\cite{roy1989esprit}. 
%techniques.
%using this estimate to obtain the signal subspace ~\cite{kumaresan1983estimating, schmidt1986multiple}
%a utilize the covariance matrix, decomposing it into signal and noise subspaces, and perform a linear search for the DoA within the subspace dominantly occupied by the received signal. 
%Examples include multiple signal classification~(MUSIC) and minimum norm~(NM)~\cite{kumaresan1983estimating, schmidt1986multiple}. 
%NM imposes constraints on the polynomial coefficients corresponding to the noise eigenvectors, assuming uniform distribution inside the unit circle, resulting in superior performance over MUSIC in the presence of multiple correlated signals. However, these methods often suffer from high computational complexity. To circumvent the need for linear search, alternative techniques like 
%To achieve this goal, root-MUSIC employs polynomial root-solving, whereas ESPRIT %divides the antenna array into subarrays and leverages 
%invokes the rotational invariance of the signal subspace. % to estimate the DoAs directly. 
Apart from Bartlett and MVDR, which are computationally intensive, all aforementioned algorithms depend on the true covariance matrix, which % of the received signal. % being available. %Unfortunately, such a covariance matrix is usually not available, particularly when the number of received snapshots is small. 
%Such a covariance matrix 
is often unavailable, especially when the number of received snapshots is limited.
A technique that alleviates the aforementioned challenges  %such as accurate estimation of the covariance matrix, and a large number of snapshots,
relies on the so-called matrix pencil method~(MPM)~\cite{yilmazer2006utilization, sarkar1995using, gaber2014study, hua1990matrix, yilmazer2008multiple}. % presents a promising solution. %However, these assumptions are challenging to guarantee in practical applications. 
%Employing the MP can significantly 
%This method relies on the structure of the received signals rather than their statistics and enables DoA estimation of multiple sources 
%even when the number of snapshots is limited.
%\blu{This method relies on the structure of the received signals rather than their statistics, enabling DoA estimation of multiple sources even with limited snapshots.}
This method relies on the received signal structure rather than statistics, enabling DoA estimation of multiple sources even with limited snapshots.

Some of the classical, statistics-based,  DoA estimation techniques originally developed for FD receivers have been extended to HAD ones. %, e.g.,~\cite{shu2018low,li2020covariance}.  %adopted for hybrid architectures are explored in~\cite{shu2018low} and~\cite{li2020covariance}. 
For instance, in~\cite{shu2018low}, root-MUSIC %with phase alignment 
is applied in PC-HAD receivers with single-phase analog combiners. This approach enables reliable estimation within a particular spatial sector but unreliable estimates outside it.  %and phase ambiguity. To address phase ambiguity,~\cite{shu2018low} searches a finite set of pseudo-solutions for the DoA that maximizes the average received power.%}
This drawback is addressed by other methods %of~\cite{shu2018low} 
 through successive tuning of the analog combiner to angles that are drawn from either a random~\cite{zhang2021direction} or a predefined~\cite{li2020covariance,guo2017millimeter, Mona2023doa} set.
%the techniques in~\cite{li2020covariance, zhang2021direction} involve a sequential adjustment of the direction of the analog combiner. This adjustment is performed using a predefined set of selected angles. 
%sing the snapshots corresponding to these angles, 
%The sample covariance matrix corresponding to these angles is computed, and the DoAs are estimated using classical techniques, e.g., MUSIC~\cite{li2020covariance, guo2017millimeter} and Maximum Likelihood~\cite{zhang2021direction}. %  is applied to estimate the DoAs.
%The signal covariance matrix is estimated by utilizing various snapshots corresponding to these angles. Subsequently, the signal covariance matrix is utilized within the MUSIC algorithm in~\cite{li2020covariance} for the estimation of DoA, whereas in~\cite{zhang2021direction}, it is applied with the Maximum Likelihood method for DoA estimation.
Despite their effectiveness,  these techniques rely on statistical averaging, which 
%suffer from two main challenges. First, they lack guidance on the optimal selection of angles to which the analog combiner is tuned. %, \blu{which may have a direct impact on the accuracy of DoA estimation.}
%Second, to maintain a certain accuracy, 
require a large number of snapshots to obtain accurate estimates of the underlying covariance matrices~\cite{zhang2021direction, li2020covariance,guo2017millimeter} and signal power~\cite{shu2018low,Mona2023doa}.
%required to estimate the signal covariance must increase
%proportionally not accurate because it refer to linear relation
%with the number of antennas, 
%the number of angles and the number of snapshots required to estimate the signal covariance grows proportionally with the number of antennas, 
Obtaining such a number of snapshots incurs potentially intolerable delay and may not be feasible under high mobility conditions. %computational complexity. 
Another class of DoA estimation techniques uses compressed sensing~\cite{molu2017low, alkhateeb2014channel, el2014spatially}. Such techniques, are computationally demanding, but are generally effective 
when the underlying channel is sparse, as in millimeter-wave %~(mm-Wave) 
communications. 
%, DoAs can be estimated using compressed sensing techniques~\cite{molu2017low, alkhateeb2014channel, el2014spatially}. 
%In addition to their inherent computational complexity, these techniques are less effective when the channel matrix is not sparse.
%Another class of techniques utilizes compressive sensing by exploiting the sparsity of the mm-Wave channel in the hybrid receiver for channel estimation~\cite{molu2017low, narayanan1996millimeter, alkhateeb2014channel, el2014spatially}. However, these methods incur significant computational complexity. 
%Apart from classical techniques, other ones rely on various artificial intelligence approaches for DoA estimation. For instance, approaches based on machine learning and deep learning have been proposed in~\cite{huang2018deep, liu2018direction, elbir2021family}. The main advantage of these techniques is that they can be applied even when the \textbf{xxxxx}. Unfortunately, these techniques require training , potentially constraining their practical utility. 
Finally, DoA estimation techniques that rely on deep learning have been proposed in~\cite{huang2018deep, liu2018direction, elbir2021family}. %The main advantage of these techniques lies in their ability to adapt to varying channel conditions through offline/online training phases. 
Unfortunately, these techniques involve extensive training overhead and their performance is questionable when employed in environments not covered by the training phases.

In this paper, we consider DoA estimation in HAD %hybrid analog/digital 
receivers %, 
%featuring a reduced number of RF chains compared to the number of receive antennas,
%for estimating the DoAs of the received signals 
under stringent constraints on the number of snapshots available for the receiver.
%a limited number of snapshots is available. %Such a small number 
These constraints render statistical averaging unreliable, and techniques that rely on exploiting %thereby calling for 
the structure embedded in the received signals, rather than their sample averages, must be sought. 
%to be exploited. %One approach to exploit this structure of the received signals and conduct analysis directly on received snapshots is the MPM. 
One technique to do so is the MPM. %This technique was applied in FD receivers,  to unravel the structure underlying the received signal without invoking its statistical attributes. %and enables direct analysis of the received snapshots. 
Unfortunately, this technique, although readily applicable in FD receivers~\cite{yilmazer2006utilization, gaber2014study,yilmazer2008multiple} does not automatically lend itself to HAD receivers.  % the HAD architecture prevents the MPM, which has been employed in FD receivers~\cite{yilmazer2006utilization, gaber2014study,yilmazer2008multiple}, from being directly applied for DoA estimation. %Despite the applicability of the MPM in FD receivers, its direct implementation in HAD receivers poses challenges. '
In particular, we will show that HAD architectures tangle the signal at the receiver output, rendering the invocation of the MPM rather unwieldy. Furthermore, we will show that the analog combiner of the HAD receiver projects the received signal vector onto a hyperplane corresponding to a particular spatial sector. This projection causes the signals arriving from particular DoAs to be attenuated or even nullified. %corresponding to certain DoAs. 
%Moreover, there are challenges arising from the analog combiner in the HAD receiver that projects the received signal vector onto a specific spatial sector, resulting in significant signal attenuation corresponding to certain DoAs. 
To address these challenges, we develop two new MPM-centered approaches. % centred around the MPM. 
% \blu{In these approaches, we follow the lead 
% % ~\cite{ermolaev1994fast, huang2018deep, shu2018low, li2020covariance, han2018dft}
% of~\cite{huang2018deep, shu2018low, li2020covariance, han2018dft}, wherein the number of sources is assumed to be known \emph{a priori}.
% %(Techniques for estimating this number can be found in~\cite{fishler2002detection, nadler2010nonparametric, wax2021detection}.)
% }

%These approaches enable the MPM to overcome the effective DoA estimation in the hybrid framework. to yield reliable DoA estimates in HAD receivers. 

In the first approach, the attenuation incurred by the analog combiner is mitigated by cycling over an exhaustive set of analog combiners that collectively span the entire space. %, each constructed from distinct columns of
%each constructed from the columns of 
Such analog combiners can be constructed from the Discrete Fourier Transform~(DFT) matrix. 
%%The set of analog combiners collectively span the entire space.
% Note for Mona >> The training sequences/ pilots are assumed to be known at the receiver side but the signals that we are using are unknown and we don't care about these signals. These periodic signals can be used as pilots for any other system. However for us, we don't care, we only care that they are periodic and deterministic to use them in the estimation process. 
The main idea of this approach is to exploit the fact that, in typical communications, the received signal comprises periodic, potentially unknown, deterministic components to assist the receiver in data recovery~\cite{tse2005fundamentals}. %\blu{~\cite{tse2005fundamentals, goldsmith2005}}.
%For instance, 
%relies on known deterministic signals being periodically received. 
Such signals include training pilots or the preamble segment of the signal frame. %, which are typically used in communication systems. %, such signals are available 
%with a known deterministic structure being received 
We will show that access to such signals can be used to combine and disentangle the signals at the outputs of the analog combiners. %To avoid the attenuation incurred by the projection induced by the analog combiner by 
%cyclic processing of the received signal using an exhaustive set of analog combiners. %significant signal attenuation through the utilization of an exhaustive set of analog combiners. 
%In particular, short training sequences are periodically transmitted. The projection of these training sequences onto the exhaustive set of analog combiners, followed by the combination of received signals after applying these combiners, facilitates the untangling of the output of the HAD receiver. 
%  Dr. Ramy >> 
Hence, this approach renders HAD receivers, either FC or PC, behave as if they were FD  insofar as their output is amenable to DoA estimation via the MPM, and the attenuation incurred by individual analog combiners is alleviated. 
% Despite its efficacy, the first approach presents two key challenges: configuring each analog combiner with different phases causing entangled outputs, particularly in the FC architecture, and the reliance on the availability of periodic signals.
Despite its efficacy, the first approach relies on extracting periodicity embedded in the received signal frames. This
%the frames of the received signals having segments that repeat periodically in every frame. In addition, it 
requires recovery of the timing of the periodic segments, adding to the complexity and processing delay of the receiver.
In the second approach, we seek to eliminate the reliance on periodic signals characteristic of the first approach. To do so, we note that the entanglement of the received signals at the output of the analog combiner results from the non-identical phases of the columns used to construct the underlying matrices. To address this problem, we construct the analog combiner using a single column from the DFT matrix. Doing so in 
%In 
FC-HAD %, using a single DFT column leads to 
results in rank-one analog combiners, causing the structure of their output to be degenerate for the MPM to be applicable.
%inapplicable. %This is because the MPM relies on the structure of the received signal vector to estimate the DoAs, but with a rank-one combiner, this structure becomes degenerate. 
%elements of the received signal vector become identical. 
%%%%%%%%%%%%%%%%%%%%%%%%%%%%%%%%%%%%%%%%%%%%%%%%%%%%%%%%%%%%%%%%
In contrast, the analog combiners in PC-HAD receivers  possess a block-diagonal structure, which  
preserves the linear independence of the columns, even when they have identical phases. 
%Hence, using columns with the same phase in PC-HAD
%Such constrt allows 
As such, these analog combiners result in their outputs to be disentangled and thereby amenable to the MPM.
%%%%%%%%%%%%%%%%%%%%%%%%%%%%%%%%%%%%%%%%%%%%%%%%%%%%
Similar to the first approach, to remedy the attenuation incurred by the analog combiner, we cycle over an exhaustive set of analog combiners. % will be considered.
%Although this approach enables the MPM and eliminates reliance on periodic signals, 
%Unfortunately, this approach introduces 
Phase ambiguity inherent in this approach can be resolved by reconfiguring the analog combiner based on the estimated DoAs and selecting the ones that maximize the received signal-to-noise ratio~(SNR).
As such, %the second approach dispenses with the periodic signals needed by the first approach. As such, 
the second approach trades periodic signals needed by the first approach for an increased number of snapshots, which is required to resolve its inherent phase ambiguity.

Root-mean-square error~(RMSE) simulations and the corresponding Cram\'{e}r-Rao lower bound~(CRLB) show  the superiority of the proposed approaches over their existing counterparts.
The rest of the paper is organized as follows. The system model is presented in Section~\ref{sec:system_model}. Section~\ref{sec:Problem formulation} states the problem formulation. The proposed approaches are presented in Section~\ref{sec:proposed techs}.  Simulation results are provided in Section~\ref{sec:simulation}, and Section~\ref{sec:conclusion} concludes the paper.
%%%%%%%%%%%%%%%%%%%%%%%%%%%%%%%%%%%%%%%%%%%%%%%%%%%%%%%%%%%%%%%%%%%%%%%%%%%%%
%To evaluate the performance of the proposed algorithms, we derive the Cram\'{e}r-Rao lower bounds~(CRLB), which provides a theoretical lower bound on the variance of any unbiased estimator. We compare the CRLB with the simulation results obtained from the proposed algorithms, and the numerical results show that the proposed algorithms achieve a significantly lower estimation error than the CRLB. Furthermore, we compare the proposed method with existing state-of-the-art methods, and the simulation results demonstrate the superiority of the proposed method in terms of estimation accuracy and efficiency.

Notations: Matrices and vectors are denoted by upper and boldface lower case letters, respectively. The transpose, %conjugate, 
left pseudo-inverse, and Hermitian are denoted by $(\cdot)^T$, %$(\cdot)^*$,
$(\cdot)^\dag$, and $(\cdot)^H$, respectively. %The identity matrix of size $M \times M$ are denoted by $\boldsymbol{I}_{\!M}$, and the trace of a matrix are denoted by $\operatorname{Tr}(\cdot)$. 
%The $M \times M$ identity matrix are denoted by $\boldsymbol{I}_{\!M}$.
The $M \times M$ identity matrix and the $M\times 1$ all one column vector are denoted by $\boldsymbol{I}_{\!M}$ and $\boldsymbol{1}_{\!M}$, respectively.
The trace and eigenvalues of a matrix are denoted by $\operatorname{Tr}(\cdot)$, and $\nu(\cdot)$, respectively.
%
%\blu{The $M \times M$ identity matrix, trace, and eigenvalues of a matrix will be denoted by $\boldsymbol{I}_{\!M}$, $\operatorname{Tr}(\cdot)$, and $\nu(\cdot)$, respectively.
%The $M\times 1$ all one column vector will be denoted by $\boldsymbol{1}_{\!M}$.}
%The all one column vector of dimension $M$ will be denoted by $\boldsymbol{1}_{\!M}$.}
%The $M \times 1$ vector of ones will be denoted by $\boldsymbol{1}_{\!M}$.}
%%%%%%%%%%%%%%%%%%%%%%%%%%%%%%%%%%%%%%%%%%%%%%%%%%%%%%
% The real and imaginary parts of a complex number will be denoted by $\Re(\cdot)$ and $\Im (\cdot)$, respectively.
%%%%%%%%%%%%%%%%%%%%%%%%%%%%%%%%%%%%%%%%%%%%%%%%%%%%%%%%%%%%%%
%the argument of a complex number.
%The real, imaginary, and argument~(phase) of a complex number will be denoted by $\Re(\cdot)$, $\Im(\cdot)$, and $\arg(\cdot)$, respectively.}
%The set of integers and complex numbers will be represented by $\mathbb{Z}$ and $\mathbb{C}$, respectively. 
%The Kronecker and Hadamard products will be denoted by $\otimes$ and $\odot$, respectively. 
The Kronecker, Hadamard products and the direct sum are denoted $\otimes$, $\odot$ and $\oplus$, respectively.

\section{System Model}\label{sec:system_model}
% What is the channel model/ what if there is a Doppler shift?
We consider a uniform linear array~(ULA) of $M$ identical antennas. % receiving far-field narrowband signals.
%by a receiver featuring $M$ identical antenna elements arranged in a uniform linear array~(ULA) geometry. 
The objective is to estimate the DoAs, $\{\theta_r\}_{r=1}^R$, corresponding to far-field narrowband signals emitted by $R$ distinct sources. The ULA steering vector corresponding to the $r$-th source, $r=1,\ldots,R$, are given by~\cite{zhang2021direction, shu2018low, li2020covariance}:
\begin{subequations}
    \label{eq:steering_vector}
    \begin{gather}
        \label{eq:a}
        \boldsymbol{a}_{r} = 
        \begin{bmatrix}
            1 & e^{\jmath\mu_r} & \cdots & e^{\jmath(M-1)\mu_r}
        \end{bmatrix}^T, \\
        \label{eq:mu}
        \mu_r = \frac{2\pi \Delta}{\lambda}\sin{\theta_r},
    \end{gather}
\end{subequations}
where $\lambda$ is the wavelength, $\Delta\leq \frac{\lambda}{2}$ is the antenna spacing,  and $\mu_r\in(-\pi,\pi]$ is the phase between adjacent antennas~\cite{krim1996two}. %For the sake of clarity, we assume $\Delta=\frac{\lambda}{2}$~\cite{sohrabi2016hybrid}, which results in $\mu_r\in(-\pi,\pi]$. 
% The steering vectors for all sources are the columns of the $M\times R$ steering matrix,  $\boldsymbol{A}$, i.e., 
% \begin{equation}
%     \boldsymbol{A} = 
%     \begin{bmatrix} 
%         \boldsymbol{a}_1 &
%         %\boldsymbol{a}_2 &
%         %\boldsymbol{a}_3 &
%         \cdots &
%         \boldsymbol{a}_R
%     \end{bmatrix}.
%     \label{eq:steering_matrix}
% \end{equation}
% %\blu{
% Let the $R$ unknown baseband signals be denoted as: 
% \begin{equation}
%     \boldsymbol{s}(t) = 
%     \begin{bmatrix}
%         s_1(t) & \cdots & s_R(t)
%         \label{eq_s_t}
%     \end{bmatrix}^T.
% \end{equation}

Let the steering vectors of all sources be contained in the  matrix  %$\boldsymbol{A}\in\mathbb{C}^{M\times R}$, 
$\boldsymbol{A} = 
    \begin{bmatrix} 
        \boldsymbol{a}_{1} 
        %\boldsymbol{a}_2 &
        %\boldsymbol{a}_3 &
        \cdots 
        \boldsymbol{a}_{R}
    \end{bmatrix} \in\mathbb{C}^{M\times R}$.
Let the $R$ unknown baseband signals be denoted as $\boldsymbol{s}(t) = 
    \begin{bmatrix}
        s_1(t)  \cdots  s_R(t)
    \end{bmatrix}^T$. 
% \begin{equation}
% \boldsymbol{s}(t) = 
%     \begin{bmatrix}
%         s_1(t)  \cdots  s_R(t)
%     \end{bmatrix}^T. 
% \label{eq_s_t}    
% \end{equation}
The received signals at the antenna array can be modelled as:
%\blu{The received signals at the antenna array are modeled as:}
\begin{equation}
    \boldsymbol{x}(t) = e^{\jmath 2\pi f_c t}\boldsymbol{A}\boldsymbol{s}(t) + \boldsymbol{z}(t),
    \label{eq:X_fd}
\end{equation}
where $f_c$ is the carrier frequency and $\boldsymbol{z}(t)\in\mathbb{C}^{M}$ is the zero-mean additive white Gaussian unit variance noise, i.e., $\sigma_z^2=1$.

%-----------------------------------------------------------------------------
% After down-conversion and sampling, 
% the $k$-th snapshot is represented as the vector $\boldsymbol{x}_k\in\mathbb{C}^M$, where
% \begin{equation}
%     \boldsymbol{x}_k = \boldsymbol{A}\boldsymbol{s}(t) + \boldsymbol{z}(t)\Big|_{t=kT_s},\qquad k=1,\ldots,K,\label{eq:ADC_x_k}
% \end{equation} 
% where $T_s$ is the sampling interval. Combining $K$ samples of noise and signals of the $R$ sources in the matrices $\boldsymbol{Z}\in \mathbb{C}^{M\times K}$ and $\boldsymbol{S}\in\mathbb{C}^{R \times K}$, respectively, we can write~\eqref{eq:ADC_x_k} as:
% \begin{equation}
%     \boldsymbol{X} = \boldsymbol{A}\boldsymbol{S}+\boldsymbol{Z},
%     \label{eq:X}
% \end{equation}
% where $[\boldsymbol{Z}]_{m,k}$ represents the $k$-th sample of noise at the $m$-th antenna, $m=1,\ldots,M$, and $[\boldsymbol{S}]_{r,k}$ represents the  $k$-th sample of signal emitted by the $r$-th source, $r=1,\ldots,R$, for   $k=1,\ldots,K$.
% For uncorrelated sources, we can write:
% \begin{equation}
% \label{eq:P_r}
%     \boldsymbol{\Phi}=\frac{1}{K}\mathbb{E}\{\boldsymbol{S}\boldsymbol{S}^H\}=\operatorname{diag}(P_1,\ldots,P_R),
% \end{equation}
% where $\{P_r\}_{r=1}^R$ represent the average power of the $R$ sources.
%---------------------------------------------------------------------------
After down-conversion, the signal is sampled at $T_s$ intervals and %sampling, %\blu{using the sampling interval $T_s$}, 
the $k$-th snapshot is given by %denoted as %the vector
$\boldsymbol{x}_k\in\mathbb{C}^M$, where
\begin{equation}
    \boldsymbol{x}_k = \boldsymbol{A}\boldsymbol{s}(t) + \boldsymbol{z}(t)\Big|_{t=kT_s},\qquad k=1,\ldots,\tilde{K},\label{eq:ADC_x_k}
\end{equation} 
%where $T_s$ is the sampling interval  and 
where  $\tilde{K}=NK$ represents the number of snapshots contained  in
%\blu{with} 
$N$ length-$K$ segments. 
%Using matrix notation, we write~\eqref{eq:ADC_x_k} can be expressed as:
In matrix notation, we can write~\eqref{eq:ADC_x_k}:
%The noise and signal components of $\boldsymbol{x}_k$ are 
%For the $n$-th segment, $K$ samples of noise and signals from $R$ sources are combined 
%into  $\boldsymbol{Z}_n\in \mathbb{C}^{M\times K}$ and $\boldsymbol{S}_n\in\mathbb{C}^{R \times K}$, respectively. Hence,~\eqref{eq:ADC_x_k} can be written as:
\begin{equation}
    \boldsymbol{X}_n = \boldsymbol{A}\boldsymbol{S}_n+\boldsymbol{Z}_n,
    \label{eq:X}
\end{equation}
where the $mk$-th entry of $\boldsymbol{X}_n$ and $\boldsymbol{Z}_n \in \mathbb{C}^{M\times K}$ 
%represents the $k$samples received at the  $M$ antenna, 
%, $[\boldsymbol{Z}_n]_{m,k}$ 
represents the $k$-th sample of the received signal and  noise at the $m$-th antenna of the $n$-th segment of snapshots, respectively, $m=1,\ldots,M$. The $rk$-th entry of 
$\boldsymbol{S}_n \in  \mathbb{C}^{R \times K}$ represents the  $k$-th sample of signal emitted by the $r$-th source, $r=1,\ldots,R$,  $k=1,\ldots,K$.
For uncorrelated sources, %\blu{we can write:}
\begin{equation}
\label{eq:P_r}
    \boldsymbol{\Phi}=\frac{1}{K}\mathbb{E}\{\boldsymbol{S}_n\boldsymbol{S}_n^H\}=\operatorname{diag}(P_1,\ldots,P_R),
\end{equation}
where $\{P_r\}_{r=1}^R$ represent the average power of the $R$ sources.

%For completeness, 
We now review %the two existing versions of HAD receivers, 
FC-HAD and PC-HAD receivers~\cite{du2018hybrid, yu2016alternating, mendez2016hybrid}.
\subsection{The HAD Receiver}\label{sec: HAD-receivers}
\begin{figure}
\centering
 %---------------------PC-------------------
% \begin{subfigure}{.40\textwidth}
%              \centering
%             \psfrag{M}{\!$M$}
%             \psfrag{MRF}[][][.60]{\!\! $M_{\mathrm{RF}}$}
%             \psfrag{cdots}{$\cdots$}
%             \psfrag{delta}{$\Delta$}
%             \psfrag{RF}[][][.75]{\! RF-Chain}
%             \psfrag{ADC}[][.75]{\!\! ADC}
%             \psfrag{qt1}{${q}_{_1}(t)$} 
%             \psfrag{qn1}{${q}_{_1}[n]$} 
%             \psfrag{qn2}{${q}_{_2}[n]$} 
%             \psfrag{qtL}{${q}_{_L}(t)$} 
%             \psfrag{qnL}{${q}_{_L}[n]$} 
%             \psfrag{AC}[][][.85]{$\boldsymbol{W}_\mathrm{A}$}
%             \psfrag{DC}[][][.85]{Digital combiner, $\boldsymbol{w}_\mathrm{D}$}
%             \psfrag{yn}[][][.85]{$\boldsymbol{y}_{_{\mathrm{HAD}}}$}
%             \includegraphics[width=50mm]{Figures/PC_uncoloured}
%     \caption{}
%         \label{fig:PC-Arch-a}
%     \end{subfigure}
%-----------------------------FC--------------------------------------------
 \begin{subfigure} {.48\columnwidth}
             \centering
            \psfrag{M}{\!\!$M$}
     %       \psfrag{MRF}[][][.55]{\!\! $M_{\mathrm{RF}}$}
            \psfrag{cdots}{\!$\cdots$}
            \psfrag{delta}{\!$\Delta$}
            \psfrag{RF}[][][.55]{ RF-Chain}
            \psfrag{ADC}[][][.65]{\!\! ADC}
            \psfrag{qt1}{\hspace{0.01cm}${q}_{_1}(t)$} 
            \psfrag{qn1}{\hspace{0.01cm}${q}_{_1}[k]$} 
            \psfrag{qn2}{\hspace{0.01cm}${q}_{_2}[k]$} 
            \psfrag{qtL}{\hspace{0.01cm}${q}_{_L}(t)$} 
            \psfrag{qnL}{\hspace{0.01cm}${q}_{_L}[k]$} 
            \psfrag{AC}[][][.85]{$\boldsymbol{W}_{\!\!\mathrm{A}}$}
            \psfrag{DC}[][][.75]{Baseband Digital Processing}
            \psfrag{yn}[][][.9]{\ $\Hat{\theta}_r$}
            \includegraphics[width=40mm]{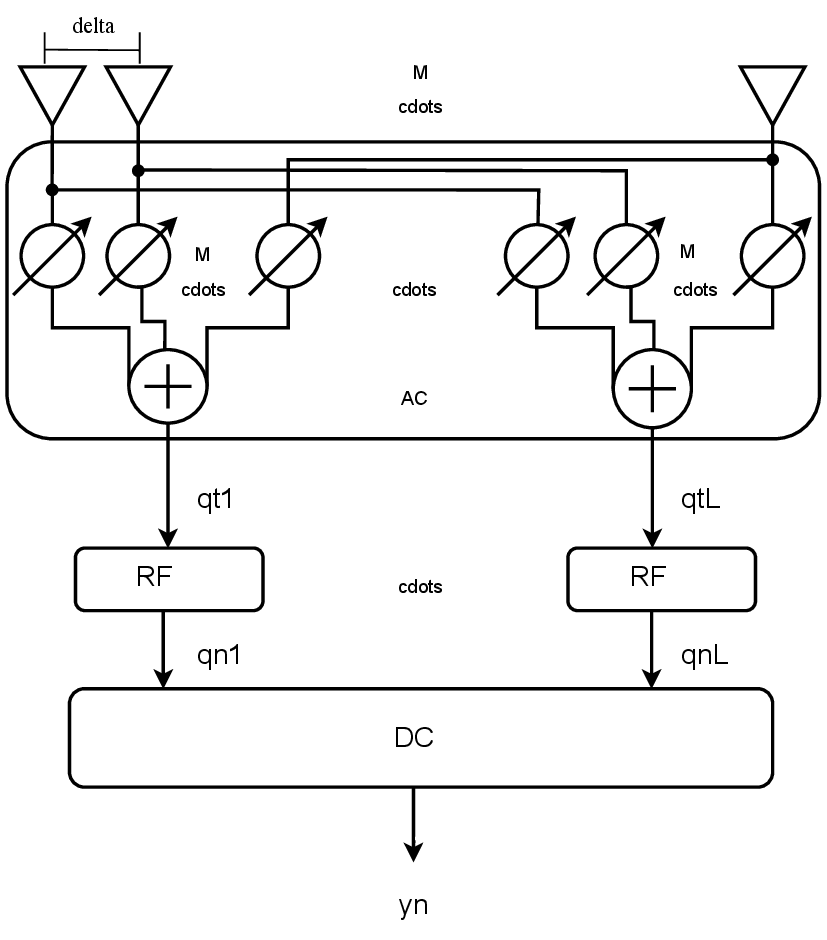}
    \caption{}
        \label{fig:FC-Arch}
    \end{subfigure}
 \hspace{0.01\columnwidth} % Horizontal spacing between subfigures
%-------------------------------------------------------------------------
    \begin{subfigure}{.48\columnwidth}
             \centering
            \psfrag{M}{\!\!$M$}
            \psfrag{MRF}[][][.55]{\!\!\! $M_{\mathrm{RF}}$}
            \psfrag{cdots}{\!$\cdots$}
            \psfrag{delta}{\!$\Delta$}
            \psfrag{RF}[][][.55]{ RF-Chain}
            \psfrag{ADC}[][][.65]{\!\! ADC}
            \psfrag{qt1}{\hspace{0.01cm}${q}_{_1}(t)$} 
            \psfrag{qn1}{\hspace{0.01cm}${q}_{_1}[k]$} 
            \psfrag{qn2}{\hspace{0.01cm}${q}_{_2}[k]$} 
            \psfrag{qtL}{\hspace{0.01cm}${q}_{_L}(t)$} 
            \psfrag{qnL}{\hspace{0.01cm}${q}_{_L}[k]$} 
            \psfrag{AC}[][][.85]{$\boldsymbol{W}_{\!\!\mathrm{A}}$}
            \psfrag{DC}[][][.75]{Baseband Digital Processing}
            \psfrag{yn}[][][.9]{\ $\Hat{\theta}_r$}
            \includegraphics[width=40mm]{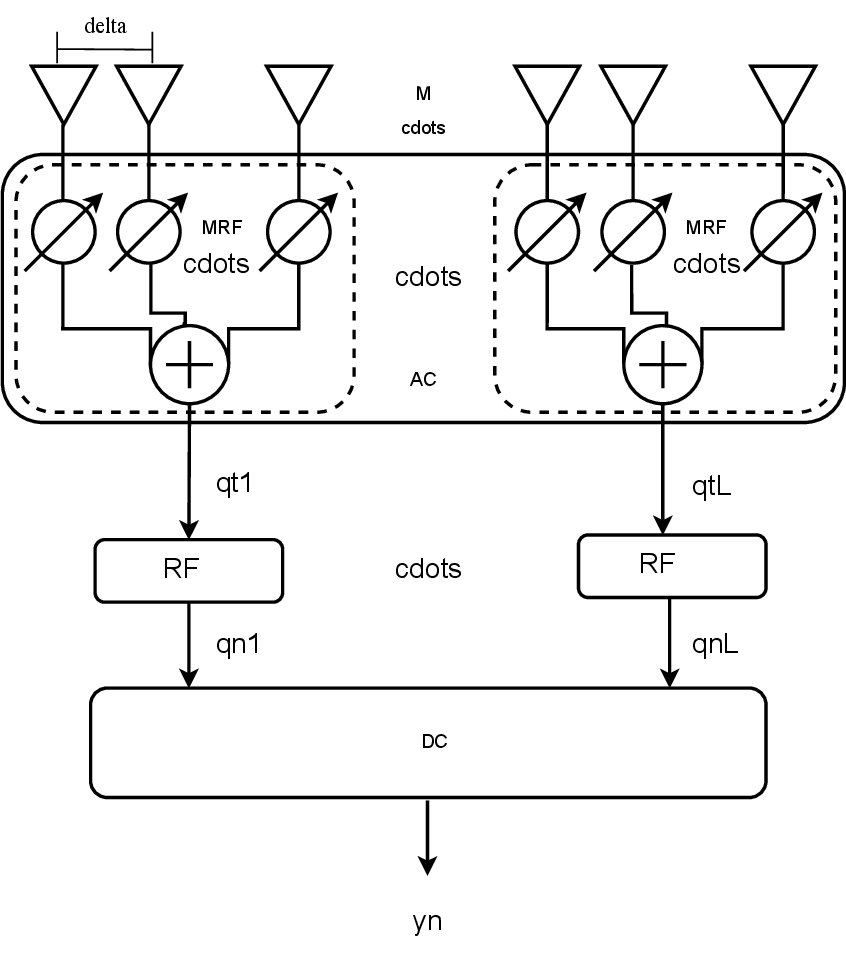}
    \caption{}
        \label{fig:PC-Arch-a}
    \end{subfigure}
     \caption{Block diagram of (a) FC-HAD, (b) PC-HAD.}
    \label{fig:HAD}
\end{figure}

%In contrast with FD receivers, in HAD ones,
%as shown in Figure~\ref{fig:PC-Arch-a}, %the antennas are partitioned into $L$ subgroups. 
In HAD receivers, the $n$-th segment of snapshots from the $M$ antennas in~\eqref{eq:X_fd} %corresponding to the $n$-th segment of snapshots 
are fed to the analog combiner,  $\boldsymbol{W}_{\mathrm{\!\!A},n}\in\mathbb{C}^{M\times L}$, which 
%The analog combiner 
phase-shifts and combines the outputs  into $L<M$ streams,  cf. Figure~\ref{fig:HAD}. %which is represented by the matrix $\boldsymbol{W}_\mathrm{\!\!A}$. %\in\mathbb{C}^{M\times L}$
Each  of the $L$ streams is composed of a subarray of  $M_{\mathrm{RF}}$ antennas %and is 
%Each stream is connected to one RF chain and one ADC, which down-converts and samples the analog signal. Denoting 
%Each of the $L$ streams is 
connected to one RF chain  %and one analog-to-digital converter~(ADC) 
for down-conversion and digitization.
%The number of antennas connected to each RF chain is denoted by $M_{\mathrm{RF}}$.
%The number of antennas connected to each RF chain is $M_{\mathrm{RF}}$.
% The analog combiner can be denoted by   $\boldsymbol{W}_{\mathrm{\!\!A}}\in\mathbb{C}^{M\times L}$.
%For the $n$-th length-$K$ segment, 
The output of %the $n$-th analog combiner,  
$\boldsymbol{W}_{\mathrm{\!\!A},n}$ is: %can be expressed as: % is given by:
\begin{equation}
\label{eq:q}
    \boldsymbol{q}(t) = \boldsymbol{W}_{\!\!\mathrm{A},n}^H\boldsymbol{x}(t).
\end{equation}

%%%%%%%%%%%%%%%%%% Updade 2024 %%%%%%%%%%%%%%%%%%%%%%%%%
% to export the esp from draw.io select all the diagrams then 
%1. Export as SVG and deselect all the options and select selected-only
%2. there is a link we can remove with Inkscape software (https://inkscape.org/release/inkscape-1.4/windows/64-bit/msi/?redirected=1) and then save as an SVG 
%3. use https://cdkm.com/svg-to-eps to convert to eps
%%%%%%%%%%%%%%%%%%%%%%%%%%%%%%%%%%%%%%%%%%%%%%%%%%%%%%%%
%%%%%%%%%%%%%%% Steps update March 21, 2025 %%%%%%%%%%%%%%%%
% Steps to Convert SVG to EPS for Use with psfrag in LaTeX
% Prepare the SVG file
% Make sure all options are deselected when exporting or saving the SVG file.
% Open the SVG in Inkscape
% Open your SVG file using Inkscape.
% Remove any links or external references if present.
% Save as Plain SVG
% Save the file as Plain SVG (not Inkscape SVG).
% Export to EPS
% Go to File > Save As and choose EPS format.
% Text Output Options (In the Save As EPS dialog)
% Select: Embed text (do not use 'Convert text to paths' or 'Omit text in PDF and create LaTeX file').
% Deselect: Rasterization.
% Set Resolution: 600 dpi.
% Save the EPS File
% Click on Save to generate the EPS file.
%%%%%%%%%%%%%%%%%%%%%%%%%%%%%%%%%%%%%%%%%%%%%%%%%%%%%%%%%%%%%

A key limitation of the analog combiner is not to 
%\blu{is that it does not}
contain active elements. Incorporating such elements is both costly and impractical for systems with a large number of antennas. Hence, the analog combiner is restricted to phase-shifting the output of each antenna without storing it or altering its power~\cite{alkhateeb2014channel}.
Down-converting and sampling $\boldsymbol{q}(t)$ yields: % \blu{the baseband column vector}:
\begin{equation}
\label{eq:q_nn}
    \boldsymbol{q}_{k} = e^{-\jmath 2\pi f_ct}\boldsymbol{q}(t)\Big|_{t=kT_s},\qquad k=1,\ldots,K.\end{equation}
%Constructing the matrix $\boldsymbol{Q}\in\mathbb{C}^{L\times K}$ from $\{ \boldsymbol{q}_{k}\}_{k=1}^{K}$ yields:
Collecting $\{\boldsymbol{q}_{k}\}_{k=1}^{K}$ in the matrix  $\boldsymbol{Q}_n \in \mathbb{C}^{L\times K}$, % be the vectors $\{\boldsymbol{q}_{k}\}$. Thus, 
we write~\eqref{eq:q_nn}:
% Using $\{\boldsymbol{q}_{k}\}_{k=1}^{K}$ as column vectors, we construct the matrix 
% Let the $k$-th column of $\boldsymbol{Q}_n \in \mathbb{C}^{L\times K}$ be $\boldsymbol{q}_{k}$. Hence 
 %$\boldsymbol{Q}=\boldsymbol{W}_{\!\!\mathrm{A}}^H\boldsymbol{X} \in\mathbb{C}^{ L \times K}$.
\begin{equation}
\boldsymbol{Q}_{n}=\boldsymbol{W}_{\!\!{\mathrm{A}},n}^H\boldsymbol{A}\boldsymbol{S}_{n}+\boldsymbol{W}_{\!\!{\mathrm{A}},n}^H\boldsymbol{Z}_n,
 \label{eq:Q_n}
\end{equation}
where $\boldsymbol{Q}_n \in \mathbb{C}^{L\times K}$ is the  output of % $n$-th segment of  snapshots corresponding to
$\boldsymbol{W}_{\mathrm{\!\!A},n}$, which is then fed to %the inout %then  %, $K$
% \begin{equation}
%     \label{eq:X_tilde} \boldsymbol{Q}=\boldsymbol{W}_{\!\!\mathrm{A}}^H\boldsymbol{X}=\boldsymbol{W}_{\!\!\mathrm{A}}^H\boldsymbol{A}\boldsymbol{S}+\boldsymbol{W}_{\!\!\mathrm{A}}^H\boldsymbol{Z}.
% \end{equation}
%where the columns of $\boldsymbol{Q}\in\mathbb{C}^{L\times N}$ are $\{\boldsymbol{q}_{n}\}_{n=1}^N$. 
% \blu{Let $\tilde{K}=NK$ be the total number of snapshots and consider $N$ analog combiners represented by $\{\boldsymbol{W}_{\mathrm{\!\! A},n}\}_{n=1}^{N}$. Let the number of snapshots processed by $\boldsymbol{W}_{\mathrm{\!\! A},n}$ be $K$, $n=1,\ldots,N$. Hence, after applying the $n$-th analog combiner, the received signal  can be rewritten as: 
% \begin{equation}
% \boldsymbol{Q}_{n}=\boldsymbol{W}_{\!\!{\mathrm{A}},n}^H\boldsymbol{A}\boldsymbol{S}_{n}+\boldsymbol{W}_{\!\!{\mathrm{A}},n}^H\boldsymbol{Z}_n,
%  \label{eq:Q_n}
% \end{equation}
% where $\boldsymbol{Q}_n \in \mathbb{C}^{ L \times K}$ is output of the $n$-segment of snapshots, $K$.} 
% $\boldsymbol{W}_{\!\!{\mathrm{A}},n}$,  $\boldsymbol{S}_{n}\in\mathbb{C}^{K\times R}$ and $\boldsymbol{Z}_{n}\in\mathbb{C}^{M\times K}$ are the signals and noise components of the $n$-th segment of  snapshots corresponding to $\boldsymbol{W}_{\mathrm{\!\!A},n}$, % $ \in\mathcal{I}$, 
% respectively.}
%Finally, the matrix $\boldsymbol{Q}_n\in\mathbb{C}^{L\times K}$ is 
%fed to 
the baseband digital DoA estimator. % digital processing unit \blu{to estimate the unknown DoAs, $\{\theta_r\}_{r=1}^R$.}
% $\boldsymbol{w}_\mathrm{D}\in\mathbb{C}^{L}$ yields:
% \begin{equation}
%     \label{eq:y_hbf}
%     \boldsymbol{y}_{_{\mathrm{HAD}}} = \boldsymbol{w}_{\mathrm{D}}^H\boldsymbol{Q}.
% \end{equation}

%The structure of $\boldsymbol{W}_\mathrm{\!\! A}$ determines whether the HAD receiver is FC or PC.

Among existing  HAD receivers are the  %structure of $\boldsymbol{W}_\mathrm{\!\! A}$ determines  the configuration of the HAD architectures, such as 
FC-HAD and PC-HAD ones.
%\footnote{The subscript, $n$, is dropped in discussions pertaining to all combiners.} %applying to all segments and corresponding combiners.}
%In particular, for
In both receivers,  all non-zero entries of $\boldsymbol{W}_\mathrm{\!\! A}$\footnote{The subscript, $n$, is dropped in discussions pertaining to all combiners.} are unit-modulus in the form $e^{\jmath \phi}$,  $\phi \in (-\pi, \pi]$.
%,  whereas in PC-HAD, $\boldsymbol{W}_\mathrm{\!\! A}$ is a block-diagonal matrix, with each diagonal block corresponding to unit-modulus phase shifters of a particular subarray. 
For FC-HAD and  PC-HAD, $\boldsymbol{W}_\mathrm{\!\! A}$ is denoted as $\boldsymbol{W}_{\mathrm{\!\! A_{FC}}}$ and  $\boldsymbol{W}_{\mathrm{\!\! A_{PC}}}$, respectively. 
%The FC-HAD and PC-HAD receivers will be considered next.

\subsubsection{The FC-HAD Receiver}\label{sec:FC-HAD}
In FC-HAD, % uses an architecture in which 
each RF chain is connected to all antenna elements via $ML$ analog phase shifters, cf. Figure~\ref{fig:FC-Arch}. As shown in this figure,  the output of each antenna is split across the $L$ phase shifters and RF chains. Hence,  $\boldsymbol{W}_{\mathrm{\!\! A_{FC}}}$ can be expressed as:
%the analog combiner of FC-HAD can be given by:
\begin{equation}\label{eq:W_A_FC}
    \boldsymbol{W}_{\mathrm{\!\! A_{FC}}} = 
        \frac{1}{\sqrt{L}}    \begin{bmatrix}
        \boldsymbol{w}_{{\mathrm{A_{FC}},1}} & \cdots & \boldsymbol{w}_{{\mathrm{A_{FC}},L}}
    \end{bmatrix},
\end{equation}
where $\boldsymbol{w}_{{\mathrm{A_{FC}},\ell}}\in\mathbb{C}^M$  contains the $M_{\mathrm{RF}}=M$ phase shifts of the $\ell$-th subarray, $\ell=1,\dots, L$. The scalar multiplier, i.e., $\frac{1}{\sqrt{L}}$,   accounts for the power splitting at the antenna outputs.
%\input{project/figures/FC-HAD}

%--------------------------------------------------------------------
\subsubsection{The PC-HAD Receiver}\label{sec:PC-HAD}
In PC-HAD,  the antenna outputs are not split across multiple phase shifters. Instead, each antenna is connected to one phase shifter before combining into one RF chain, cf. Figure~\ref{fig:PC-Arch-a}. In other words, in this receiver, the $M$ receive antennas are partitioned into $L$ identical subarrays, whereby each subarray contains $M_{\mathrm{RF}}=M/L$ antennas. %, where
For convenience, we assume that $M$ is a multiple integer of $L$. 
%\in\mathbb{Z}
%Such an architecture is shown in Figure~\ref{fig:PC-Arch-a}. 
%Unlike FC-HAD, which uses $ML$ analog phase shifters, 
As such, PC-HAD receivers use $M$ analog phase shifters, in contrast with the $ML$ ones used in FC-HAD.
Since in PC-HADs, the $L$ subarrays are connected to distinct antenna elements, %the analog combiner of the PC-HAD, 
$\boldsymbol{W}_{\mathrm{\!\! A_{PC}}}$ features a block diagonal structure, wherein the $\ell$-th  diagonal block is the vector $\boldsymbol{w}_{{\mathrm{A_{PC}},\ell}}\in\mathbb{C}^{M_{\mathrm{RF}}}$, representing the $M_{\mathrm{RF}}$ phase shifters of   $\ell$-th subarray. % of $\boldsymbol{W}_{A}$ 
Hence, %$\boldsymbol{W}_{\mathrm{\!\! A_{PC}}}$ can be represented as: 
\begin{equation}
\label{eq:W_A_PC}
    \boldsymbol{W}_{\mathrm{\!\! A_{PC}}} 
    =  \bigoplus_{\ell =1}^{L}
        \boldsymbol{w}_{{\mathrm{A_{PC}},\ell}},
\end{equation}
where, similar to their FC counterparts, the entries of  $\boldsymbol{w}_{{\mathrm{A_{PC}},\ell}}\in\mathbb{C}^{M_{\mathrm{RF}}}$ are %non-zero and 
in the form $e^{\jmath\phi}$, $\ell=1,\dots,L$.

%Despite the considerable versatility of hybrid receivers, their 
Although versatile and effective, the performance of HAD receivers is constrained by the unit-modulus analog phase shifters and suffers from a notable weakness due to the reduced dimensionality incurred by the analog combiner. 
In particular, from~\eqref{eq:q}, the analog combiner, $\boldsymbol{W}_{\mathrm{\!\!A}}\in \mathbb{C}^{L\times M}$, projects the received signals onto $L<M$ dimensions. % instead of the full-rank $M$ dimensions. 
This projection results in severe attenuation or even nullification of the signals received from particular DoAs. 
Two approaches for addressing this weakness will be developed in Section~\ref{sec:proposed techs}.
\section{Problem formulation}\label{sec:Problem formulation}
%We have a ULA antenna array with $M$ antennas, $L$ RF chain, and $R$ source signals arriving at angles $\{\theta_r\}_{r=1}^{R}$.
In this section, we consider the practical scenario in which the number of snapshots available to the receiver is too small for statistical averaging to be reliable. In such cases, it might be possible to infer DoA information from the structure embedded in the received signals. One technique that exposes this structure is the MPM, which has been employed in FD receivers~\cite{yilmazer2006utilization, gaber2014study,yilmazer2008multiple}. %~\cite{yilmazer2006utilization, gaber2014study,yilmazer2008multiple}
However, extending it to HAD receivers is fraught with difficulties, which will be addressed in Section~\ref{sec:proposed techs} below. We begin by providing a brief review of matrix pencils,  the application of the MPM to FD receivers and the difficulties encountered in applying it to HAD receivers.

\subsection{Review of the MPM in FD Receiver} \label{sec: MP in FD}
%A matrix pencil refers to 
Consider a pair of square matrices, $(\boldsymbol{A}, \boldsymbol{B})$, 
%where $\boldsymbol{A}$ and $\boldsymbol{B}$ are typically square matrices. 
%These matrices represent a system of linear equations of the form 
satisfying $\boldsymbol{A}\boldsymbol{u}= \zeta \boldsymbol{B}\boldsymbol{u}$, where the non-zero vector, $\boldsymbol{u}$, and the scalar, $\zeta$, are, respectively, a generalized eigenvector and a generalized eigenvalue of the pair of matrices~\cite{golub2013matrix, swillam2009efficient}.
% % In the context of matrix pencil theory, the matrices involved can be either fat (more columns than rows) or tall (more rows than columns), depending on the particular problem you're working with.
% \blu{extension to non square~\cite{boutry2005generalized}}. 
A matrix pencil of $(\boldsymbol{A}, \boldsymbol{B})$ is given by   $\boldsymbol{A} - \zeta \boldsymbol{B}$.
%, where $\boldsymbol{A}$ and $\boldsymbol{B}$ are typically square matrices. This pencil satisfies $\boldsymbol{A}\boldsymbol{v}= \zeta \boldsymbol{B}\boldsymbol{v}$, where the non-zero vector, $\boldsymbol{v}$, and the scalar, $\zeta$, are a generalized eigenvector and a generalized eigenvalue of the matrix pencil, respectively~\cite{golub2013matrix, swillam2009efficient}. 
A scalar, $\zeta$, is a generalized eigenvalue of the pencil if and only if this value reduces the rank of the pencil. Although, originally defined for square pencils, the generalized eigenvalue problem extends to the case in which % to non-square pencils, where 
$\boldsymbol{A}$ and $\boldsymbol{B}$ are tall matrices~\cite{boutry2005generalized}.
%rectangular  $m \times n$  matrices with  $m > n$~\cite{boutry2005generalized}.}
%, thus expanding the range of applications~\cite{boutry2005generalized}.
Matrix pencils appear in many practical applications, including system identification~\cite{sarkar1995using} and DoA estimation in FD receivers~\cite{yilmazer2006utilization, gaber2014study}. 
In these applications, % the unknown parameters 
%after mathematical manipulation, 
the linear equations describing the system are represented by matrix pencils, with generalized eigenvalues in the form of complex exponentials representing the unknown phase parameters. 
To exemplify the philosophy of the MPM, we consider its application to DoA estimation in FD receivers.

%Moreover, the structure of the MP in such applications enables these eigenvalues to be readily inferred using a standard singular value decomposition. To elaborate, to exploit the structure of the MP arising in the aforementioned applications, we consider the case of estimating the DoAs in FD receivers. 

%To estimate the DoA using the MPM in the FD receiver, we construct 
%\blu{%In FD receivers, DoA estimation using the  MPM  
To expose the pencil matrix structure embedded in the baseband signals of an FD receiver, we begin by considering a single snapshot, say 
the $k$-th one, in the noise-free case, whereupon the $m$-th entry of $\boldsymbol{x}_k\in\mathbb{C}^{M}$ in~\eqref{eq:ADC_x_k} is given by:
\begin{equation}
   x_{k}[m]= %\boldsymbol{W}_A^H\boldsymbol{x}[k]=%e^{\jmath 2\pi f_c t}
    \sum_{r=1}^R  s_r[k] e^{\jmath(m-1)\mu_r}, \quad m=1,\ldots,M.
    \label{eq:x_k}
\end{equation}
These entries are parsed into $\xi+1$ overlapped shifted
%overlapped
segments of length 
$M-\xi$ each, where $\xi$ is a design parameter as discussed below~\cite{sarkar1995using}. 
Using these segments,  
%begin by expressing the baseband received signal of the 
a  Hankel matrix,  $\boldsymbol{H}_{k}^{{\mathrm{FD}}}\in \mathbb{C}^{(M-\xi) \times(\xi+1)}$,
%$\boldsymbol{H}_{k}^{{\mathrm{FD}}}$, 
is constructed as: 
\begin{align}
\boldsymbol{H}_{k}^{{\mathrm{FD}}} & =
\scalebox{0.85}{$\begin{bmatrix}
x_k[1]  & \ldots &x_k[\xi+1] \\
% \boldsymbol{x}_k[2] & \boldsymbol{x}_k[3] & \ldots & \boldsymbol{x}_k[\xi+2] \\
\vdots  & \ddots & \vdots \\
x_k[M-\xi]  & \ldots &x_k[M]
\end{bmatrix}$}.
\label{eq:x_n_first}
\end{align}
Substituting for $\{x_k[m]\}_{m=1}^{M}$ from~\eqref{eq:x_k} in~\eqref{eq:x_n_first}, it can be shown that  $\boldsymbol{H}_{k}^{{\mathrm{FD}}}$ possesses a special structure, which can be used to determine the unknown DoAs.
%%%%%%%%%%%%%%%%%%%%%%%%%%%%%%%%%%%%%%%%%%%%%%%%%%%%%%%%%%%
% To show this, we use the notation $\boldsymbol{H}_{k, -\ell}^{\mathrm{FD}}\in\mathbb{C}^{(M-\xi) \times \xi}$ to represents a matrix identical to $\boldsymbol{H}_{k}^{\mathrm{FD}}$ but with the $\ell$-th column removed.
%%%%%%%%%%%%%%%%%%%%%%%%%%%%%%%%%%%%%%%%%%%%%%%%%%%%%%%%%%%%%%%%%%%%%%%
To show this, %wards that end, 
% explain why H_1 formed in Pi lambda Pi >> resembles decomposition but it is not Jordan decomposition as Pi1 is not inverse pi2, we got this structure from the received signals structure and Hankel matrix H,......elaborate 
%By observing the received signals structure in~\eqref{eq:x_k}, the Hankel matrix, as formed in~\eqref{eq:x_n_first}, organizes the complex exponential received signals by the antenna array into a structured matrix. 
%where each element in the matrix corresponds to a time-delayed version of the antenna readings.  
%for each matrix $\boldsymbol{H}_{k}^{{\mathrm{FD}}}$, 
two matrices, $\boldsymbol{H}_{k,1}^{{\mathrm{FD}}}$ and $\boldsymbol{H}_{k, 2}^{{\mathrm{FD}}}$ $\in \mathbb{C}^{(M-\xi) \times \xi}$, are obtained from each matrix $\boldsymbol{H}_{k}^{{\mathrm{FD}}}$. 
In particular, $\boldsymbol{H}_{k,1}^{{\mathrm{FD}}}$ and $\boldsymbol{H}_{k,2}^{{\mathrm{FD}}}$ are obtained by deleting the last and first columns of $\boldsymbol{H}_{k}^{{\mathrm{FD}}}$, respectively. These matrices are used to construct a matrix pencil, which will subsequently yield the unknown DoAs.
Now, it can be readily verified that $\boldsymbol{H}_{k,1}^{{\mathrm{FD}}}$ can be written as: % Next, the 
%
%
%is structured as:
%formed to construct a matrix pencil, from which the generalized eigenvalues are obtained, with  
%the rank of the $\boldsymbol{H}_{k}^{{\mathrm{FD}}}$,  $\boldsymbol{H}_{k,1}^{{\mathrm{FD}}}$ and $\boldsymbol{H}_{k,2}^{{\mathrm{FD}}}$ is determined by the number of independent signal sources $R$.
%These matrices are then used to construct a matrix pencil, from which the generalized eigenvalues are computed.
%The matrices $\boldsymbol{H}_{k,1}^{{\mathrm{FD}}}$ and $\boldsymbol{H}_{k,2}^{{\mathrm{FD}}}$ are obtained by deleting the last and first columns of $\boldsymbol{H}_{k}^{{\mathrm{FD}}}$ respectively. The matrix $\boldsymbol{H}_{k,1}^{{\mathrm{FD}}}$ is constructed as:
%The 
%matrix  $\boldsymbol{H}_{k,1}^{{\mathrm{FD}}}$  is obtained by deleting the last  column of $\boldsymbol{H}_{k}^{{\mathrm{FD}}}$ and is structured as:
\begin{equation}
 \boldsymbol{H}_{k,1}^{{\mathrm{FD}}}=\boldsymbol{\Pi}_1 \boldsymbol{\Lambda}_{k}\boldsymbol{\Pi}_2,\label{eq:H_k_1}
\end{equation}
where $\boldsymbol{\Pi}_1 \in \mathbb{C}^{(M-\xi) \times R}$,  $\boldsymbol{\Pi}_2 \in \mathbb{C}^{R \times \xi}$, and $\boldsymbol{\Lambda}_{k} \in \mathbb{C}^{R \times R}$ are: %given by:
\begin{align}
\boldsymbol{\Pi}_1 & =
\scalebox{0.85}{$
\begin{bmatrix}
1 & \ldots &  1 \\
e^{\jmath  \mu_1 }  & \ldots &  e^{\jmath  \mu_{R} } \\
\vdots & \ddots & \vdots \\
e^{\jmath (M- \xi-1)  \mu_1 } &  \ldots &  e^{\jmath  (M- \xi-1)  \mu_{R}} 
\end{bmatrix}$} ,
\label{eq:Z_1_fd}
\end{align}
%------------------------------------------------------------------
\begin{align}
\boldsymbol{\Pi}_2 &=
\scalebox{0.85}{$
\begin{bmatrix}
1 & e^{\jmath  \mu_1 } & \ldots &  e^{\jmath (\xi-1)  \mu_1 } \\
\vdots & \vdots & \ddots & \vdots \\
1 & e^{\jmath   \mu_{R} } & \ldots &  e^{\jmath (\xi-1)  \mu_{R} } 
\end{bmatrix}$},
\label{eq:Z_2_fd}
\end{align}
%-------------------------------------------------------------------
\begin{align}
& \boldsymbol{\Lambda}_{k}\!\!=\operatorname{diag}\bigl(s_{1}[k] , \ldots, s_{R}[k] \bigr).
\label{eq:R_n_fd}
\end{align}
It is noteworthy that expression in~\eqref{eq:H_k_1} %The matrix $\boldsymbol{H}_{k,1}^{{\mathrm{FD}}}$ 
resembles a singular value decomposition~(SVD)  of $\boldsymbol{H}_{k,1}^{{\mathrm{FD}}}$. However, $\boldsymbol{\Pi}_1$ and $\boldsymbol{\Pi}_2$ have a Vandermonde, rather than a unitary structure,  and $\boldsymbol{\Lambda}_{k}$ has complex, rather than positive real diagonal entries. 
%Using a similar approach
Similar to $\boldsymbol{H}_{k,1}^{{\mathrm{FD}}}$, the matrix $\boldsymbol{H}_{k,2}^{{\mathrm{FD}}}$
can be expressed as:
\begin{align}
{\boldsymbol{H}_{k,2}^{{\mathrm{FD}}}=\boldsymbol{\Pi}_1\boldsymbol{\Lambda}_{k}\boldsymbol{\Pi}_0\boldsymbol{\Pi}_2,}
\label{eq:H_k_2}
\end{align}
where 
%ich shares the same matrices as $\boldsymbol{H}_{k,1}^{{\mathrm{FD}}}$ with the addition of a diagonal matrix $\boldsymbol{\Pi}_0 \in \mathbb{C}^{R \times R}$, is defined as:
\begin{equation}
 \boldsymbol{\Pi}_0 =\operatorname{diag}\big(e^{\jmath \mu_1 },\ldots, e^{\jmath  \mu_{R} }\big).\label{eq:Z_0_fd}
\end{equation}
%will be used to extract the DoAs from the received signals
This matrix %captures the \blu{difference} between 
accounts for the deleted columns from
$\boldsymbol{H}_{k,1}^{{\mathrm{FD}}}$ and $\boldsymbol{H}_{k,2}^{{\mathrm{FD}}}$, and will enable the DoAs, $\{\mu_r\}_{r=1}^R$, to be expressed as the generalized eigenvalues of the matrix pencil constructed from these matrices. In particular, using~\eqref{eq:H_k_1} and~\eqref{eq:H_k_2}, the matrix pencil $\boldsymbol{H}_{k,2}^{{\mathrm{FD}}} - \zeta \boldsymbol{H}_{k,1}^{{\mathrm{FD}}}$ can be written as:
\begin{equation}
\boldsymbol{H}_{k,2}^{{\mathrm{FD}}} - \zeta \boldsymbol{H}_{k,1}^{{\mathrm{FD}}}
= \boldsymbol{\Pi}_1 \boldsymbol{\Lambda}_{k} \bigl( \boldsymbol{\Pi}_0 - \zeta \boldsymbol{I}_{\mathrm{R}} \bigr) \boldsymbol{\Pi}_2. 
\label{eq:generalized_eignvalue_fd}
\end{equation}
% \begin{equation}
% \big[\boldsymbol{H}^{(k)}_2\big] - \lambda \big[\boldsymbol{H}^{(k)}_1\big]
% = \big[\boldsymbol{\Pi}_1\big] \big[\boldsymbol{R}^{(k)}\big] \Bigl\{ \big[\boldsymbol{\Pi}_0\big] - \lambda \big[\boldsymbol{I}\big] \Bigr\} \big[\boldsymbol{\Pi}_2 \big],
% \label{eq:generalized_eignvalue}
% \end{equation}
% If you obtain fewer than R generalized eigenvalues 
% , it implies that the matrix pencil has a reduced rank. This typically happens because The matrix pencil is not capture all 
% R independent sources due to noise or insufficient dimensionality.
% Some sources may be correlated or have insufficient separation, causing the matrix rank to drop below R, which leads to fewer eigenvalues.
%---------------------------------------------------------------------------------------
% The rank of the $\boldsymbol{H}_{k}^{{\mathrm{FD}}}$,  $\boldsymbol{H}_{k,1}^{{\mathrm{FD}}}$ and $\boldsymbol{H}_{k,2}^{{\mathrm{FD}}}$ is determined by the number of independent signal sources $R$, provided that $R\leq \xi_{_\mathrm{FD}} \leq M-R$ \cite{hua1990matrix}.
%-----------------------------------------------------------------------------------------
%\blu{
For distinct DoAs, and $\xi\in [R,M-R]$~\cite{hua1990matrix}, the rank of the Vandermonde matrices,  $\boldsymbol{\Pi}_1$ and  $\boldsymbol{\Pi}_2$, and %for independent source signals, 
the rank of $\boldsymbol{\Lambda}_{k}$ is $R$. Hence, $\operatorname{rank}(\boldsymbol{H}_{k,2}^{{\mathrm{FD}}} - \zeta \boldsymbol{H}_{k,1}^{{\mathrm{FD}}})=R$,  unless %, provided that $R\leq \xi \leq M-R$ \cite{hua1990matrix}.  
$\zeta =e^{\jmath  \mu_{r}}$, for some  $r\in\{1,\ldots, R\}$. In that case, the rank of the matrix pencil will be reduced by one, $\operatorname{rank}(\boldsymbol{H}_{k,2}^{{\mathrm{FD}}} - \zeta \boldsymbol{H}_{k,1}^{{\mathrm{FD}}})=R-1$. 
Using this observation, it can be readily seen that the diagonal elements of $\boldsymbol{\Pi}_0$ are the generalized eigenvalues of the matrix pair  $(\boldsymbol{H}_{k,2}^{{\mathrm{FD}}},\boldsymbol{H}_{k,1}^{{\mathrm{FD}}})$, which %. These eigenvalue 
are the eigenvalues of %the matrix
%solving the following eigenvalue problem:
% \begin{equation}
%     \boldsymbol{H}_{k,1}^{\dag} \boldsymbol{H}_{k,2}= (\boldsymbol{H}_{k,1}^H\boldsymbol{H}_{k,1})^{-1}\boldsymbol{H}_{k,1}^H\boldsymbol{H}_{k,2}
% \end{equation}
\begin{equation}
    (\boldsymbol{H}_{k,1}^{\mathrm{FD}})^{\dag} \boldsymbol{H}_{k,2}^{\mathrm{FD}} = \Bigl( (\boldsymbol{H}_{k,1}^{\mathrm{FD}})^{H} \boldsymbol{H}_{k,1}^{\mathrm{FD}} \Bigr)^{-1} (\boldsymbol{H}_{k,1}^{\mathrm{FD}})^{H} \boldsymbol{H}_{k,2}^{\mathrm{FD}}.
    \label{eq:eignval-prob}
\end{equation}
% the phases of the diagonal elements of $\boldsymbol{\Pi}_0$ correspond to the phases of the unknown DoAs. 
%Equivalently, the problem of solving for $z_i$ can be cast as an ordinary eigenvalue problem,
%\begin{equation}
%    \Bigl\{\left[\boldsymbol{Y}_1\right]^{\dag}\left[\boldsymbol{Y}_2\right]-\lambda[\boldsymbol{I}]\Bigr\},
%\end{equation}
Using this notation, the $r$-th eigenvalue of  $(\boldsymbol{H}_{k,1}^{\mathrm{FD}})^{\dag} \boldsymbol{H}_{k,2}^{\mathrm{FD}}$ is given by $[\boldsymbol{\Pi}_0]_{r,r}$, which, using~\eqref{eq:mu}, yields that: 
%Following that, the  DoAs for FD receiver can be estimated by: % there are no grating lobes in this method.
\begin{equation}
   \Hat{\theta}_r=\arcsin \biggl( \frac{\lambda \Im \{\log{[\boldsymbol{\Pi}_0}]_{r,r}\}}{2 \pi \Delta} \biggr).
   % \arcsin \Biggl \{ \frac{\lambda \operatorname{\Im }\{\log {e^{\jmath  \mu_{r} }}\}}{2 \pi \Delta} \Biggr\}.
   \label{eq:theta_est_fd_1}
\end{equation}

%\textbf{conclude that we can estimate the DoA using one snapshot, then introduce the estimation with the existing noise.}
The aforementioned discussion shows that, in the absence of noise,  MPM enables the DoAs of multiple sources to be extracted from a single snapshot, i.e., $K=1$. Next, we 
%From~\eqref{eq:theta_est_fd_1}, 
%From the previous analysis, it is readily shown that DoA can be estimated using a single snapshot, i.e., $k=1$, through the MPM. 
%We now 
discuss using MPM for DoA estimation in the presence of noise.
\subsubsection*{Using MPM in the Presence of Noise}\label{sec:SVD}
% Add a paragraph for the signal in 11 with noise. Calculate the SNR for it to be compared with the SNR of PMPM approach
In the absence of noise,  the rank of $\boldsymbol{H}_{k}^{{\mathrm{FD}}}$ is $R$.  However, in the presence of independent noise components, its rank becomes $\min(\xi+1, M-\xi)$, where $\xi\in [R,M-R]$. (Choosing $\xi \in [\tfrac{M}{3}, \tfrac{M}{2}]$ has been recommended in~\cite{hua1990matrix} to minimize noise variance.) % and e estimated parameters~\cite{hua1990matrix}.
To reduce the effect of noise, the SVD of  $\boldsymbol{H}_{k}^{\mathrm{FD}}$ can be used to %into its signal subspace and noise subspace, and subsequently 
discard the components associated with the noise subspace~\cite{sarkar1995using, yilmazer2006utilization, gaber2014study}.
In particular,  $\boldsymbol{H}_{k}^{\mathrm{FD}}$ can be decomposed as  $\boldsymbol{U} \boldsymbol{\Sigma} \boldsymbol{V}^H$, where $\boldsymbol{U}$ and $\boldsymbol{V}$ are unitary matrices, and $\boldsymbol{\Sigma}$ is a diagonal matrix of singular values.
The left and right singular vectors in $\boldsymbol{U}$ and $\boldsymbol{V}$, corresponding to the $R$ dominant singular values, in $\boldsymbol{\Sigma}$, are retained and the remaining singular
vectors and the corresponding singular values are discarded~\cite{sarkar1995using}. These matrices are denoted by $\hat{\boldsymbol{U}}$, $\hat{\boldsymbol{V}}$ and $\hat{\boldsymbol{\Sigma}}$. 
Using these matrices, %the matrix 
$\hat{\boldsymbol{H}}_{k}^{ \mathrm{FD}}$ is obtained as $\hat{\boldsymbol{U}} \hat{\boldsymbol{\Sigma}} \hat{\boldsymbol{V}}^{H}$, from which the  DoAs can be estimated
using  the approach in Section~\ref{sec: MP in FD}.
We now consider using MPM for DoA estimation in HAD receivers.

\subsection{Applying the MPM in HAD Receivers}\label{sec: MP in HAD} 
% Use of 
To expose the pencil structure of the baseband signals of the HAD receiver, we consider the analog combiner output signal $\boldsymbol{Q}_n$ in~\eqref{eq:Q_n}. %, after applying the analog combiner. We write 
The $k$-th column of $\boldsymbol{Q}_n$ %in~\eqref{eq:Q_n}, which 
corresponds to the $k$-th snapshot of the $n$-th segment and is given by: 
%the $k$-th snapshot $\boldsymbol{Q}$ in~\eqref{eq:X_tilde} as:
%in the noise-free case,
%The received signal vector after applying the analog combiner, 
%we write the $k$-th column of $\boldsymbol{Q}$ in~\eqref{eq:X_tilde} as:
\begin{equation}
    \boldsymbol{q}_{k}= %\boldsymbol{W}_A^H\boldsymbol{x}[k]=%e^{\jmath 2\pi f_c t}
    \sum_{r=1}^R\boldsymbol{W}_{\!\!\mathrm{A},n}^H\boldsymbol{a}_r s_r[k] + \boldsymbol{W}_{\!\!\mathrm{A},n}^H\boldsymbol{z}[k].
    \label{eq:q_n}
\end{equation}
The $\ell$-th entry of the vector $\boldsymbol{q}_{k}$,  $q_{k}[\ell]$, is given by: 
\begin{equation}
    q_{k}[\ell] = %e^{\jmath 2\pi f_c t}
    \sum_{r=1}^R\boldsymbol{w}_{\mathrm{A},\ell}^H\boldsymbol{a}_r s_r[k] + \boldsymbol{w}_{\mathrm{A},\ell}^H\boldsymbol{z}[k],
    \label{eq:q_n_ell}
\end{equation}
where $\boldsymbol{w}_{\mathrm{A},\ell}$ is the $\ell$-th column of $\boldsymbol{W}_{\!\!\mathrm{A},n}$ either  FC-HAD or PC-HAD. %, i.e. $\boldsymbol{W}_{\!\!\mathrm{A}}$. 
% For PC-HAD, the $\ell$-th column is chosen from the $\ell$-th column of the $M_{\mathrm{RF}}$-point DFT matrix, i.e., $\boldsymbol{w}_{\mathrm{A},\ell}=[\boldsymbol{0}_{(\ell-1)M_{\mathrm{RF}}}^T \boldsymbol{v}_\ell^T\boldsymbol{0}_{M-\ell M_{\mathrm{RF}}}^T]^T$, where 
The $m$-th entry of $\boldsymbol{w}_{\!\mathrm{A},\ell}$ is given by $e^{\jmath (m-1)\phi_\ell}$, $\ell,m=1,\ldots,M_{\mathrm{RF}}$ and $\phi_\ell$ denotes the phase shift of the analog phase shifters in the $\ell$-th subarray.
%is phase shift related to each RF chain which is based on the design of the analog combiner.
Using this notation, we have:
\begin{align}
\!\!q_{k}[\ell] \!&=\! 
  \sum_{r=1}^R\sum_{m=1}^{ M_{\mathrm{RF}}} s_r[k]e^{\jmath(\mu_r(m-1)+\psi_{r,\ell})}e^{-\jmath\phi_\ell(m-1)}
    + \boldsymbol{w}_{\mathrm{A},\ell}^H\boldsymbol{z}[k] \nonumber\\
 & \!\!\!\!\!\!\!\!\!\!\!\!\!=\!  
    \alpha\sum_{r=1}^R s_r[k] g(\mu_r-\phi_\ell)
   e^{\jmath\psi_{r,\ell}}
+\boldsymbol{w}_{\mathrm{A},\ell}^H\boldsymbol{z}[k], %\ell=1,\ldots,L,
%\nonumber \\ &\qquad \qquad \qquad \qquad \qquad \qquad \qquad  \ 
  \ell=1,\ldots,L,
\label{eq:q_n_ell_sec}
\end{align}
where
%\begin{equation}
%    f_\ell(\mu)=\frac{\sin{(\frac{ M_{RF}}{2}(\mu-\phi_\ell))}}{\sin{(\frac{1}{2}(\mu-\phi_\ell))}},
%  \label{eq:f_ell_mu}
%\end{equation}
\begin{equation}
    g(\mu-\phi_\ell)
    % Exponential Sum or Geometric Series Sum
    %= \sum_{m=1}^{ M_{\mathrm{RF}}} e^{\jmath(m-1)(\mu-\phi_\ell)}
    =\frac{\sin{(\frac{ M_{{\mathrm{RF}}}}{2}(\mu-\phi_\ell))}}{\sin{(\frac{1}{2}(\mu-\phi_\ell))}}e^{\jmath\bigl(\frac{(M_{{\mathrm{RF}}}-1)(\mu-\phi_\ell)}{2}\bigr)},
  \label{eq:f_ell_mu}
\end{equation}
%%%%%%%%%%%%%%%%%%%%%%%%%%%%%%%%%%%%%%%%%%%%%%%%%%%%%%%
% \begin{align}
% &\! q_{k}[\ell] \!  = %e^{\jmath 2\pi f_c t}
%     \!\alpha\!\sum_{r=1}^R\sum_{m=1}^{ M_{{\mathrm{RF}}}} s_r[k]e^{\jmath(\mu_r(m-1)+\psi_{r,\ell})}e^{-\jmath\phi_\ell(m-1)}%\nonumber\\&\quad\quad\quad\quad\quad\quad
%    \! +\! \boldsymbol{w}_{\!{\mathrm{A}},\ell}^H\boldsymbol{z}[k]\nonumber\\
%   &\!= %e^{\jmath 2\pi f_c t}
%     \!\alpha\!\sum_{r=1}^Rs_r[k]f_\ell(\mu_r)
%    % \frac{\sin{(\frac{M_{RF}}{2}(\mu_r-\phi_\ell))}}{\sin{(\frac{1}{2}(\mu_r-\phi_\ell))}}
%    e^{\jmath\bigl(\frac{(M_{{\mathrm{RF}}}-1)(\mu_r-\phi_\ell)}{2}+\psi_{r,\ell}\bigr)}
%    %\nonumber\\
%    % &
% %\qquad
% \!+\!\boldsymbol{w}_{\!{ {\mathrm{A}},\ell}}^H\boldsymbol{z}[k],\label{eq:q_n_ell_sec}
% \end{align} 
% where
% \begin{equation}
%     \label{eq:f_ell_mu}
%     f_\ell(\mu)=\frac{\sin{(\frac{ M_{{\mathrm{RF}}}}{2}(\mu-\phi_\ell))}}{\sin{(\frac{1}{2}(\mu-\phi_\ell))}},
% \end{equation}
%%%%%%%%%%%%%%%%%%%%%%%%%%%%%%%%%%%%%%%%%%%%%%%%%%%%%%%%%%%%%%%
where $\mu_r=\mu(\theta_r)$, cf.~\eqref{eq:mu}.
%%, and $\psi_{r,\ell}$ depends on the HAD configuration, as will be shown in the next section. 
% The scalar $\alpha=1$ for PC-HAD and $\alpha=\frac{1}{\sqrt{L}}$ in FC-HAD, due to the power splitting inherent in this architecture. 
% $\psi_\ell=0$ for FC-HAD
% $\psi_{r,\ell}=(\ell-1)\mu_r M_{\mathrm{RF}}$ for PC-HAD. 
%For PC-HAD, the scalar $\alpha = 1$ and $\psi_{r,\ell} = (\ell - 1) \mu_r M_{\mathrm{RF}}$, while for FC-HAD, $\alpha = \frac{1}{\sqrt{L}}$ due to power splitting in the architecture, and $\psi_\ell = 0$.
For PC-HAD, the scalar $\alpha = 1$  and the phase $\psi_{r,\ell} = (\ell - 1) \mu_r M_{\mathrm{RF}}$, whereas for FC-HAD, the scalar $\alpha = \frac{1}{\sqrt{L}}$ due to power splitting, % inherent in this architecture, 
and the phase $\psi_{r,\ell} = 0$.
%, while for FC-HAD, $\alpha = \frac{1}{\sqrt{L}}$ due to power splitting inherent in this architecture. For PC-HAD, $\psi_{r,\ell} = (\ell - 1) \mu_r M_{\mathrm{RF}}$, whereas for FC-HAD, $\psi_\ell = 0$.
%, and $\psi_{r,\ell}$ depends on the HAD configuration, as will be shown in the next section. 

%Two main challenges arise in the HAD receivers:  the first one when applying the MPM to extract the DoAs
%, and the second one is the limitations introduced by the analog combiner.
%the second is the limitations introduced by the analog combiner used in these receivers.
%First, to proceed with 
We now elaborate on the difficulties encountered when
%consider the potential of 
%First, when 
applying MPM for DoA estimation in HAD receivers. Towards that end, one would be tempted to follow a path analogous to that used for FD receivers. To show that HAD receivers are not amenable to this path, we note that, in FD receivers, the DoA-bearing signals $\{x_{k}[m]\}_{m=1}^M$ in~\eqref{eq:x_k} are directly available for baseband processing. In contrast, in HAD receivers, only the signals $\{q_{k}[\ell]\}_{\ell=1}^L$ in~\eqref{eq:q_n_ell_sec}, the output of the analog combiner, are accessible for baseband processing.
Comparing the structure of $x_{k}[m]$ with that of $q_{k}[\ell]$, we observe that, whereas the baseband signals, $\{s_r\}_{r=1}^R$, and the DoAs embedded in the phases $\{\mu_r\}_{r=1}^R$  of the $R$ sources are directly available in  $\{x_{k}[m]\}_{m=1}^M$, these signals and phases are not directly recoverable from the signals  %\blu{$q_{k}[\ell]$}. 
$\{q_{k}[\ell]\}_{\ell=1}^L$. 
To see that, we note that, in  $q_{k}[\ell]$, the baseband signals, $\{s_r\}_{r=1}^R$, are scaled  by $\alpha |g(\mu_r - \phi_{\ell})|$ and the phases  $\{\mu_r\}_{r=1}^R$   are shifted by  
the phase shifts,  $\{\phi_{\ell}\}_{\ell=1}^L$, of the analog combiner.
As such, following the path used to apply the MPM to FD receivers in their HAD counterparts would require constructing a Hankel matrix with $q_{k}[\ell]$ signals in~\eqref{eq:q_n_ell_sec} analogous to the FD one in~\eqref{eq:x_n_first}.  
This matrix must be amenable to a decomposition analogous to the one in~\eqref{eq:x_n_first}
and the unknown phases %\blu{${\mu_r}$} 
$\{\mu_r\}_{r=1}^R$
must be amenable to be isolated in a diagonal matrix analogous to $\boldsymbol{\Pi}_0$ in~\eqref{eq:Z_0_fd}. Unfortunately, when the phase shifts of the analog combiner are not identical, the Hankel matrix constructed from the %\blu{$q_{k}[\ell]$} 
$\{q_{k}[\ell]\}_{\ell=1}^L$. 
signals cannot be decomposed in a manner analogous to the ones in~\eqref{eq:H_k_1} and~\eqref{eq:H_k_2}. 
In particular, the signal on which MPM is applied depends 
not only on the unknown phases $\{\mu_r\}_{r=1}^R$ but also on the phases $\{\phi_{\ell}\}_{\ell=1}^L$.  
%These phases distort the structure of the matrices, making it infeasible to obtain the proper arrangement for constructing a valid matrix pencil.
These phases disrupt the matrix structure, rendering the 
construction of a pencil akin to the one used in FD receivers infeasible. % in the required form.
Hence, using matrix pencils to extract the unknown DoAs in HAD receivers requires fundamentally different approaches.
%in HAD receivers novel approaches must be used to extract the unknown DoA information in HAD receivers. % differs fundamentally from the corresponding technique used in FD receivers.
%fundamentally differs from the approach used in FD receivers.
% , 
% such as the ones in~\eqref{eq:H_k_1} and~\eqref{eq:H_k_2} for constructing a valid matrix pencil.
% pencil that depends, not only on 
% ]To appreciate the implication of this difference, we note  implies that the signal on which MPM is applied depends 
% not only on the unknown phases $\{\mu_r\}_{r=1}^R$ but also on the phases $\{\phi_{\ell}\}_{\ell=1}^L$.

%In addition to the difficulties encountered in applying MPM to HAD receivers, 
In addition to the difficulties of applying MPM to HAD receivers,
the HAD structure suffers from another limitation~\cite{Mona2023doa}, which follows from 
%Second, in the HAD receiver, the received signal is initially
processing  the received signals  by the $M \times L$ analog combiner matrix $\boldsymbol{W}_{\!\!\mathrm{A},n}$, cf.~\eqref{eq:Q_n}. Since in these receivers $L < M$, it can be seen that the $\boldsymbol{W}_{\!\!\mathrm{A},n}$ projects the received signal vector on a particular subspace spanned by the column vectors of $\boldsymbol{W}_{\!\!\mathrm{A},n}$.
Thus, an analog combiner defines spatial sectors for
which the DoAs can be estimated.  Outside these sectors, the received signal would be heavily attenuated or even nullified. 
To address these challenges,  we next propose two approaches centered around the MPM.

%Proposed Approaches for DoA Estimation Enabling MPM in HAD Receivers
%Proposed Approaches for DoA Estimation to Enable MPM in HAD Receivers
\section{Proposed MPM-based Approaches for DoA Estimation in HAD Receivers}\label{sec:proposed techs}
In this section, we develop two MPM-based approaches for DoA estimation in HAD receivers. These approaches will result in Hankel matrices that can used to extract the DoAs using the MPM. These approaches will also eliminate the potential of heavy attenuation which the received signals may be subjected to, otherwise. 
We begin by considering the noiseless case with a single snapshot and then extend these approaches to multiple snapshots in the presence of noise.
%%%%%%%%%%%%%%%%%%%%%%%%%%%%%%%%%%%%%%%%%%%%%%%%%%%%%%%%%%

%-------------------------------------------------------------------------
\subsection {Full Coverage and Periodicity-Based MPM~(PMPM)}\label{sec:PMP-set-analog-combiner-alg} %(MPTS- FSC)
% Training-Enabled Matrix Pencil: Full Space Coverage (TEMP-FSC)
% Matrix Pencil with Training: Full-Space Combiner Coverage (MPT-FSCC)
% Matrix Pencil and Exhaustive Analog Combiners Algorithm
%%%%%%%%%%%%%%%%%%%%%%%%%%%%%%%%%%%%%%%%%%%%%%%%%%%%%%%%%%%%%%%%%%%%%%%

% MP and Exhaustive set of Analog Combiners along with side information~(MPACs)
%Matrix Pencil with a set of Analog Combiners
%Multi-Sector Aggregation Algorithm
% write about the different cases for the PC-HAD for L \leq M_{RF} & L > M_{RF}
% Use the projection matrix for each analog combiner and add all of these signals by using the preamble signals.
%As discussed in Section~\ref{sec:Problem formulation}, the analog combiner in the HAD receiver projects received signals onto a subspace, causing significant attenuation for particular DoAs. In addition, to remedy the entanglement to the received signals caused by the HAD receivers once applying the MPM.
% In this section, we introduce an approach that addresses the attenuation incurred by the analog combiner by cycling over an exhaustive set of analog combiners, 
%each constructed from the columns of the DFT matrix.
%that resembles the one in~\blu{~\cite{Mona2023doa}}. 
%The set of analog combiners 
The first proposed approach addresses the attenuation incurred by the analog combiner by cycling over an exhaustive set of analog combiners that collectively span the entire space.
This approach then uses the periodic, potentially unknown, deterministic components, such as training pilots or preamble segments of the received signal frames, typically used in communication protocols~\cite{tse2005fundamentals}.
%, to assist in data recovery~\cite{tse2005fundamentals}.
We will show that access to such signals can be used to combine and disentangle the signals at the outputs of the analog combiners. Hence, HAD receivers, either FC or PC, can feature behavior that resembles that of FD receivers, and DoA can be estimated using the MPM. 
We begin with a scheme to ensure full spatial coverage.

%%%%%%%%%%%%%%%%%%%%%%%%%%%%%%%%%%%%%%%%%%%%%%%%%%%%%%%%%%%%%%%%%%%%%%%
% for each analog combiner we will define the spatial sectors.
\subsubsection{A Set of Exhaustive  Analog Combiners}\label{sec:set-analog-combiner}
% In our previous work \cite{Mona2023doa}, it was demonstrated that, for PC-HAD receivers with $L \ge M_{\mathrm{RF}}$, one analog combiner is sufficient to cover all DoAs within $\bigl[\frac{-\pi}{2},\frac{\pi}{2}\bigr]$. 
%Conversely, in the case of PC-HAD 
%In the HAD receivers with $L < M_{\mathrm{RF}}$, only the DoAs that lie in the subspace spanned by the analog combiner,
In HAD receivers, the analog combiner represents a spatial sector within which the DoAs can be estimated. Received signals with DoAs 
outside this sector will be attenuated, or even nullified.
To avoid this limitation, we follow the scheme advocated in~\cite{Mona2023doa}. In this scheme, an exhaustive set of analog combiners
%, $\{\boldsymbol{W}_{\!\!{\mathrm{A}},n}\}_{n=1}^N$, 
is used to collectively span the entire space. 
% Each $\boldsymbol{W}_{\mathrm{\!\! A},n}$ processes the $n$-th segment of $K$ snapshots, $n=1,\ldots,N$, 
% where $N= \tfrac{M}{L}$, and the total number of snapshots is $NK$.
It was argued in~\cite{Mona2023doa} that, in order to span a subspace with the largest possible dimension, the matrix $\boldsymbol{W}_{\!\!\mathrm{A}}$  must be full column rank, i.e., $L$. 
% Each $\boldsymbol{W}_{\mathrm{\!\! A},n}$ will processes the $n$-th segment of $K$ snapshots, $n=1,\ldots,N$, where $N= \tfrac{M}{L}$, and the total number of snapshots is $NK$.
Towards that end, we construct the analog combiners of FC-HAD receivers from distinct columns of the DFT matrix, and we construct those of  PC-HAD receivers using a single column of the  DFT matrix for each of those combiners. This construction of PC-HAD analog combiners contrasts that proposed in~\cite{Mona2023doa}, wherein the  diagonal blocks %of the PC-HAD analog combiners proposed in~\cite{Mona2023doa} 
are constructed from distinct, rather than identical, columns of the DFT matrix. Using identical columns of the DFT matrix in the  PC-HAD analog combiners will enable using the  MPM to extract the DoAs embedded in the received signals, cf. Sections~\ref{sec:PMP}--\ref{sec:SPC} below.
%In , are each constructed from a single column of the  DFT matrix.\footnote{ 
%The rationale for this design  will be clarified  in Section~\ref{sec:PMP}.}
To elaborate, we consider the $M_{\mathrm{RF}}$-point DFT matrix which is denoted by $\boldsymbol{V}_{\!\!M_{\mathrm{RF}}}= [\boldsymbol{v}_1,\ldots,\boldsymbol{v}_{\!M_{\mathrm{RF}}}]$, where 
%the $m$-th entry of $\boldsymbol{v}_\ell\in\mathbb{C}^{M_{\mathrm{RF}}}$ is given by $e^{\frac{\jmath 2\pi(\ell-1)(m-1)}{M_{\mathrm{RF}}}}$, 
the $m$-th entry of $\boldsymbol{v}_\ell$ is given by 
$e^{\jmath (m-1)\phi_\ell}$, $\phi_\ell=\frac{2\pi(\ell-1)}{M_{\mathrm{RF}}}$, $\ell,m=1,\ldots,M_{\mathrm{RF}}$. Using $\boldsymbol{V}_{\!\!M_{\mathrm{RF}}}$ to construct the HAD analog combiners ensures not only that all non-zero entries are unit modulus, but also that analog combiners are full column rank, %. This choice ensures that the columns are orthogonal, 
which will facilitate subsequent mathematical manipulation. %ty among columns. 
% Such a choice has several advantages, which include being well-conditioned, and the ease with which the spatial sector spanned by each analog combiner can be determined.

For FC-HAD, $M_{\mathrm{RF}}=M$. In this case, an exhaustive set of analog combiners, $\{\boldsymbol{W}_{\mathrm{\!\! A_{FC}},n}\}_{n=1}^N$, $N= \tfrac{M}{L}$, can be constructed 
%from the $M_{\mathrm{RF}}$-DFT matrix such that each analog combine $L$ distinct columns of the  DFT matrix. In particular,  
by choosing each % matrix
$\boldsymbol{W}_{\mathrm{\!\! A_{FC}},n}$ to contain $L$ distinct columns of the $M_{\mathrm{RF}}$-point DFT matrix, for $n\in\{1,\ldots,N\}$, i.e., 
\begin{equation}
    \boldsymbol{W}_{\mathrm{\!\! A_{FC}},n} = \frac{1}{\sqrt{L}}\begin{bmatrix}
        \boldsymbol{v}_{{(n-1)L+1}} & \cdots & \boldsymbol{v}_{nL}
    \end{bmatrix}.
    \label{eq:W_A_FC_k}
\end{equation}
%\eqref{eq:W_A_FC_k}.
%$(n-1)L+1$ to $nL$. 
The matrix $\boldsymbol{W}_{\mathrm{\!\! A_{FC}},n}$  processes the $n$-th segment containing $K$ snapshots, assuming that the total number of snapshots is $NK$.
% , $n=1,\ldots,N$, where $N= \tfrac{M}{L}$, and the total number of snapshots is $NK$.
% For each $\boldsymbol{W}_{\mathrm{\!\! A_{FC}},n}$, the $\ell$-th column of the analog combiner  %for FC-HAD, cf., \eqref{eq:W_A_FC}, 
% is selected to be a scaled version of the $\ell$-th column of the $M_{\mathrm{RF}}$-point DFT matrix, i.e., $\org{\boldsymbol{w}_{{\mathrm{A_{FC}},\ell}}}=\alpha\boldsymbol{v}_{{\!\ell}}$, where  the $m$-th entry of $\boldsymbol{v}_\ell$ is given by 
% $e^{\jmath (m-1)\phi_\ell}$, $\phi_\ell=\frac{2\pi(\ell-1)}{M_{\mathrm{RF}}}$, $\ell,m=1,\ldots,M_{\mathrm{RF}}$.
%%%%%%%%%%%%%%%%%%%%%%%%%%%%%%%%%%%%%%%%%%%%%%%%%%%%%%%%%
The set  of phases used by $\boldsymbol{W}_{\mathrm{\!\!A_{FC}},n}$, is 
% \begin{equation}\label{eq:V_n}
%    \mathcal{V}_{n}=\!\begin{cases}
%        \Bigl\{\frac{2\pi(n-1)L}{M_{\mathrm{RF}}},\dots,\frac{2\pi(nL-1)}{M_{\mathrm{RF}}}\Bigr\}, n\in\{1,\ldots,\frac{N}{2}\}, \nonumber \\
%         \Bigl\{\frac{2\pi(n-1)L}{M_{\mathrm{RF}}}-2\pi,\dots,\frac{2\pi(n L-1)}{M_{\mathrm{RF}}}-2\pi\Bigr\}, n\in\{\frac{N}{2}+1,\ldots,N\}.
%     \end{cases}
%    \end{equation} 
%. It can be  verified that
$\mathcal{V}_{n} =\Bigl\{\frac{2\pi(n-1)L}{M_{\mathrm{RF}}},\dots,\frac{2\pi(nL-1)}{M_{\mathrm{RF}}}\Bigr\}$ for $n\in\{1,\ldots,\frac{N}{2}\}$ and $\mathcal{V}_{n}=\Bigl\{\frac{2\pi(n-1)L}{M_{\mathrm{RF}}}-2\pi,\dots,\frac{2\pi(n L-1)}{M_{\mathrm{RF}}}-2\pi\Bigr\}$ for $n\in\{\frac{N}{2}+1,\ldots,N\}$. 
% Using~\eqref{eq:W_A_FC}, the $n$-th  analog combiner, $ \boldsymbol{W}_{\mathrm{\!\! A_{FC}},n}$, %is given by: %in~\eqref{eq:W_A_FC}, i.e.,
% for $n=1,\ldots,N$. %For the $n$-th analog combiner, substituting $ \boldsymbol{W}_{\mathrm{\!\! A_{FC}},n}$ in~\eqref{eq:q_n} yields $\psi_{r,\ell}=0$ in~\eqref{eq:q_n_ell_sec}.
%Using the orthogonality of the DFT matrix $\boldsymbol{V}_{\!\!M_{\mathrm{RF}}}$, it can be readily verified that the $n$-th analog combiner lies in the nullspace of the other analog combiners, i.e., 
%$\boldsymbol{W}_{\mathrm{\!\! A_{FC}},n}\in\mathcal{N}\bigl\{\{\boldsymbol{W}_{\mathrm{\!\! A_{FC}},j}\}_{j\neq n}\bigr\}$.
This set of  analog combiners %in the set $\{\boldsymbol{W}_{\mathrm{\!\! A_{FC}},n}\}_{n=1}^{N}$ 
spans the entire space. % thereby 
%thereby all DoAs in $\bigl[\frac{-\pi}{2},\frac{\pi}{2}\bigr]$ can be estimated.

%%-----------------------------------------------------------------------

%\subsubsection*{PC-HAD}\label{sec:PC-HAD-DFT}

For PC-HAD, $M_{\mathrm{RF}}=\tfrac{M}{L}$ and the analog combiners possess a block diagonal structure. Hence, in this case,  it suffices for an exhaustive set of analog combiners, $\{\boldsymbol{W}_{\mathrm{\!\! A_{PC}},n}\}_{n=1}^N$, $N= \tfrac{M}{L}$, to be such that each of the matrices in the set is constructed from a single column of the  DFT matrix. 
%In particular,  the  $L$ columns of the matrix $\boldsymbol{W}_{\mathrm{\!\! A_{PC}},n}$ are configured using the same phase that corresponds to a column of the $M_{\mathrm{RF}}$-point DFT matrix. 
In other words, 
$\boldsymbol{W}_{\mathrm{\!\! A_{PC}},n}$ takes the form in~\eqref{eq:W_A_PC}, but with $  \boldsymbol{w}_{{\mathrm{A_{PC}},\ell}}=\boldsymbol{v}_{n}$, for $n,\ell\in\{1,\ldots, M_{\mathrm{RF}}\}$, i.e.,   
%we define the $n$-th analog combiner, $\boldsymbol{W}_{\mathrm{\!\! A_{PC}},n}$, as: 
\begin{equation}
\boldsymbol{W}_{\mathrm{\!\! A_{PC}},n} 
   = \bigoplus_{\ell=1}^{L}\boldsymbol{v}_{n}.
    %   =\begin{bmatrix} 
    %     \boldsymbol{v}_{n} & 
    %     \cdots & \boldsymbol{0}_{M_{\mathrm{RF}}} \\
    %     \vdots & %\vdots & 
    %    \ddots & \vdots \\
    %    \boldsymbol{0}_{M_{\mathrm{RF}}} & 
    %    \cdots & \boldsymbol{v}_{n}
    % \end{bmatrix}, 
   \label{eq:W_A_PC_k}
\end{equation}
The $m$-th entry of $\boldsymbol{v}_{n}$ is $ e^{\jmath (m-1)\phi_n}$, and  %$\phi_n$ is given by: 
% $\gamma_n$ be the phase used by each $\boldsymbol{W}_{\mathrm{\!\! A_{PC}},n}$. It can be verified that:
 %----------------------------------------------------------------
% $\gamma_n=\frac{2\pi(n-1)}{M_{\mathrm{RF}}}$ for $n\in\{1,\ldots,\frac{M_{\mathrm{RF}}}{2}\}$ and $\gamma_n=\frac{2\pi(n-1)}{M_{\mathrm{RF}}}-2\pi$ for $n\in\{\frac{M_{\mathrm{RF}}}{2}+1,\ldots,M_{\mathrm{RF}}\}$.
%--------------------------------------------------------------
\begin{equation}\label{eq:gamma_n}
   \phi_n=\!\begin{cases}
        \frac{2\pi(n-1)}{M_{\mathrm{RF}}}, \qquad \quad \ n\in\{1,\ldots,\frac{M_{\mathrm{RF}}}{2}\}, \\
        \frac{2\pi(n-1)}{M_{\mathrm{RF}}}-2\pi, \quad n\in\{\frac{M_{\mathrm{RF}}}{2}+1,\ldots,M_{\mathrm{RF}}\}.
    \end{cases}
   \end{equation} 
% , for $n,\ell\in\{1,\ldots, M_{\mathrm{RF}}\}$,
%where $\boldsymbol{b}_{\ell}=\boldsymbol{v}_{n}, for all $n,\ell\in\{1,\ldots,\}$
%for $n=1,\ldots,N$. 
As in the case of FC-HAD, in PC-HAD, the matrix $\boldsymbol{W}_{\mathrm{\!\! A_{PC}},n}$  processes the $n$-th segment containing $K$ snapshots, and  %, assuming that the total number of snapshots is $NK$.
%Noting that $m$-th entry of the $\ell$-th column of the $M_{\mathrm{RF}}$-point DFT matrix is given by $ e^{\jmath (m-1)\phi_\ell}$, $\phi_\ell=\frac{2\pi(\ell-1)}{M_{\mathrm{RF}}}$
%For the $n$-th analog combiner, substituting $ \boldsymbol{W}_{\mathrm{\!\! A_{PC}},n}$ in~\eqref{eq:q_n} yields $\psi_{r,\ell}=(\ell-1)\mu_r M_{\mathrm{RF}}$ in~\eqref{eq:q_n_ell_sec}. 
%Similar to FC-HAD, 
%the $n$-th PC-HAD analog combiner, $\boldsymbol{v}_{n}\in\mathcal{N}\bigl\{\{v_{j}\}_{j\neq n}\bigr\}$, and  
the set $\{\boldsymbol{W}_{\mathrm{\!\! A_{PC}},n}\}_{n=1}^{N}$  %all $N$ combiners 
spans the entire $M_{\mathrm{RF}}$-dimensional space.

\begin{figure}
\centering
\begin{subfigure}{.23\textwidth}
        \psfrag{subfig1}[][][3]{}
        \psfrag{XXX}[][][2.5]{$\mu$}
        \psfrag{YYY}[][][2.8]{$|g(\mu-\phi_\ell)|/\sqrt{L}$}
        \psfrag{phi1}[][][2.2]{$\phi_5$}
        \psfrag{phi2}[][][2.2]{$\phi_6$}
        \psfrag{phi3}[][][2.2]{$\phi_7$}
        \psfrag{phi4}[][][2.2]{$\phi_8$}
        \psfrag{phi5}[][][2.2]{$\phi_1$}
        \psfrag{phi6}[][][2.2]{$\phi_2$}
        \psfrag{phi7}[][][2.2]{$\phi_3$}
        \psfrag{phi8}[][][2.2]{$\phi_4$}
        \psfrag{phi9}[][][2.2]{$\phi_5$}
        %\psfrag{0}[][][2.2]{$\phi_1$}
        \resizebox{\textwidth}{!}{\includegraphics{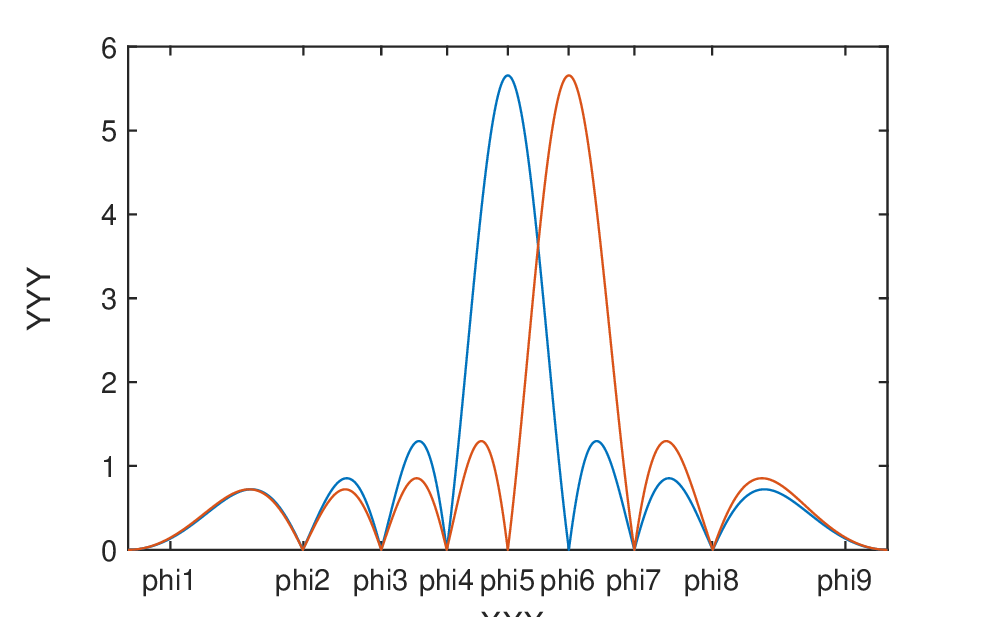}}
        \caption{FC, %$\phi_1$ and $\phi_2$ for 
        %$\boldsymbol{W}_{\mathrm{\!\! A},1}$ with $M_{\mathrm{RF}}=8$, $L=2$.
        %$M_{\mathrm{RF}}=8$,  $\{\phi_\ell\}_{\ell=1}^L$.
        $\phi_\ell$ for $\ell=1,\ldots, L$.
        } 
    \label{fig:FC_1}
  \end{subfigure}
  % \quad
  \begin{subfigure}{.23\textwidth}
        \centering
        \psfrag{subfig2}[][][3]{}
        \psfrag{XXX}[][][2.5]{$\mu$}
        \psfrag{YYY}[][][2.8]{$|g(\mu-\phi_\ell)|/\sqrt{L}$}
        \psfrag{phi1}[][][2.2]{$\phi_5$}
        \psfrag{phi2}[][][2.2]{$\phi_6$}
        \psfrag{phi3}[][][2.2]{$\phi_7$}
        \psfrag{phi4}[][][2.2]{$\phi_8$}
        \psfrag{phi5}[][][2.2]{$\phi_1$}
        \psfrag{phi6}[][][2.2]{$\phi_2$}
        \psfrag{phi7}[][][2.2]{$\phi_3$}
        \psfrag{phi8}[][][2.2]{$\phi_4$}
        \psfrag{phi9}[][][2.2]{$\phi_5$}
        %\psfrag{1}[][][2.2]{$\phi\phi$}
        \resizebox{\textwidth}{!}{\includegraphics{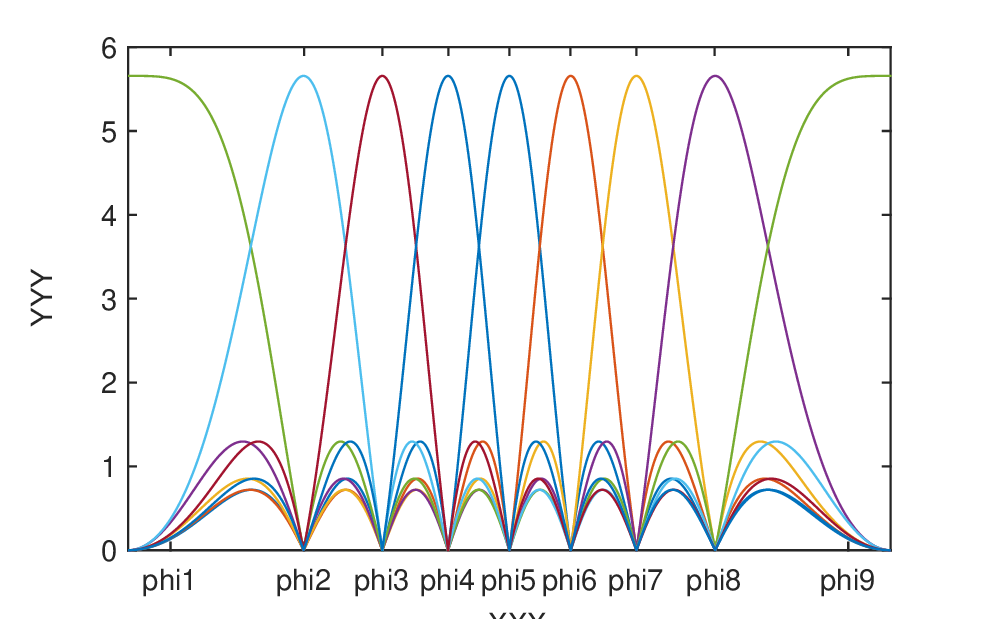}}
        \caption{FC, %$\{\phi_\ell\}_{\ell=1}^L$ for 
        %$\{\boldsymbol{W}_{\mathrm{\!\! A},n}\}_{n=1}^N$ with
        %$M_{\mathrm{RF}}=8$, $\{\phi_\ell\}_{\ell=1}^{M_{\mathrm{RF}}}$.
        $\phi_\ell$ for $\ell=1,\ldots, M_{\mathrm{RF}}$.
        } 
        %and $\phi_\ell$ for $\ell=1,\ldots,M_{\mathrm{RF}}$.} %, %$\ell=1,\ldots,L$.} 
        \label{fig:FC_2}
  \end{subfigure}
   \quad
     \begin{subfigure}{.23\textwidth}
        \centering
        \psfrag{subfig3}[][][3]{}
        \psfrag{XXX}[][][2.5]{$\mu$}
        \psfrag{YYY}[][][2.8]{$|g(\mu-\phi_n)|$}
        \psfrag{AC1}[][][1.5]{ $|g(\mu-\phi_1)|$}
        \psfrag{AC2}[][][1.5]{ $|g(\mu-\phi_2)|$}
        \psfrag{AC3}[][][1.5]{ $|g(\mu-\phi_3)|$}
        \psfrag{AC4}[][][1.5]{ $|g(\mu-\phi_4)|$}
        \psfrag{phi1}[][][2.2]{$\phi_3$}
        \psfrag{phi2}[][][2.2]{$\phi_4$}
        \psfrag{phi3}[][][2.2]{$\phi_1$}
        \psfrag{phi4}[][][2.2]{$\phi_2$}
        \psfrag{phi5}[][][2.2]{$\phi_3$}
        \psfrag{mu}[][][2]{\ \  $\mu_r\in(\phi_1,\phi_2)$}
        %\psfrag{0}[][][2.2]{$\phi_1$}
        \resizebox{\textwidth}{!}{\includegraphics{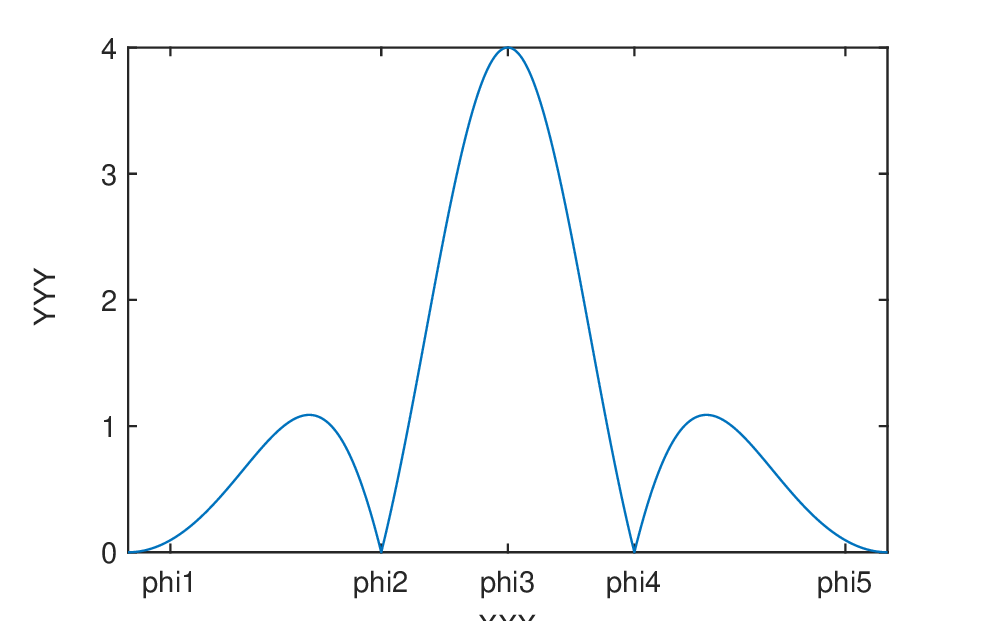}}
        \caption{PC, %$\phi_1$ for 
        %$\boldsymbol{W}_{\mathrm{\!\! A},1}$ with 
        %$M_{\mathrm{RF}}=4$, 
        $\{\phi_\ell\}_{\ell=1}^L=\phi_1$.
       % $\phi_\ell=\phi_1$ for $\ell=1,\ldots, L$.
        } 
        \label{fig:PC_1}
   \end{subfigure}
    %\quad
 \begin{subfigure}{.23\textwidth}
        \centering
        \psfrag{subfig4}[][][3]{}
        \psfrag{XXX}[][][2.5]{$\mu$}
        \psfrag{YYY}[][][2.8]{$|g(\mu-\phi_n)|$}
        \psfrag{AC1}[][][1.5]{ $|g(\mu-\phi_1)|$}
        \psfrag{AC2}[][][1.5]{ $|g(\mu-\phi_2)|$}
        \psfrag{AC3}[][][1.5]{ $|g(\mu-\phi_3)|$}
        \psfrag{AC4}[][][1.5]{ $|g(\mu-\phi_4)|$}
        \psfrag{phi1}[][][2.2]{$\phi_3$}
        \psfrag{phi2}[][][2.2]{$\phi_4$}
        \psfrag{phi3}[][][2.2]{$\phi_1$}
        \psfrag{phi4}[][][2.2]{$\phi_2$}
        \psfrag{phi5}[][][2.2]{$\phi_3$}
        \psfrag{mu}[][][2]{\ \  $\mu_r\in(\phi_1,\phi_2)$}
        \psfrag{mu1}[][][2]{\ \  }
        \psfrag{mu2}[][][2]{\ \  }
        \psfrag{mu3}[][][2.1]{\ $\mu_r$ }
        %\psfrag{0}[][][2.2]{$\phi_1$}
        \resizebox{\textwidth}{!}
        %{\includegraphics{Figures/Figure_amplitude_response_PC_2_v3.eps}}
        {\includegraphics{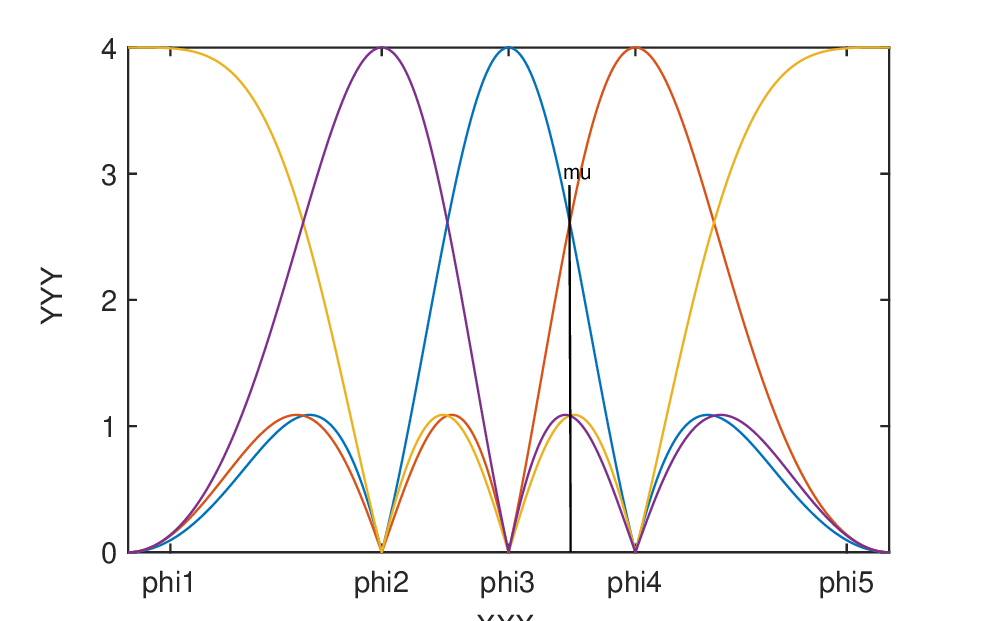}}
        \caption{PC, %$M_{\mathrm{RF}}=4$, 
        %$\{\phi_n\}_{n=1}^{M_{\mathrm{RF}}}$.
        $\{\phi_n\}_{n=1}^{N}$, %for $n=1,\ldots, M_{\mathrm{RF}}$ 
        $\mu_r=\phi_1+\frac{\pi}{4}$.}
       % } %$n=1,\ldots,M_{\mathrm{RF}}$.} %, $\ell=1,\ldots,L$.} 
    \label{fig:PC_2}
   \end{subfigure}
    \caption{Illustration of~\eqref{eq:f_ell_mu} 
    with $M=8$, $L=2$, $N=4$ and  %$N=\frac{M_{\mathrm{RF}}}{L}$ 
     $M_{\mathrm{RF}}=8$ in FC-HAD and $M_{\mathrm{RF}}=4$ in PC-HAD.}
    %for various $\phi_\ell$,
    %$\{\boldsymbol{W}_{\mathrm{\!\! A},n}\}_{n=1}^N$, for $n=1,\ldots, M_{\mathrm{RF}}$ 
    \label{fig:amplitude_response_FC_PC}
\end{figure}

%%%%%%%%%%%%%%%%%%%%%%%%%%%%%%%%%%%%%%%%%%%%%%%%%%%%%%%%%%%%%
% define Fc explain first a, b then c and d 
% analogosy 
% explain why 
Figures~\ref{fig:FC_1} and~\ref{fig:FC_2} illustrate $\frac{|g(\mu_r-\phi_\ell)|}{\sqrt{L}}$, defined in~\eqref{eq:f_ell_mu}, for FC-HAD with $M_{\mathrm{RF}}=8$ and  $L=2$. The scaling $\frac{1}{\sqrt{L}}$ arises from power splitting. In particular,  Figure~\ref{fig:FC_1} corresponds to $\boldsymbol{W}_{\mathrm{\!\! A_{FC}},n}$ constructed using the first two columns of $\boldsymbol{V}_{\!\!8}$, %i.e., $\ell=1,\ldots, 2$, 
whereas Figure~\ref{fig:FC_2} corresponds to $\{\boldsymbol{W}_{\mathrm{\!\! A_{FC}},n}\}_{n=1}^N$ constructed 
using all columns of $\boldsymbol{V}_{\!\!8}$. %, i.e.,  $\ell=1,\ldots, 8$.
Similarly, Figures~\ref{fig:PC_1} and~\ref{fig:PC_2} illustrate $|g(\mu- \phi_n)|$  for PC-HAD, with $M_{\mathrm{RF}}=4$ and $L=2$.
%, where the phase shifters connected to $L$ RF chains of the $n$-th analog combiner are configured using $\phi_n$, cf.~\eqref{eq:W_A_PC_k} and~\eqref{eq:gamma_n}. 
%%%%%%%%%%%%%%%%%%%%%%%%%%%%%%%%%%%%%%%%%%%%%%%%%%%%%%%%%%%%%%%%%%%%%%%%%%
%, where the $L$ scaling arises from configuring phase shifters connected to $L$ RF chains using the same phase $\phi_n$, cf.~\eqref{eq:W_A_PC_k} and~\eqref{eq:gamma_n}. % coherent sum of contributions from all L RF chains
Figure~\ref{fig:PC_1} shows $|g(\mu- \phi_n)|$ corresponding to $\boldsymbol{W}_{\mathrm{\!\! A_{PC}},n}$ constructed using the first column of $\boldsymbol{V}_{{\!\!4}}$, whereas Figure~\ref{fig:PC_2} corresponds to  $\{\boldsymbol{W}_{\mathrm{\!\! A_{PC}},n}\}_{n=1}^N$ constructed using all columns of $\boldsymbol{V}_{\!\!4}$. %, i.e.,  $n=1,\ldots, 4$.

%%%%%%%%%%%%%%%%%%%%%%%%%%%%%%%%%%%%%%%%%%%%%%%%%%%%%%%%%%%%%%%%%%%%%%%
% \blu{Figures~\ref{fig:FC_1} and~\ref{fig:FC_2} illustrate $\frac{|g(\mu_r-\phi_\ell)|}{\sqrt{L}}$ for FC-HAD, while Figures~\ref{fig:PC_1} and~\ref{fig:PC_2} illustrate $L|g(\mu- \phi_n)|$  for PC-HAD, with $M=8$, $L=2$ and $N=4$. 
% The $\frac{1}{\sqrt{L}}$ scaling in FC-HAD results from inherent power-splitting, whereas the $L$ scaling in PC-HAD arises from configuring phase shifters connected to $L$ RF chains using the same phase $\phi_n$.
% For instance, Figure~\ref{fig:FC_1} shows $\frac{|g(\mu_r-\phi_\ell)|}{\sqrt{L}}$ for  $\boldsymbol{W}_{\mathrm{\!\! A_{FC}},n}$ constructed from the first two columns of $\boldsymbol{V}_{{\!\!8}}$, extended to $N$ analog combiners, $\{\boldsymbol{W}_{\mathrm{\!\! A_{FC}},n}\}_{n=1}^N$, in Figure~\ref{fig:FC_2}. Similarly, Figure~\ref{fig:PC_1} shows $L|g(\mu- \phi_n)|$ for $\boldsymbol{W}_{\mathrm{\!\! A_{PC}},n}$ constructed using the first column of $\boldsymbol{V}_{{\!\!4}}$, extended to the $N$ analog combiners, $\{\boldsymbol{W}_{\mathrm{\!\! A_{PC}},n}\}_{n=1}^N$,  in Figure~\ref{fig:PC_2}.}
%%%%%%%%%%%%%%%%%%%%%%%%%%%%%%%%%%%%%%%%%%%%%%%%%%%%%%%%%%%%%

The fact that both $\{\boldsymbol{W}_{\mathrm{\!\! A_{FC}},n}\}_{n=1}^{N}$  and $\{\boldsymbol{W}_{\mathrm{\!\! A_{PC}},n}\}_{n=1}^{N}$ span the entire space 
ensures that signals arriving from all DoAs in $\bigl[\frac{-\pi}{2},\frac{\pi}{2}\bigr]$ are not heavily attenuated or nullified by either the FC-HAD or the  PC-HAD analog combiner.
%, ensuring that all DoAs can be reliably estimated.
%ensuring that all DoAs in $\bigl[\frac{-\pi}{2},\frac{\pi}{2}\bigr]$ can be estimated. 

%After discussing the spatial partitioning caused by analog combiners into $N$ overlapping sectors, we next propose an approach to enable using MP in the HAD receiver.
% \blu{After our discussion on the design of the exhaustive set of $N$ orthogonal analog combiners, we next propose an approach to enable the use of MP in the HAD receiver.}

Having described the set of $N$ orthogonal analog combiners,  we next introduce our first approach to address the difficulties in applying the MPM in HAD receivers.
%, which utilizes the set of orthogonal analog combiners. %as a key element. 

%------------------------------------------------------------------------------
\subsubsection{PMPM Approach}\label{sec:PMP}
In Section~\ref{sec: MP in HAD}, we showed that applying the MPM in HAD receivers entangles the output signals, rendering MPM invocation unwieldy.
%rendering the invocation of the MPM unwieldy.
One approach to remedy this difficulty is presented in the following theorem.

\begin{theorem}\label{Thr:PMPM} 
Let  $\{\boldsymbol{W}_{{\mathrm{\!\! A}},n}\}_{n=1}^{N}$ be a set of analog combiners, each satisfying  $\boldsymbol{W}_{{\mathrm{\!\! A}},n}^H\boldsymbol{W}_{{\mathrm{\!\! A}},n}=\alpha^2 M_{{\mathrm{RF}}}\boldsymbol{I}_{\!L}$. Let the number of snapshots processed by $\boldsymbol{W}_{{\mathrm{\!\! A}},n}$ be $K$
and let the signal $\boldsymbol{S}_n$ in~\eqref{eq:Q_n} be periodic, i.e., $\boldsymbol{S}_n=\boldsymbol{S}$. 
Let $\{\boldsymbol{W}_{\!\mathrm{D},n}\}_{n=1}^N$ be a set of $N$ baseband digital combiners, 
each  processes the output of the $n$-th analog combiner, $\boldsymbol{Q}_n$ in~\eqref{eq:Q_n}. %, for $n=1,\ldots N$. 
Setting $\boldsymbol{W}_{\!\mathrm{D},n}=\frac{1}{\alpha^2 M_{\mathrm{RF}}}\boldsymbol{W}^H_{\mathrm{\!\! A},n}\in \mathbb{C}^{L \times M}$ yields an aggregate output of the $N$ combiners, $\boldsymbol{Y}_{\!\!\mathrm{PMPM}} \in\mathbb{C}^{M\times K}$, which is given by:
\begin{align}
\boldsymbol{Y}_{\!\!\mathrm{PMPM}} 
=\boldsymbol{A}\boldsymbol{S}+\sum_{n=1}^{N}  \boldsymbol{P}_{\boldsymbol{W}_{\!\!{\mathrm{A}},n}} \boldsymbol{Z}_{n},  
\label{eq:y_PMPM}
\end{align}
where 
$\boldsymbol{P}_{\boldsymbol{W}_{\!\!{\mathrm{A}},n}}=\frac{1}{\alpha^2 M_{\mathrm{RF}}}\boldsymbol{W}_{\!\!{\mathrm{A}},n} \boldsymbol{W}_{\!\!{\mathrm{A}},n}^H$.
The signal $\boldsymbol{Y}_{\!\!\mathrm{PMPM}}$ is equivalent to the $K$-snapshot input signal, $\boldsymbol{X}$, of the FD receiver in~\eqref{eq:X}. The SNR of both $\boldsymbol{Y}_{\!\!\mathrm{PMPM}}$  and  $\boldsymbol{X}$ is 
%operating at %  i.e., 
%$\mathrm{SNR}=
$\tfrac{1}{M}\Tr( \boldsymbol{A}\boldsymbol{\Phi}\boldsymbol{A}^H)$. In particular, 
%\blu{and operating at a $10\log M$~dB lower SNR}, i.e.,
% PMPM/FD>> and negative sign removed with lower
% \begin{equation}\label{eq:snr_PMPM_FD}
% \boldsymbol{Y}_{\mathrm{PMPM}} \Big|_{\substack{\!\!\!\!\tilde{K}=NK \\ \mathrm{SNR_{PMPM}}}} 
% = \boldsymbol{X} \Big|_{\substack{\tilde{K}=K \\ \mathrm{SNR_{\mathrm{FD}}}=M \times \mathrm{SNR_{PMPM}}}}.
% \end{equation}
%%%%%%%%%%%%%%%%%%%%%%%%%%%%%%%%%%%%%%%%%%%%%%%%%%%%%%%%%%%%%%%%%%%%%%%%
% Note: hspace/vspace>> elastic the space may not be as needed (rubber)
% \! will do space of 1/4 character, more specific.
\begin{equation}\label{eq:snr_PMPM_FD}
\boldsymbol{Y}_{\!\!\mathrm{PMPM}} \Big|_{\substack{\hspace{0.01cm}\tilde{K}=NK 
%\\ \mathrm{SNR}= \tfrac{1}{M}\Tr( \boldsymbol{A}\boldsymbol{\Phi}\boldsymbol{A}^H) 
}} 
= \boldsymbol{X} \Big|_{\substack{\hspace{0.01cm} \tilde{K}=K 
%\\ \mathrm{SNR}=\tfrac{1}{M}\Tr( \boldsymbol{A}\boldsymbol{\Phi}\boldsymbol{A}^H)
}}.
\end{equation}
\end{theorem}
\begin{IEEEproof}
See Appendix \ref{sec: proof-threorm1}.
\end{IEEEproof}
From this theorem, it can be seen that  % any set of 
DoAs %,  $\theta_r\in\bigl[\frac{-\pi}{2},\frac{\pi}{2}\bigr]$, $r=1,\ldots, R$, 
that are recoverable by the FD receiver, are also recoverable by the HAD receiver using the PMPM approach, at the same the signal power %is $M$-fold higher and} 
but with  $N$-fold the number of snapshots. In particular, PMPM  ensures that no signal is nullified by projection, irrespective of its DoA. This is achieved in PMPM by using $N$, rather than one snapshot as in the FD case.
We note that the periodicity requirement in Theorem~\ref{Thr:PMPM} can be satisfied in commonly deployed communication systems, e.g., 5G+, in which the frame contains repeated preamble segments or training pilots. It is also worth noting that the periodic snapshots may not be contiguous for the PMPM approach to apply. For instance, the periodic preamble segments may be separated by non-periodic payloads. 
% is to show that, by using the periodicity of the received signals and a set of full column rank analog combiners, the PMPM approach processes the received signals from HAD receivers, either FC or PC,  to be equivalent to those of an FD receiver, but with $N$ snapshots instead of a single one.  This periodicity can be found either in the preamble, where the $N$ snapshots are not contiguous~(with gaps due to the frame payload), or in the training data segments, such as beacons, where the $N$ snapshots are contiguous. 
Furthermore, these periodic signals do not need to be known \emph{a priori}, nor do they need to be detected for the PMPM approach to be applied. However, exploiting signal periodicity requires the HAD receiver to be synchronized.

Having developed a technique for making the output of the digital combiners of the HAD receiver equivalent to that of its FD counterpart, we are now in a position to apply the MPM to estimate the DoAs in a manner analogous to FD receivers. %/ ,using a small number of snapshots. 

To apply the MPM in HAD receivers, we begin by considering the $k$-th 
column of 
$\boldsymbol{Y}_{\!\!\mathrm{PMPM}}$,  $\boldsymbol{y}_k \in \mathbb{C}^M$.\footnote{The $k$-th column of $\boldsymbol{Y}_{\!\!\mathrm{PMPM}}$,  $\boldsymbol{y}_k$, represents a virtual snapshot for HAD receivers using PMPM  and an actual  
%corresponds to the $k$-th 
snapshot, %of $\boldsymbol{X}$
$\boldsymbol{x}_k$, 
in the FD receiver, cf.~\eqref{eq:ADC_x_k}.}
We construct 
the Hankel matrix $\boldsymbol{H}_{k}^{{\mathrm{PMPM}}} \in \mathbb{C}^{(M-\xi) \times(\xi+1)}$,   
\begin{align}
\boldsymbol{H}_{k}^{{\mathrm{PMPM}}} \!\!
& =
\scalebox{0.90}{$\begin{bmatrix}
y_k[1] &  \ldots & y_k[\xi+1] \\
%\boldsymbol{y}_k[2]  & \ldots & \boldsymbol{y}_k[\xi_{_\mathrm{HAD}}+2] \\
\vdots  & \ddots & \vdots \\
y_k[M-\xi]  & \ldots & y_k[M]
\end{bmatrix}$}.
\label{eq:G_PMPM}
\end{align}
%where $\xi$ is the matrix parameter for PMPM. ($\xi_{_\mathrm{PMP}}$ can equal to $\xi$.)
Analogous to the FD case, a matrix pencil in the proposed  PMPM approach can be synthesized from the two matrices, $\boldsymbol{H}_{k,1}^{{\mathrm{PMPM}}}$ and $\boldsymbol{H}_{k,2}^{{\mathrm{PMPM}}}$, which are obtained by deleting the last and first columns of $\boldsymbol{H}_{k}^{{\mathrm{PMPM}}}$, respectively.
% %obtaining two matrices from each matrix $\boldsymbol{H}_{k}^{{\mathrm{PMPM}}}$.
% %, we  define two matrices,
% %For each $\boldsymbol{H}_{k}^{{\mathrm{PMPM}}}$, we define two matrices,
% % $\boldsymbol{H}_{k,1}^{{\mathrm{PMPM}}}$ and $\boldsymbol{H}_{k,2}^{{\mathrm{PMPM}}}$ $\in \mathbb{C}^{(M-\xi) \times \xi}$. 
% %In particular, $\boldsymbol{H}_{k,1}^{{\mathrm{PMPM}}}$ and $\boldsymbol{H}_{k,2}^{{\mathrm{PMPM}}}$ 
% %are obtained by deleting the last and first columns of $\boldsymbol{H}_{k}^{{\mathrm{PMPM}}}$, respectively. 
%both with dimensions $(M-\tilde{\xi_{_\mathrm{HAD}}}) \times (\tilde{\xi_{_\mathrm{HAD}}})$. 
%Using analogs of~\eqref{eq:Z_1_fd},~\eqref{eq:Z_2_fd},~\eqref{eq:R_n_fd}, and~\eqref{eq:Z_0_fd}, 
As in the FD case, we write  $\boldsymbol{H}_{k,1}^{{\mathrm{PMPM}}}= \boldsymbol{\Pi}_1 \boldsymbol{\Lambda}_{k}\boldsymbol{\Pi}_2$ and $\boldsymbol{H}_{k,2}^{{\mathrm{PMPM}}} = \boldsymbol{\Pi}_1 \boldsymbol{\Lambda}_{k} \boldsymbol{\Pi}_0 \boldsymbol{\Pi}_2$, where $\boldsymbol{\Pi}_1 \in \mathbb{C}^{(M-\xi) \times R}$, $\boldsymbol{\Pi}_2 \in \mathbb{C}^{R \times \xi}$, $\boldsymbol{\Lambda}_{k}$, and $\boldsymbol{\Pi}_0\in \mathbb{C}^{R\times R}$ are given in~\eqref{eq:Z_1_fd},~\eqref{eq:Z_2_fd},~\eqref{eq:R_n_fd}, and~\eqref{eq:Z_0_fd}, respectively.
%, which  can be obtained as~\eqref{eq:Z_1_fd},~\eqref{eq:Z_2_fd},~\eqref{eq:R_n_fd} and~\eqref{eq:Z_0_fd}, respectively. 
Analogous to FD, %cf. \eqref{eq:eignval-prob}, 
%it can be readily seen that 
the diagonal elements of $\boldsymbol{\Pi}_0$ are the generalized eigenvalues of the matrix pair  $(\boldsymbol{H}_{k,2}^{{\mathrm{PMPM}}},\boldsymbol{H}_{k,1}^{{\mathrm{PMPM}}})$, which 
are the eigenvalues of $(\boldsymbol{H}_{k,1}^{\mathrm{PMPM}})^{\dag} \boldsymbol{H}_{k,2}^{\mathrm{PMPM}}$.
In particular, the $r$-th diagonal entry of  $\boldsymbol{\Pi}_0$,  $[\boldsymbol{\Pi}_0]_{r,r}$, is equal to the $r$-th eigenvalue of $(\boldsymbol{H}_{k,1}^{\mathrm{PMPM}})^{\dag} \boldsymbol{H}_{k,2}^{\mathrm{PMPM}}$. Substituting this eigenvalue in~\eqref{eq:theta_est_fd_1} yields the   $r$-th DoA, $r=1,\ldots,R$.
%can be readily %extracted using ~\eqref{eq:mu} as in~\eqref{eq:theta_est_fd_1}.}
%Hence, the $r$-th eigenvalue   %$(\boldsymbol{H}_{k,1}^{\mathrm{PMPM}})^{\dag} \boldsymbol{H}_{k,2}^{\mathrm{PMPM}}$ 
%is given by $[\boldsymbol{\Pi}_0]_{r,r}$, which,  using~\eqref{eq:mu}, allows extracting the $r$-th DoA, $\Hat{\theta}_r$, as in~\eqref{eq:theta_est_fd_1}.
% that yields extracting the $r$-th DoA, $\Hat{\theta}_r$, as in~\eqref{eq:theta_est_fd_1}. 
%As a result, DoAs can be extracted using~\eqref{eq:eignval-prob} and \eqref{eq:theta_est_fd_1}.
%%%%%%%%%%%%%%%%%%%%%%%%%%%%%%%%%%%%%%%%%%%%%%%%%%%%%%%%%%%%%%%%%%%%%%%%%%%%
% Following the same steps as outlined in Section~\ref{sec:RMP-a} and Section~\ref{sec:RMP-b}, the DoAs can be estimated using MP. Instead of using \eqref{eq:theta_est_1}, By solving the generalized eigenvalue problem, we can estimate the DoA as follows: 
% \begin{equation}
%    \Hat{\theta}_r=\sin ^{-1} \biggl \{ \frac{\lambda \operatorname{\Im}\{\log {e^{\jmath \mu_{r} }}\}}{2 \pi  \Delta} \biggr\}.
%    \label{eq:theta_est_2}
% \end{equation}
%\end{IEEEproof}
%------------------------------------------------------------------------------

Having considered one column of $\boldsymbol{Y}_{\!\!\mathrm{PMPM}}$ in the noise-free case, we now consider applying the PMPM to multiple columns of  $\boldsymbol{Y}_{\!\!\mathrm{PMPM}}$ in the presence of noise.
%Next, we consider DoA estimation using multiple snapshots in the presence of noise.
\subsubsection{Multiple Snapshots in the Presence of Noise}\label{sec:multiple-snapshots} 
In this section, we apply the MPM when multiple (actual, as in FD, or virtual as in HAD) snapshots are considered simultaneously. 
\paragraph{Noiseless Case}\label{sec:multiple-snapshots-Noiseless}
The MPM hinges on the decompositions in~\eqref{eq:H_k_1} and~\eqref{eq:H_k_2}. Hence, to apply it when multiple snapshots are available to the receiver, a Hankel-based construction is needed to enable a matrix pencil with generalized eigenvalues corresponding to the desired DoAs.  
One approach to do so was alluded to in~\cite{yilmazer2008multiple}. In particular, each actual or virtual snapshot is used to construct a Hankel matrix and a collection of such matrices is concatenated in one augmented matrix, which we denote by $\boldsymbol{H}_{\!\mathrm{A}} \in \mathbb{C}^{(M-\xi)\times K_{\mathrm{A}} (\xi+1)}$. This collection contains $K_{\mathrm{A}}$ matrices, where $K_{\mathrm{A}}=\tilde{K}$  for FD and $K_{\mathrm{A}}=\frac{\tilde{K}}{N}$  for HAD using PMPM. 
% That is, for FD receivers, the augmented matrix can be given by:
% \begin{equation}
% \boldsymbol{H}^{\mathrm{FD}}_\mathrm{\!A} = \begin{bmatrix}
% \boldsymbol{H}^{\mathrm{FD}}_{\!1} & \ldots & \boldsymbol{H}^{\mathrm{FD}}_{\!k} & \ldots & \boldsymbol{H}^{\mathrm{FD}}_{\!\tilde{K}=NK}
% \end{bmatrix},
% \label{eq:Y_E_FD}
% \end{equation}
%column of $\boldsymbol{X}_n$ for FD ra
%%%%%%%%%%%%%%%%%%%%%%%%%%%%%%%%%%%%%%%%%%%%%%%%%%%%%%%%%%%%%%
%columns of $\boldsymbol{Y}_{\!\!\mathrm{PMPM}}$
%Sections~\ref{sec: MP in FD}--\ref{sec:PMP}, we developed a technique for applying the MPM to HAD architectures when a single noise-free column of   is considered for extracting the DoAs. Hence, the Hankel matrices in~\eqref{eq:x_n_first} and~\eqref{eq:G_PMPM} for FD and PMPM are constructed once. H
% owever,  when multiple snapshots are considered, 
% \blu{we are guided by~\cite{yilmazer2008multiple}. According to this approach,} 
% a Hankel matrix is constructed for each snapshot, with a total number of snapshots $\tilde{K}$. 
% %(For the PMP approach $\tilde{K}=K$.)  
% In particular, for the $k$-th snapshot, the corresponding Hankel matrix is formed, for $k=1,\ldots, \tilde{K}$. 
The augmented matrix is: 
\begin{equation}
\boldsymbol{H}_\mathrm{\!A} = \begin{bmatrix}
\boldsymbol{H}_{\!1} & \cdots %& %\boldsymbol{H}_{\!k} & \cdots 
& \boldsymbol{H}_{\!K_{\mathrm{A}}}
\end{bmatrix}.
\label{eq:Y_E}
\end{equation}

% \begin{equation}
% \boldsymbol{H}_\mathrm{\!A} = \begin{bmatrix}
% \boldsymbol{H}_{\!1} & \cdots & \boldsymbol{H}_{\!K} & \cdots & \boldsymbol{H}_{\!K_A-K+1} & \cdots 
% & \boldsymbol{H}_{\!K_{\mathrm{A}}}\\
% \underbrace_{\text{Snapshots corresponding to} W_{A,1}} &\cdots & \underbrace_{\text{Snapshots corresponding to} W_{A,N}}
% \end{bmatrix}.
% \label{eq:Y_E_SPC}
% \end{equation}

% \begin{equation}
% \boldsymbol{H}_\mathrm{A} = 
% \begin{bmatrix}
% \boldsymbol{H}_{1} & \cdots & \boldsymbol{H}_{K} & \cdots & \boldsymbol{H}_{K_A - K + 1} & \cdots & \boldsymbol{H}_{K_{\mathrm{A}}} \\
% \underbrace{\hspace{4em}}_{\text{Snapshots corresponding to } \boldsymbol{W}_{\!\!\mathrm{A},1}} & \cdots & 
% \underbrace{\hspace{4em}}_{\text{Snapshots corresponding to } \boldsymbol{W}_{\!\!\mathrm{A},N}}
% \end{bmatrix}.
% \label{eq:Y_E_SPC}
% \end{equation}

%%%%%%%%%%%%%%%%%%%%%%%%%%%%%%%%%%%%%%%%%%%%%%%%%%%%%%%%%%%%%%%%%%%%%
% We need to add a rationale as to why we need to use concatenation: Hankel matrix and decomposable 
Towards using $\boldsymbol{H}_\mathrm{\!A}$ to extract the DoAs, we introduce the notation $\boldsymbol{H}_{\mathrm{\!A}, -\ell}$ 
%can be estimated. 
%For ease of exposition, we use the notation 
%$\hat{\boldsymbol{H}}_{\mathrm{\!A}, -\ell}$
%\in\mathbb{C}^{(M-\xi) \times K_{\mathrm{A}} (\xi +1)}
to represents a matrix identical to $\boldsymbol{H}_{\mathrm{\!A}}$ but with the $\ell$-th column removed.
In particular,  to construct a matrix pencil, two Hankel-based matrices $\boldsymbol{H}_{\mathrm{\!A}, -\mathcal{I}}$ and $\boldsymbol{H}_{\mathrm{\!A}, -\mathcal{J}} \in \mathbb{C}^{(M-\xi)\times K_{\mathrm{A}} \xi}$ are obtained from $\boldsymbol{H}_{\mathrm{\!A}}$  as follows: 
\begin{align}
\boldsymbol{H}_{\mathrm{\!A},-\mathcal{I}} &= \begin{bmatrix}
\boldsymbol{H}_{\!1, -(\zeta+1)} 
%& \cdots & \boldsymbol{H}_{\!k, -(\zeta+1)} 
& \cdots & \boldsymbol{H}_{\!K_{\mathrm{A}}, -(\zeta+1)}
\end{bmatrix}, \label{eq:H_A_1}\\
 \boldsymbol{H}_{\mathrm{\!A},-\mathcal{J}} &= \begin{bmatrix}
\boldsymbol{H}_{\!1, -1} 
%& \cdots & \boldsymbol{H}_{\!k, -1} 
& \cdots & \boldsymbol{H}_{\!K_{\mathrm{A}}, -1}
\end{bmatrix},  \label{eq:H_A_2}    
\end{align}
where the sets 
%$\mathcal{I} = \{\xi + 1, 2(\xi + 1), 3(\xi + 1), \ldots, K_{\mathrm{A}}(\xi + 1)\}$ and $\mathcal{J} = \{1, \xi + 2, 2(\xi + 1) + 1, \ldots, (K_{\mathrm{A}} - 1)(\xi + 1) + 1\}$ 
% $\mathcal{I} = \{ i(\xi + 1) \mid i = 1, 2, \ldots, K_{\mathrm{A}}\}$
% and 
% $\mathcal{J} = \{ (j-1)(\xi + 1) + 1 \mid j = 1, 2, \ldots, K_{\mathrm{A}}\}$
$\mathcal{I} = \{ i(\xi + 1)\}_{i=1}^{K_{\mathrm{A}}}$
and 
$\mathcal{J} = \{ (j-1)(\xi + 1) + 1\}_{j=1}^{K_{\mathrm{A}}}$
identify  the column indices to be removed from $\boldsymbol{H}_\mathrm{\!A}$ to construct $\boldsymbol{H}_{\mathrm{\!A}, -\mathcal{I}}$ and $\boldsymbol{H}_{\mathrm{\!A}, -\mathcal{J}}$, respectively.
% \begin{align}
% \mathcal{I} &= \{\xi + 1, 2(\xi + 1), 3(\xi + 1), \ldots, K_{\mathrm{A}}(\xi + 1)\}, \\
% \!\!\!\!\!\!\mathcal{J} &= \{1, \xi + 2, 2(\xi + 1) + 1, \ldots, (K_{\mathrm{A}} - 1)(\xi + 1) + 1\}.
% \end{align}
%%%%%%%%%%%%%%%%%%%%%%%%%%%%%%%%%%%%%%%%%%%%%%%%%%%%%%%%%%%%%%%%%%%%%%%%
Using the embedded Hankel structure, we can write: % $\boldsymbol{H}_{\!\mathrm{A}, -\mathcal{I}}$ and $\boldsymbol{H}_{\!\mathrm{A}, -\mathcal{J}}$ can be expressed as:
\begin{align}
\boldsymbol{H}_{\mathrm{\!A},-\mathcal{I}}  &= \boldsymbol{\Pi}_{\mathrm{A},1} \boldsymbol{\Lambda}_{\mathrm{A}}\boldsymbol{\Pi}_{\mathrm{A},2}, \label{eq:Y_E_1} \\
\boldsymbol{H}_{\mathrm{\!A},-\mathcal{J}}  &= \boldsymbol{\Pi}_{\mathrm{A},1} \boldsymbol{\Lambda}_{\mathrm{A}} %\boldsymbol{\Pi}_{\mathrm{A},0}
\bigl(\boldsymbol{I}_{\!K_{\mathrm{A}}}  \otimes  \boldsymbol{\Pi}_{0}\bigr)
\boldsymbol{\Pi}_{\mathrm{A},2}, 
\label{eq:Y_E_2}
\end{align}
where $\boldsymbol{\Pi}_{\mathrm{A},1}= \boldsymbol{1}_{\!K_{\mathrm{A}}}^T  \otimes  \boldsymbol{\Pi}_1 \in \mathbb{C}^{(M-\xi)\times (K_{\mathrm{A}} \times R)}$, $\boldsymbol{\Pi}_{\mathrm{A},2}= \boldsymbol{I}_{\!K_{\mathrm{A}}}   \otimes \boldsymbol{\Pi}_2 \in \mathbb{C}^{ (K_{\mathrm{A}} \times R) \times(K_{\mathrm{A}} \times \xi)}$, 
%$\boldsymbol{\Pi}_{\mathrm{A}, 0} = \boldsymbol{I}_{\!K_{\mathrm{A}}}  \otimes  \boldsymbol{\Pi}_{0} \in \mathbb{C}^{ (K_{\mathrm{A}} \times R) \times (K_{\mathrm{A}} \times R)}$, 
and  $\boldsymbol{\Lambda}_{\mathrm{A}} =  \bigoplus_{k=1}^{K_{\mathrm{A}}}  \boldsymbol{\Lambda}_k \in \mathbb{C}^{ (K_{\mathrm{A}} \times R) \times (K_{\mathrm{A}} \times R)}$. 
Similar to FD, the matrix $\boldsymbol{\Pi}_{0}$ accounts for the deleted columns from %captures the difference between 
$\boldsymbol{H}_{\mathrm{\!A},-\mathcal{I}}$ and $\boldsymbol{H}_{\mathrm{\!A},-\mathcal{J}}$, and enables the DoAs, $\{\mu_r\}_{r=1}^R$, to be expressed as the generalized eigenvalues of the matrix pencil constructed from these matrices. In particular, using~\eqref{eq:Y_E_1} and~\eqref{eq:Y_E_2}, the matrix pencil $\boldsymbol{H}_{\mathrm{\!A},-\mathcal{J}} - \zeta \boldsymbol{H}_{\mathrm{\!A},-\mathcal{I}}$, which has a rank $R$,  can be written as:
\begin{equation}
\!\!\!\boldsymbol{H}_{\mathrm{\!A},-\mathcal{J}} - \zeta \boldsymbol{H}_{\mathrm{\!A},-\mathcal{I}}
= \boldsymbol{\Pi}_{\mathrm{A},1} \boldsymbol{\Lambda}_{\mathrm{A}} \bigl(\boldsymbol{I}_{\!K_{\mathrm{A}}}  \otimes  \bigl( \boldsymbol{\Pi}_{0}  - \zeta \boldsymbol{I}_{\!R} \bigr)\bigr) \boldsymbol{\Pi}_{\mathrm{A},2}. 
\label{eq:generalized_eignvalue_E}
\end{equation}
It can be readily seen that
%the rank of this matrix in~\eqref{eq:generalized_eignvalue_E} is $R$ and that
the diagonal elements of $\boldsymbol{\Pi}_{0}$ are the generalized eigenvalues of the matrix pair  $(\boldsymbol{H}_{\mathrm{\!A},-\mathcal{J}},\boldsymbol{H}_{\mathrm{\!A},-\mathcal{I}})$, corresponding to the eigenvalues of  $(\boldsymbol{H}_{\mathrm{\!A},-\mathcal{I}})^{\dag} \boldsymbol{H}_{\mathrm{\!A},-\mathcal{J}}$, that is, 
%Denoting 
the $r$-th eigenvalue of  $(\boldsymbol{H}_{\mathrm{\!A},-\mathcal{I}})^{\dag} \boldsymbol{H}_{\mathrm{\!A},-\mathcal{J}}$  equals $[\boldsymbol{\Pi}_{0}]_{r,r}$. %Hence, using~\eqref{eq:mu}, we have $\hat{\theta}_r$ as in~\eqref{eq:theta_est_fd_1}. 
%, replacing $\boldsymbol{\Pi}_{0}$ with $\boldsymbol{\Pi}_{\mathrm{A}, 0}$.
% \begin{equation}
%    \Hat{\theta}_r=\arcsin \biggl \{ \frac{\lambda \operatorname{\Im }\{\log{[\boldsymbol{\Pi}_{\mathrm{A}, 0}}]_{r,r}\}}{2 \pi \Delta} \biggr\}.
%    \label{eq:theta_est_aug}
% \end{equation}
%In the multiple snapshots case, the generalized eigenvalues are repeated $K_{\mathrm{A}}$ times.
It is worth noting that MPM requires only a single snapshot to extract the DoAs.
%, so multiple snapshots add no extra information in the noiseless case but  are needed to average out noise, as discussed next.
In the noiseless case, multiple snapshots provide no extra information.  However, in the presence of noise, multiple snapshots can be exploited to improve estimation accuracy, as discussed next.

%%%%%%%%%%%%%%%%%%%%%%%%%%%%%%%%%%%%%%%%%%%%%%%%%%%%%%%%%%%%%%%%%
% One to two Statements to show that the concatenation is the best based on "data processing inequality", averaging we lose information
%%%%%%%%%%%%%%%%%%%%%%%%%%%%%%%%%%%%%%%%%%%%%%%%%%%%%%%%%%%%%%%%%

%%%%%%%%%%%%%%%%%%%%%  Reduce noise effect
\paragraph{Noisy Case} \label{sec:multiple-snapshots-Noisy}
Analogous to the FD case, the noise subspace can be discarded by invoking the 
%reduce the noise effects, the 
SVD of $\boldsymbol{H}_\mathrm{\!A}$ in~\eqref{eq:Y_E}, cf. Section~\ref{sec:SVD}.
In particular, $\boldsymbol{H}_\mathrm{\!A}$ can be decomposed as $\boldsymbol{U}_\mathrm{\!A} \boldsymbol{\Sigma}_\mathrm{A} \boldsymbol{V}_\mathrm{\!\!A}^H$, where $\boldsymbol{U}_\mathrm{\!A} \in \mathbb{C}^{(M-\xi)\times(M-\xi)}$ and $\boldsymbol{V}_\mathrm{\!\!A} \in \mathbb{C}^{K_{\mathrm{A}} (\xi+1)\times (M-\xi)}$ are unitary matrices, and $\boldsymbol{\Sigma}_\mathrm{A} \in \mathbb{C}^{(M-\xi)\times (M-\xi)} $ is a diagonal matrix containing the singular values of $\boldsymbol{H}_\mathrm{\!A}$.
% Next, we use the filtered matrices, $\boldsymbol{U}_\mathrm{\!A}^{\prime} \in \mathbb{C}^{(M-\xi)\times R}$,  $\boldsymbol{V}_\mathrm{\!\!A}^{\prime} \in \mathbb{C}^{K_{\mathrm{A}} (\xi+1)\times R}$ and $\boldsymbol{\Sigma}_\mathrm{A}^{\prime} \in \mathbb{C}^{R \times R} $
The left and right-singular vectors in $\boldsymbol{U}_\mathrm{\!A}$ and $\boldsymbol{V}_\mathrm{\!\!A}$ corresponding to  the largest  $R$ singular values, 
%largest singular values 
in $\boldsymbol{\Sigma}_\mathrm{A}$, %corresponding to the largest singular values, 
are retained and the remaining singular vectors and the corresponding singular values %$\rank\boldsymbol{H}_{\!\mathrm{A}}= M-\zeta-R$ 
are discarded. 
These matrices represent the signal components in the signal subspace after eliminating the noise subspace and are denoted by $\hat{\boldsymbol{U}}_\mathrm{\!A} \in \mathbb{C}^{(M-\xi)\times R}$,  $\hat{\boldsymbol{V}}_\mathrm{\!\!A} \in \mathbb{C}^{K_{\mathrm{A}} (\xi+1)\times R}$ and $\hat{\boldsymbol{\Sigma}}_\mathrm{A} \in \mathbb{C}^{R \times R}$.
% Using $\hat{\boldsymbol{U}}_\mathrm{\!A}$, $\hat{\boldsymbol{V}}_\mathrm{\!A}$, and  $\hat{\boldsymbol{\Sigma}}_\mathrm{A}$, 
Using these matrices, the matrix  $\hat{\boldsymbol{H}}_\mathrm{\!A}$ is obtained as $\hat{\boldsymbol{U}}_\mathrm{\!A} \hat{\boldsymbol{\Sigma}}_\mathrm{A} \hat{\boldsymbol{V}}_\mathrm{\!\!A}^{H}$. 
%\blu{Notably, the use of multiple snapshots increases the singular values associated with the signal relative to those associated with noise, improving the separation between signal and noise subspaces and thereby enhancing DoA estimation accuracy.}
Using $\hat{\boldsymbol{H}}_\mathrm{\!A}$,  we follow the procedure outlined in the noiseless case to obtain the two matrices $\hat{\boldsymbol{H}}_{\mathrm{\!A}, -\mathcal{I}}$ and  $\hat{\boldsymbol{H}}_{\mathrm{\!A}, -\mathcal{J}}$, cf.~\eqref{eq:H_A_1} and~\eqref{eq:H_A_2}.
The matrices $\hat{\boldsymbol{H}}_{\mathrm{\!A}, -\mathcal{I}}$ and  $\hat{\boldsymbol{H}}_{\mathrm{\!A}, -\mathcal{J}}$, being synthesized from the singular vectors and the $R$ largest singular values, no longer possess the Hankel structure. However, 
%Although this results in a degenerate Hankel structure for these matrices, following the lead of~\cite{hua1991svd}, 
the generalized eigenvalues of the matrix pair $(\hat{\boldsymbol{H}}_{\mathrm{\!A},-\mathcal{J}},\hat{\boldsymbol{H}}_{\mathrm{\!A},-\mathcal{I}})$, i.e., 
% the eigenvalues of $(\hat{\boldsymbol{H}}_{\mathrm{\!A}, -\mathcal{I}})^{\dag} \hat{\boldsymbol{H}}_{\mathrm{\!A}, -\mathcal{J}}$,
% constitute an approximation of the diagonal entries of $\boldsymbol{\Pi}_0$~\cite{hua1991svd}.
%$r$-th eigenvalue of  $(\hat{\boldsymbol{H}}_{\mathrm{\!A}, -\mathcal{I}})^{\dag} \hat{\boldsymbol{H}}_{\mathrm{\!A}, -\mathcal{J}}$, denoted as $\nu_r$ for $r=1,\ldots,R$,  constitute an approximation of $[\boldsymbol{\Pi}_{0}]_{r,r}$~\cite{hua1991svd}.
the eigenvalues of  $(\hat{\boldsymbol{H}}_{\mathrm{\!A}, -\mathcal{I}})^{\dag} \hat{\boldsymbol{H}}_{\mathrm{\!A}, -\mathcal{J}}$,  $\{\nu_r\}_{r=1}^R$,  constitute an approximation of the diagonal elements of $\boldsymbol{\Pi}_0$~\cite{hua1991svd}.
Using this approximation, DoA estimates, $\{\hat{\theta}_r\}_{r=1}^R$, can be obtained as: %in~\eqref{eq:generalized_eignvalue_fd}.
%till corresponding to 
%he eigenvalues of $(\hat{\boldsymbol{H}}_{\mathrm{\!A}, -\mathcal{I}})^{\dag} \hat{\boldsymbol{H}}_{\mathrm{\!A}, -\mathcal{J}}$, thereby %enabling the estimation of the DoAs as follows:
%enabling the estimation of the DoAs, $\{\hat{\theta}_r\}_{r=1}^R$, as follows:
\begin{equation}
\Hat{\theta}_r \!= \!\arcsin \biggl(\frac{\lambda  \Im \left\{\log\bigl( \nu_r\bigl((\hat{\boldsymbol{H}}_{\!\mathrm{A}, -\mathcal{I}})^{\dag} \hat{\boldsymbol{H}}_{\!\mathrm{A}, -\mathcal{J}} \bigr) \right\}}{2 \pi \Delta} \biggr).
\label{eq:theta_est_aug_noise}
\end{equation}
Increasing the number of snapshots increases the spread of the singular values corresponding to the signal subspace and those corresponding to the noise subspace. This subsequently results in enhancing the DoA estimation accuracy, as will be demonstrated numerically in Section~\ref{sec:simulation} below.
%%%%%%%%%%%%%%%%%%%%%%%%%%%%%%%%%%%%%%%%%%%%%%%%%%%%%%%%%%%%%%%%%%
% When estimating the DOA using techniques such as SVD on a Hankel matrix, the use of multiple snapshots offers significant advantages over a single snapshot. Given a Hankel matrix constructed from multiple snapshots,  \( H_{multi} \), the increased data diversity improves the statistical properties of the matrix. This results in a wider spectral gap between the signal-related singular values \( \sigma_1, \sigma_2, \ldots, \sigma_k \) and the noise-related singular values \( \sigma_{k+1}, \ldots, \sigma_n \), where \( \sigma_1 > \sigma_2 > \ldots > \sigma_n \). The gap is often quantified as  Singular Value Spread (NSS) as \( \sigma_k - \sigma_{k+1} \), which indicates better separability and thus more accurate DOA estimation. Conversely, a Hankel matrix constructed from a single snapshot, \( H_{single} \), contains less information, making it more challenging to distinguish between signal and noise components, resulting in reduced estimation accuracy and higher susceptibility to noise effects. Therefore, multiple snapshots provide better robustness and improved subspace separation for DOA estimation.
%%%%%%%%%%%%%%%%%%%%%%%%%%%%%%%%%%%%%%%%%%%%%%%%%%%%%%%%%%%%%%%%%

Despite its effectiveness, the PMPM approach depends on segments of the received signal being periodic, which requires synchronization and may not be always possible.
%periodic signals 
%Having discussed the multiple snapshots and the noise in the signals within the PMPM approach, 
%it is evident that, despite its effectiveness, PMPM depends on periodic signals and requires synchronization of these segments. 
To address these limitations, a second approach that does not rely on signal periodicity is proposed next.
%Although the PMPM is effective, it requires synchronization of these periodic segments, which adds to the complexity and processing delay of the receiver.}

%%%%%%%%%%%%%%%%%%%%%%%%%%%%%%%%%%%%%%%%%%%%%%%%%%%%%%%%%%%%%%%%%%%%%%%%

\subsection {Single-Phase PC Analog Combiners for MPM~(SPC-MPM)}\label{sec:SPC} %~(MPESC-FPSC)
%Expanded Set of Combiners-Enabled MP~(ESC-MP): Full/Precise Spatial Coverage
%MP with an Expanded Set of Combiners: Full and Precise Spatial Coverage~(MPESC-FPSC)
%MP with a Larger Set of Combiners: Full-Precise Space Coverage~(MPLSC-FPSC)
% Sequential Steering: Fixed-Direction Analog Combiner
% "Matrix Pencil with Expanded Combiners: Precision Spatial Coverage" (MPEC-PSC)
% "Enhanced Spatial Precision: Matrix Pencil with Larger Combiners" (ESP-MPLC)
%%%%%%%%%%%%%%%%%%%%%%%%%%%%%%%%%%%%%%%%%%%%%%%%%%%%%%%%%%%%%%%%%%%
% MP and finer set of analog combiners~(MPFACs) 
%In Section~\ref{sec:PMP-set-analog-combiner-alg}, we assumed the availability of periodic signals. 
To eliminate the contingency on periodic signals characteristic of the approach proposed in Section~\ref{sec:PMP-set-analog-combiner-alg}, we now propose another approach that capitalizes on the block diagonal structure of the analog combiners of PC-HAD receivers. The second approach is introduced in the following theorem. 
%-----------------------------------------------------------------
\begin{theorem} \label{Thr:SPC}
%%%%%%%%%%%%%%%%%%%%%%%%%%%%%%%%%%%%%%%%%%%%%%%%%%%%%%%%%%%%%%%%%%%%%%%%
% Related to the FD the best and the worst cases: best when DoA lies in the columns of the DFT columns and worst when it lies between two analog combiners.
% Let  Let Let,...... then ..... Crisp :)
%%%%%%%%%%%%%%%%%%%%%%%%%%%%%%%%%%%%%%%%%%%%%%%%%%%%%%%%%%%%%%%%%%%%%%%%
Let  $\{\boldsymbol{W}_{\mathrm{\!\! A_{PC}},n}\}_{n=1}^{N}$ be a set of PC-HAD analog combiners defined in~\eqref{eq:W_A_PC_k},
such that  $\boldsymbol{W}_{\mathrm{\!\! A_{PC}},n}^H\boldsymbol{W}_{\mathrm{\!\! A_{PC}},n}=M_{{\mathrm{RF}}}\boldsymbol{I}_{\!L}$, $n=1,\ldots,N$, and 
 $\sum_{n=1}^N\frac{1}{ M_{\mathrm{RF}}}\boldsymbol{W}_{\!\!{\mathrm{A}},n} \boldsymbol{W}_{\!\!{\mathrm{A}},n}^H=\boldsymbol{I}_{\!M}$.
If the number of snapshots processed by the set of  analog combiners is denoted by  $NK$, then 
%For the $n$-th analog combiner $\boldsymbol{W}_{\mathrm{\!\! A_{PC}},n} \in \mathbb{C}^{M \times L}$, 
%with a block diagonal structure that preserves the linear independence of its columns,
%Let %the $M$ phase shifters connected to the $L$ columns of 
%$\boldsymbol{W}_{\mathrm{\!\! A_{PC}},n}$ be given by~\eqref{eq:W_A_PC_k}, for $n=1,\ldots,N$.
%steered to a fixed direction, i.e., $\{\boldsymbol{w}_{{\mathrm{A_{PC}},\ell}}\}_{\ell=1}^L=\boldsymbol{v}_{n}$, where the $m$-th entry of $\boldsymbol{v}_{n}$ is $ e^{\jmath (m-1)\phi_n}$,  where $\phi_n$ is defined in~\eqref{eq:gamma_n}. 
% \blu{Cycling}
% Let $\theta_r$ and $\boldsymbol{a}_r$ denote the DoA and the corresponding steering vector (cf.~\eqref{eq:a}) of the $r$-th source, $r=1,\ldots,R$, respectively.
%%%%%%   Upper bound and lower bound compared to FD 
%%%%%%%%%%%%%%%%%%%%%%%%%%%%%%%%%%%%%%%%%%%%%%%%%%%By cycling over $\{\boldsymbol{W}_{\mathrm{\!\! A_{PC}},n}\}_{n=1}^{N}$, let the  $r$-th source fall within the spatial sector of $\boldsymbol{W}_{\mathrm{\!\! A_{PC}},n}$ or at the edges of the spatial sectors between $\boldsymbol{W}_{\mathrm{\!\! A_{PC}},n}$ and $\boldsymbol{W}_{\mathrm{\!\! A_{PC}},n+1}$.
%Let the DoA of the $r$-th source, $\theta_r\in [)$, then
%Let the $r$-th source fall within the spatial sector of one of the $N$ combiners, $\boldsymbol{W}_{\mathrm{\!\! A_{PC}},n}$,  
%Let the number of sources be $R=1$. Then, 
 the output of $\{\boldsymbol{W}_{\mathrm{\!\! A_{PC}},n}\}_{n=1}^N$, $\{\boldsymbol{Q}_n\}_{n=1}^N$ of the $M$-antenna PC-HAD receiver is equivalent to the input of an FD receiver that: 1) has $L$ antennas with spacing $\Delta_{\mathrm{SPC}}=M_{\mathrm{RF}}\Delta$,   %spacing scaled by  $M_{\mathrm{RF}}$, 
2) uses $K$ snapshots,  % NK
and 3) requires an SNR of the $r$-th source that is at least $10\log M_{\mathrm{RF}} %(\sin{\pi/(2M_{\mathrm{RF}}}))$~dB
\sin^2{(\frac{\pi}{2M_{\mathrm{RF}}})}$~dB higher than the corresponding SNR of the PC-HAD receiver.   %the FD receiver %, i.e., 
In particular,
%Assuming a single source, the output of the analog combiner, cf. $\boldsymbol{Q}_n$ in~\eqref{eq:Q_n}, compared to that of an FD receiver, cf. $\boldsymbol{X}$ in~\eqref{eq:X}, can be expressed as:
\begin{equation}\label{eq:snr_SPC_FD}
\!\!\{\boldsymbol{Q}_n\}_{n=1}^N \Big|_{\substack{\hspace{-4.3cm}\tilde{K}=NK \\ \mathrm{SNR}_{\mathrm{SPC}, r} \in \bigl[\frac{P_r}{M_{\mathrm{RF}}\sin^2({\pi/(2M_{\mathrm{RF}}}))},      M_{\mathrm{RF}}P_r\bigr] }}  \!\!\!\!\!
=\! \boldsymbol{X} \Big|_{\substack{\hspace{-1.cm} \tilde{K}=K \\ \mathrm{SNR}_{\mathrm{FD},r}=P_r}}.
\end{equation}  
%where the $k$-th snapshot, i.e., the $k$-th column, of the received signals, has a dimension of $L$ in the SPC approach and a dimension of $M$ in the FD receiver.
\end{theorem}
\begin{IEEEproof}
See Appendix \ref{sec: proof-threorm2}.
\end{IEEEproof}

\begin{remark}\label{remark:SNR-SPC}
The lower bound of $\mathrm{SNR}_{\mathrm{SPC}, r}$ in~\eqref{eq:snr_SPC_FD} increases monotonically with $M_{\mathrm{RF}}$. Hence, this bound % corresponds 
asymptotically approaches  %at least 
$10\log \frac{4}{\pi^2} M_{\mathrm{RF}}P_r$~dB as $M_{\mathrm{RF}}\to\infty$. \hfill~$\Box$
% in lim M_rf goes to inf the answer can be function of M_rf if it is domentaed by M_rf as we have in this case. In Taylor expansion of func= x+x^2 +x^3 if x goes to zero it will be dominated by x. 
%----------------------------------------------------
% in the asymptotic worst case.
%$0.4MP_r= -3.9$~dB.
% This result follows from $\lim_{M_{\mathrm{RF}} \to \infty} M^2_{\mathrm{RF}} \sin^2 (\frac{\pi}{2 M_{\mathrm{RF}}}) = \lim_{x \to 0} \frac{\pi^2}{4}(\frac{\sin x}{x})^2$, with $x=\frac{\pi}{2M_{\mathrm{RF}}}$. 
% For a fixed $M$, decreasing $L$ increases $M_{\mathrm{RF}} = \frac{M}{L}$, leading asymptotically to $\pi^2/4$. %$\frac{\pi^2}{4}$
%leading asymptotically to
%the asymptotic behavior toward $\pi^2/4$.
% and achieving the lower bound.
%driving the asymptotic behavior toward 
%$M_{\mathrm{RF}}=\frac{\pi}{2x}$
%$\lim_{M_{\mathrm{RF}} \to \infty} M^2_{\mathrm{RF}}\sin^2(\frac{\pi}{2 M_{\mathrm{RF}}})= \frac{\pi^2}{4}(\frac{\sin x}{x})^2= \frac{\pi^2}{4}$, where $x=\frac{\pi}{2M_{\mathrm{RF}}}$. % when $M$ is fixed and $L$
%This is by denoting $M_{\mathrm{RF}}=\frac{\pi}{u}$
%\blu{Remark on the the worst SNR of SPC.}    
\end{remark}

The equivalence of $\boldsymbol{Q}_n$ and $\boldsymbol{X}$  elucidated in Theorem~\ref{Thr:SPC} shows that setting $\boldsymbol{w}_{{\mathrm{A_{PC}},\ell}}=\boldsymbol{v}_n$ enables the MPM 
%readily verified that the MPM 
to extract the DoAs in the PC-HAD architecture, albeit with %\blu{SNR} and 
the snapshot penalty and SNR gain outlined in the theorem.
% the output signals $\boldsymbol{q}_k$ in~\eqref{eq:q_n_ell_sec} after being processed by $\boldsymbol{W}_{\mathrm{\!\! A_{PC}},n}$ depend only on a single index $r$, corresponding to the $r$-th source.
%This is because the phases  $\{\phi_\ell\}_{\ell=1}^{L}$  are constant at $\phi_n$ and the  $R$ source phases $\{\mu_r\}_{r=1}^R$  are shifted by 
%$\phi_n$, i.e., $\{\mu_r-\phi_n\}_{r=1}^R$ in~$\{g(\mu_r-\phi_n)\}_{r=1}^R$, cf.~\eqref{eq:f_ell_mu}. %, in  the PC-HAD receiver.
%across the vector. 
%As a result, the output of $\boldsymbol{W}_{\mathrm{\!\! A_{PC}},n}$, i.e., $\boldsymbol{Q}_n$ in~\eqref{eq:Q_n}, is amenable to the MPM to extract the DoAs in PC-HADs.
% Hence, the MPM can extract the DoAs in the considered PC-HAD architecture. 

To apply the MPM, we assume that $\theta_r\in\Theta_n$, the spatial sector covered by $\boldsymbol{W}_{\mathrm{\!\! A_{PC}},n}$. We begin by considering a single snapshot of the signal arriving from this DoA, 
%say the $k$-th snapshot,
%the $k$-th snapshot, 
i.e, the $k$-th column of $\boldsymbol{Q}_n$ in~\eqref{eq:Q_n},  $\boldsymbol{q}_k \in \mathbb{C}^{L \times 1}$, %the output of $\boldsymbol{W}_{\mathrm{\!\! A_{PC}},n}$, 
%we begin by considering a single snapshot of the output of the $n$-th analog combiner %, say the $k$-th snapshot,
in the noise-free case. 
%In the noise-free case, 
The Hankel matrix, $\boldsymbol{H}_{k}^{{\mathrm{SPC}}}\in \mathbb{C}^{(L-\xi) \times(\xi+1)}$, is constructed as:  %from $\boldsymbol{q}_k$,  as follows: 
\begin{align}
&\!\!\!\!\!\!\!\!\boldsymbol{H}_{k}^{\mathrm{SPC}} = \nonumber \\
\!\!\!\sum_{r=1}^R & \scalebox{0.85}{$
\begin{bmatrix}
 s_r[k] g(\mu_r - \phi_n) e^{\jmath \psi_{r,1}} 
 & \cdots &  s_r[k] g(\mu_r - \phi_n)e^{\jmath \psi_{r,(\xi +1)}} \\
\vdots & \ddots & \vdots \\
 s_r[k] g(\mu_r - \phi_n)e^{\jmath \psi_{r,{L-\xi}}} & \cdots &  s_r[k] g(\mu_r - \phi_n)e^{\jmath \psi_{r,{L}}}
\end{bmatrix}
$}.
\label{eq:H_SPC}
\end{align}
%%%%%%%%%%%%%%%%%%%%%%%%%%%%%%%%%%%%%%%%%%%%%%%%%%%%%%%%%%%%%%%%%%%%%%%%
%where $\xi_{_\mathrm{SPC}}$ is the MPM parameter for SPC-MPM approach.
Analogous to FD and PMPM, a matrix pencil in the SPC-MPM approach can be synthesized from the two matrices, $\boldsymbol{H}_{k,1}^{{\mathrm{SPC}}}$ and $\boldsymbol{H}_{k,2}^{{\mathrm{SPC}}} \in \mathbb{C}^{(L-\xi) \times \xi}$, which are obtained by deleting the last and first columns of $\boldsymbol{H}_{k}^{{\mathrm{SPC}}}$, respectively.
% For each $\boldsymbol{H}_{k}^{{\mathrm{SPC}}}$, we introduce two matrices, denoted as $\boldsymbol{H}_{k,1}^{{\mathrm{SPC}}}$ and $\boldsymbol{H}_{k, 2}^{{\mathrm{SPC}}} \in \mathbb{C}^{(L-\xi) \times \xi}$.
% %, each having dimensions $(L-\xi) \times \xi$. % For the sake of brevity, we omit the time index $n$. 
% These matrices are be obtained by deleting the last and first columns of $\boldsymbol{H}_{k}^{{\mathrm{SPC}}}$.
Hence,  $\boldsymbol{H}_{k,1}^{{\mathrm{SPC}}}$ and $\boldsymbol{H}_{k,2}^{{\mathrm{SPC}}}$ can be expressed as: % follows:
\begin{align}
& {\boldsymbol{H}_{k,1}^{{\mathrm{SPC}}} =\boldsymbol{\Pi}_1^{{\mathrm{SPC}}} \boldsymbol{\Lambda}_{k}^{\mathrm{SPC}}\boldsymbol{\Pi}_2^{{\mathrm{SPC}}} }, \label{eq:H_SPC_1}\\ &
\boldsymbol{H}_{k,2}^{{\mathrm{SPC}}} =\boldsymbol{\Pi}_1^{{\mathrm{SPC}}} \boldsymbol{\Lambda}_{k}^{\mathrm{SPC}}\boldsymbol{\Pi}_0^{{\mathrm{SPC}}} \boldsymbol{\Pi}_2^{{\mathrm{SPC}}}, \label{eq:H_SPC_2}
\end{align}
%\begin{align}
%& {\boldsymbol{Y}^{(n)}_1 =\boldsymbol{\Pi}_1 %\boldsymbol{R}^{(n)}\boldsymbol{\Pi}_2},\\ &
%{\boldsymbol{Y}^{(n)}_2=\boldsymbol{\Pi}_1 \boldsymbol{R}^{(n)} \boldsymbol{\Pi}_0 \boldsymbol{\Pi}_2,} 
%\end{align}
where $\boldsymbol{\Pi}_1^{{\mathrm{SPC}}}  \in \mathbb{C}^{(L-\xi) \times R} $,  $\boldsymbol{\Pi}_2^{{\mathrm{SPC}}}  \in \mathbb{C}^{R \times \xi} $, $\boldsymbol{\Pi}_0^{{\mathrm{SPC}}}$ and  $\boldsymbol{\Lambda}_{k}^{\mathrm{SPC}} \in \mathbb{C}^{R\times R}$ are given by:
\begin{align}
\!\!\!\boldsymbol{\Pi}_1^{{\mathrm{SPC}}} & \!\!= \!\!  
%\left[\begin{smallmatrix}
\scalebox{0.85}{$
\begin{bmatrix}
1 & \ldots &  1 \\
e^{\jmath  M_{\mathrm{RF}} \mu_1 }  & \ldots &  e^{\jmath M_{\mathrm{RF}} \mu_{R} } \\
\vdots & \ddots & \vdots \\
e^{\jmath (L- \xi-1) M_{\mathrm{RF}} \mu_1 } &  \ldots &  e^{\jmath  (L- \xi-1) M_{\mathrm{RF}} \mu_{R}} 
\end{bmatrix} $},
%\end{smallmatrix}\right],
\label{eq:Z_1_SPC}
\end{align}
% \begin{align}
% \boldsymbol{\Pi}_1 & =  
% \scalebox{0.8}{$
% \left[
% \begin{array}{ccc}
% 1 & \cdots &  1 \\
% e^{\jmath  M_{\mathrm{RF}} \mu_1 }  & \cdots &  e^{\jmath M_{\mathrm{RF}} \mu_{R} } \\
% \vdots & \ddots & \vdots \\
% e^{\jmath (L- \xi-1) M_{\mathrm{RF}} \mu_1 } &  \cdots &  e^{\jmath  (L- \xi-1) M_{\mathrm{RF}} \mu_{R}} 
% \end{array}
% \right]
% $},
% \label{eq:Z_1}
% \end{align}
%--------------------------------------------------------------
\begin{align}
\boldsymbol{\Pi}_2^{{\mathrm{SPC}}}  &=
\scalebox{0.85}{$
\begin{bmatrix}
%\left[\begin{smallmatrix}
1 & e^{\jmath  M_{\mathrm{RF}} \mu_1 } & \ldots &  e^{\jmath (\xi-1) M_{\mathrm{RF}} \mu_1} \\
\vdots & \vdots & \ddots & \vdots \\
1 & e^{\jmath  M_{\mathrm{RF}} \mu_{R} } & \ldots &  e^{\jmath (\xi-1) M_{\mathrm{RF}} \mu_{R}} 
\end{bmatrix}$},
%\end{smallmatrix}\right],
\label{eq:Z_2_SPC}
\end{align}
% \begin{align}
% \boldsymbol{\Pi}_2 &=
% \scalebox{0.8}{$
% \left[
% \begin{array}{cccc}
% 1 & e^{\jmath  M_{\mathrm{RF}} \mu_1 } & \cdots &  e^{\jmath (\xi-1) M_{\mathrm{RF}} \mu_1} \\
% \vdots & \vdots & \ddots & \vdots \\
% 1 & e^{\jmath  M_{\mathrm{RF}} \mu_{R} } & \cdots &  e^{\jmath (\xi-1) M_{\mathrm{RF}} \mu_{R}} 
% \end{array}
% \right]
% $},
% \label{eq:Z_2}
% \end{align}
%-----------------------------------------------------------------
\begin{align}
& \boldsymbol{\Pi}_0^{{\mathrm{SPC}}}  =\operatorname{diag}\big(e^{\jmath M_{\mathrm{RF}}\mu_1 },\ldots, e^{\jmath M_{\mathrm{RF}} \mu_{R} }\big), \label{eq:Z_0_SPC}\\ 
%-------------------------------------------------------------------------------------
&\!\!\!\boldsymbol{\Lambda}_{k}^{\mathrm{SPC}}\!\!\!= \!\operatorname{diag}\bigl(s_{1}[k] g(\mu_1 - \phi_n), \ldots, s_{R}[k] g(\mu_{R} - \phi_n)\bigr).
\label{eq:R_n_SPC}
\end{align}
% By considering the matrix pencil  %generalized eigenvalue problem, 
% %\begin{equation}
% $\boldsymbol{H}_{k,2}^{{\mathrm{SPC}}}  - \zeta \boldsymbol{H}_{k,1}^{{\mathrm{SPC}}} 
% = \boldsymbol{\Pi}_1^{{\mathrm{SPC}}} \boldsymbol{\Lambda}_{k}^{\mathrm{SPC}} \bigl( \boldsymbol{\Pi}_0^{{\mathrm{SPC}}} - \zeta \boldsymbol{I}_{\!R} \bigr) \boldsymbol{\Pi}_2^{{\mathrm{SPC}}}$.
%\label{eq:generalized_eignvalue}
%\end{equation}
%Analogous to FD, 
The diagonal elements of  $\boldsymbol{\Pi}_0^{{\mathrm{SPC}}}$ are the generalized eigenvalues of the matrix pair  $(\boldsymbol{H}_{k,2}^{{\mathrm{SPC}}},\boldsymbol{H}_{k,1}^{{\mathrm{SPC}}})$, i.e., the eigenvalues of $(\boldsymbol{H}_{k,1}^{{\mathrm{SPC}}})^{\dag} \boldsymbol{H}_{k,2}^{{\mathrm{SPC}}}$. 
%In particular, the $r$-th diagonal entry of  $\boldsymbol{\Pi}_0^{{\mathrm{SPC}}}$,  is equal to the $r$-th eigenvalue of $(\boldsymbol{H}_{k,1}^{\mathrm{SPC}})^{\dag} \boldsymbol{H}_{k,2}^{\mathrm{SPC}}$. 
Hence, using~\eqref{eq:mu} yields the following estimate of the  $r$-th DoA: 
%can be extracted using~\eqref{eq:mu} as f:
% of the matrix pair $\left\{\boldsymbol{H}_{k,2}^{{\mathrm{SPC}}},\boldsymbol{H}_{k,1}^{{\mathrm{SPC}}} \right\}$. 
%Thus, the  DoAs can be extracted by: % there are no grating lobes in this method.
\begin{equation}
   \Hat{\theta}_r=\arcsin \Biggl( \frac{\lambda \Im \{\log {[\boldsymbol{\Pi}_0^{{\mathrm{SPC}}}}]_{r,r}\}}{2 \pi M_{\mathrm{RF}} \Delta} \Biggr).
   % =\arcsin \Biggl \{ \frac{\lambda \operatorname{\Im }\{\log {e^{\jmath M_{\mathrm{RF}} \mu_{r} }}\}}{2 \pi M_{\mathrm{RF}} \Delta} \Biggr\}.
   \label{eq:theta_est_SPC}
\end{equation} 
%It is worth noting that, in the SPC-MPM approach, the Hankel matrix is constructed using a received signal vector of length $L$ rather than $M$, as in the FD and the PMPM approach, leading to comparatively lower accuracy.
%-----------------------------------------------------------
% \begin{itemize}
%     \item \blu{Write a note that the $g(\mu_r-\phi_n)$ contains the DoA information but we didn't use it in SPC-MPM.}
% \end{itemize}
%%%%%%%%%%%%%%%%%%%%%%%%%%%%%%%%%%%%%%%%%%%%%%%%%%%%%%%%%%%%%%%%%%%%%%%%%%%%%%%%%%%%%%%
% \begin{remark}\label{remark:SNR-SPC2}
% A limitation of the SPC-MPM approach is that it relies only on the phase information $\psi_{r,\ell} = (\ell - 1) \mu_r M_{\mathrm{RF}}$ embedded in $\boldsymbol{Q}_n$ for estimating DoAs, while ignoring the amplitude and phase  DoA-related information in $g(\mu_r-\phi_n)$.
% Although this could potentially improve the estimation accuracy, SPC-MPM  exhibits a small gap relative to the corresponding CRLB, %with negligible impact on results,
% as demonstrated numerically in Section~\ref{sec:simulation} below.
% \hfill~$\Box$
% \end{remark}
%%%%%%%%%%%%%%%%%%%%%%%%%%%%%%%%%%%%%%%%%%%%%%%%%%%%%%%%%%%%%%%%%%%%%%%%%%%%%%%%%%%%%%%%

Since the spatial sector in which $\theta_r$ lies is not known \emph{a priori}, $r=1,\ldots, R$, the entire space must be covered by an exhaustive set of analog combiners. One such set is defined in~\eqref{eq:W_A_PC_k}.
%Towards that end, the PC-HAD receiver cycles over the exhaustive set of $N$ analog combiners, $\{\boldsymbol{W}_{\mathrm{\!\! A_{PC}},n}\}_{n=1}^N$, %$N= \tfrac{M}{L}$ 
%defined in~\eqref{eq:W_A_PC_k}. 
Now, for $K$ snapshots to be received from each DoA, the total number of snapshots must be $\tilde{K}=NK$; $\tilde{K}$ should be, at least, equal to $N$. 
% \blu{It can be noted that the $k$-th  snapshot of the output signals after applying any analog combiner from the set  $\{\boldsymbol{W}_{\mathrm{\!\! A_{PC}},n}\}_{n=1}^{N}$, given by $\boldsymbol{q}_k$ in~\eqref{eq:q_n_ell_sec}, differ across these $N$ combiners only by $\{s_{r}[k] g(\mu_r - \phi_n)\}_{r=1}^R$  for $n=1,\ldots, N$.}
To use $K$ snapshots per analog combiner to extract the DoAs,  a technique analogous to the one outlined in Section~\ref{sec:multiple-snapshots-Noiseless} can be used with the 
SPC-MPM approach.
%can be used to %, for $\tilde{K}= NK$ with $K \geq 1$, extract the DoAs using a technique analogous to the one outlined in Section~\ref{sec:multiple-snapshots-Noiseless}. 
In particular, a Hankel matrix, i.e., $\boldsymbol{H}_{k}^{{\mathrm{SPC}}}$ in~\eqref{eq:H_SPC}, is constructed for each snapshot. The collection of $K_A=\tilde{K}$ matrices corresponding to $\tilde{K}$ snapshots are concatenated into the augmented matrix $\boldsymbol{H}_\mathrm{\!A}$ in~\eqref{eq:Y_E}.  The Hankel matrices constructed from the $K$ snapshots of the $n$-th analog combiner, i.e., $\boldsymbol{H}_\mathrm{\!A} \in \mathbb{C}^{(L-\xi)\times K_{\mathrm{A}} (\xi+1)}$ can be rewritten as:
\begin{equation}
   \!\!\!\!\boldsymbol{H}_\mathrm{\!A} \!\!= \!
    \left[
    \begin{array}{ccccccc}
    \boldsymbol{H}_{\!1} \!\!& \cdots & \!\! \boldsymbol{H}_{\!K} \!\!& \cdots & \!\! \boldsymbol{H}_{\!K_A - K + 1}  \!\!& \cdots & \!\! \boldsymbol{H}_{\!K_{\mathrm{A}}}
    \end{array}
    \right]. \label{eq:Y_E_SPC} % Equation number aligned here
\end{equation}
\vspace{-2.2em} % Adjust spacing between the two parts
\[
\begin{array}{ccccccc}
\hspace{-1.1em} % Horizontal alignment for the first underbrace
\underbrace{\hphantom{\boldsymbol{H}_{\!1} \;\; \cdots \;\; \boldsymbol{H}_{\!K}}}_{\substack{\text{correspond to } \boldsymbol{W}_{\mathrm{\!\! A_{PC}},1}}}
& & & \hspace{-0.8em} % Horizontal alignment for the second underbrace
\underbrace{\hphantom{\boldsymbol{H}_{\!K_A - K + 1} \;\; \cdots \;\; \boldsymbol{H}_{\!K_{\mathrm{A}}}}}_{\substack{\text{correspond to } \boldsymbol{W}_{\mathrm{\!\! A_{PC}},N}}}
\end{array}
\]
%%%%%%%%%%%%%%%%%%%%%%%%%%%%%%%%%%%%%%%%%%%%%%%%%%%%%%%%%%%%%%%%%%%%%%%%%%%%%%%%%
%Each This augmented matrix contains $K_{\mathrm{A}} = NK$ matrices,  where every group of $K$ 
%consecutive Hankel matrices is constructed from the columns of $\boldsymbol{Q}_n$ in~\eqref{eq:Q_n}, i.e., $\{\boldsymbol{q}_k\}_{k=1}^K$, for all $n$-th analog combiners in the set $\{\boldsymbol{W}_{\mathrm{\!\! A_{PC}},n}\}_{n=1}^{N}$.
% the outputs, i.e., the columns of $\boldsymbol{Q}_n$, i.e., $\{\boldsymbol{q}_k\}_{k=1}^K$, of one analog combiner from the set
% $\{\boldsymbol{W}_{\mathrm{\!\! A_{PC}},n}\}_{n=1}^{N}$.
% which consists of
% comprising $K_{\mathrm{A}} = NK$ matrices,
% with every consecutive $K$ Hankel matrices corresponding to outputs of one analog combiner from $\{\boldsymbol{W}_{\mathrm{\!\! A_{PC}},n}\}_{n=1}^{N}$.
%%%%%%%%%%%%%%%%%%%%%%%%%%%%%%%%%%%%%%%%%%%%%%%%%%%%%%%%%%%%%%%%%%%%%%%%%%%%%%%%%%%%%%%%%%

A matrix pencil is constructed using two Hankel-based matrices $\boldsymbol{H}_{\mathrm{\!A}, -\mathcal{I}}$ and $\boldsymbol{H}_{\mathrm{\!A}, -\mathcal{J}} \in \mathbb{C}^{(L-\xi)\times K_{\mathrm{A}} \xi}$, which are obtained from $\boldsymbol{H}_\mathrm{\!A}$ in~\eqref{eq:Y_E_SPC} as in~\eqref{eq:H_A_1} and~\eqref{eq:H_A_2}.  
%Expressing $\boldsymbol{H}_{\mathrm{\!A}, -\mathcal{I}}$ and $\boldsymbol{H}_{\mathrm{\!A}, -\mathcal{J}}$ as in~\eqref{eq:Y_E_1} and~\eqref{eq:Y_E_2}, 
Expressing these matrices in the forms of~\eqref{eq:Y_E_1} and~\eqref{eq:Y_E_2}
enables extracting the DoAs using the MPM outlined in Section~\ref{sec:multiple-snapshots-Noiseless} 
with $\boldsymbol{\Pi}_1^{{\mathrm{SPC}}}$, $\boldsymbol{\Pi}_2^{{\mathrm{SPC}}}$, $\boldsymbol{\Pi}_0^{{\mathrm{SPC}}}$ and $\boldsymbol{\Lambda}_{k}^{\mathrm{SPC}}$ defined in~\eqref{eq:Z_1_SPC}, \eqref{eq:Z_2_SPC}, \eqref{eq:Z_0_SPC} and~\eqref{eq:R_n_SPC}, respectively.
Note that, unlike FD and %\blu{PMPM} 
the PMPM approach 
in HAD receivers, the number of antennas is virtually equal to $L$ rather than $M$. %,a dbt ehantenna 
Finally, in the presence of noise, the technique in Section~\ref{sec:multiple-snapshots-Noisy} can be used to estimate 
the DoAs  using~\eqref{eq:theta_est_aug_noise} but with $\Delta \leftarrow M_{\mathrm{RF}}\Delta$.

% \begin{equation}
% \Hat{\theta}_r \!= \!\arcsin \biggl(\frac{\lambda  \Im \left\{\log\bigl( \nu_r\bigl((\hat{\boldsymbol{H}}_{\!\mathrm{A}, -\mathcal{I}})^{\dag} \hat{\boldsymbol{H}}_{\!\mathrm{A}, -\mathcal{J}} \bigr) \right\}}{2 \pi M_{\mathrm{RF}}\Delta} \biggr).
% \label{eq:theta_est_aug_noise_2}
% \end{equation}
% Key obsevation paragraph
\begin{remark}\label{remark:SNR-SPC2}
%To apply the MPM  
%A potential 
A limitation of the SPC-MPM DoA estimation approach is that it relies only on the phase information, $\psi_{r,\ell} = (\ell - 1) \mu_r M_{\mathrm{RF}}$, embedded in $\boldsymbol{Q}_n$, but does not exploit the amplitude and phase information embedded in $g(\mu_r-\phi_n)$.
Although exploiting this information could potentially improve the estimation accuracy, such improvement is likely marginal because the gap between the variance of the estimates generated by the SPC-MPM  approach and the corresponding CRLB is typically small, %exhibits a small gap relative to the corresponding CRLB, %with negligible impact on results,
as %will be numerically 
demonstrated in Section~\ref{sec:simulation} below.
% only the phases $\{M_{\mathrm{RF}} \mu_r\}_{r=1}^R$  
% in $\boldsymbol{\Pi}_0^{{\mathrm{SPC}}}$ %in~\eqref{eq:Z_0_SPC}
% are used to extract the DoAs, while $\boldsymbol{\Lambda}_{k}^{\mathrm{SPC}}$ in~\eqref{eq:R_n_SPC}  contains additional unexploited DoA-related information  $\{g(\mu_r-\phi_n)\}_{r=1}^R$ that could potentially enhance estimation accuracy. 
\hfill~$\Box$
%------------------------------------------------------------------------
% In the SPC-MPM approach,  only the phases $\{M_{\mathrm{RF}} \mu_r\}_{r=1}^R$  
% in $\boldsymbol{\Pi}_0^{{\mathrm{SPC}}}$
% %in~\eqref{eq:Z_0_SPC}
% are used to extract the DoAs. However, $\boldsymbol{\Lambda}_{k}^{\mathrm{SPC}}$ in~\eqref{eq:R_n_SPC} contains $\{g(\mu_r-\phi_n)\}_{r=1}^R$,  providing additional unused DoA-related information that could improve estimation accuracy if incorporated.
%--------------------------------------------------------------------------
%providing additional DoA-related information that, though unused in the SPC-MPM approach,  can be exploited to improve estimation accuracy if incorporated.
%providing additional information about the DoAs. This information is unused in the current formulation but could enhance estimation performance if incorporated.
% Using $\boldsymbol{\Lambda}_{k}^{\mathrm{SPC}}$ in~\eqref{eq:R_n_SPC}, we note that $\{g(\mu_r-\phi_n)\}_{r=1}^R$ contains information about the phases  $\{\mu_r\}_{r=1}^R$ of the $R$ sources. However, in the SPC-MPM approach, only the phases $\{M_{\mathrm{RF}} \mu_r\}_{r=1}^R$  
% in $\boldsymbol{\Pi}_0^{{\mathrm{SPC}}}$
% %in~\eqref{eq:Z_0_SPC}
% are considered for extracting the DoAs.
\end{remark}
%%%%%%%%%%%%%%%%%%%%%%%%%%%%%%%%%%%%%%%%%%%%%%%%%%%%%%%%%%%%%%%%%%%%%%%%%%%%
\begin{remark}\label{remark:SNR-SPC3}
% \blu{When $\theta_r$ falls at the edges of the spatial sectors of 
% $\boldsymbol{W}_{\!\!\mathrm{A_{PC}},n}$ and $\boldsymbol{W}_{\!\!\mathrm{A_{PC}},n+1}$, i.e.,
% $\theta_r \in \Theta_n \cap \Theta_{n+1}$, using outputs from all combiners in $\{\boldsymbol{W}_{\mathrm{\!\! A_{PC}},n}\}_{n=1}^{N}$ improves DoA estimation accuracy. Thus, the lower bound of $\mathrm{SNR}_{\mathrm{SPC}, r}$ in Theorem~\ref{Thr:SPC} derived based on one analog combiner, becomes less tight.} %  becomes loose
%--------------------------------------------------------------------
% When the DoA lies at the boundary between  
% $\boldsymbol{W}_{\!\!\mathrm{A_{PC}},n}$ and $\boldsymbol{W}_{\!\!\mathrm{A_{PC}},n+1}$, using  snapshots  from all analog combiners in $\{\boldsymbol{W}_{\mathrm{\!\! A_{PC}},n}\}_{n=1}^{N}$ improves DoA estimation accuracy. Thus, the lower bound of $\mathrm{SNR}_{\mathrm{SPC}, r}$ in Theorem~\ref{Thr:SPC} derived based on one analog combiner, becomes less tight. 
%-----------------------------------------------------------------------------
The lower bound on $\mathrm{SNR}_{\mathrm{SPC}, r}$ in Theorem~\ref{Thr:SPC} is derived when the output of only one analog combiner is used for DoA estimation. Since the output of all analog combiners, $\{\boldsymbol{W}_{\mathrm{\!\! A_{PC}},n}\}_{n=1}^{N}$, contain DoA information, a strictly higher SNR can be achieved by concatenating the corresponding Hankel matrices as in~\eqref{eq:Y_E_SPC}. 
% Using  $K$  snapshots for each analog combiner in 
% %per combiner in
% $\{\boldsymbol{W}_{\mathrm{\!\! A_{PC}},n}\}_{n=1}^{N}$, along with the outputs from all combiners, improves the accuracy of DoA estimation.
% %enhances the accuracy of DoA estimation by utilizing outputs from all combiners. 
% Hence, when the DoA lies at the boundary between the spatial sectors,  $\boldsymbol{W}_{\!\!\mathrm{A_{PC}},n}$ and $\boldsymbol{W}_{\!\!\mathrm{A_{PC}},n+1}$, the lower bound of $\mathrm{SNR}_{\mathrm{SPC}, r}$ in Theorem~\ref{Thr:SPC}, derived from a single combiner, becomes less tight. 
%  becomes loose
%loosening the $\mathrm{SNR}_{\mathrm{SPC}, r}$ lower bound derived based on one analog combiner in Theorem~\ref{Thr:SPC} less tight.  
\hfill~$\Box$
\end{remark}
%%%%%%%%%%%%%%%%%%%%%%%%%%%%%%%%%%%%%%%%%%%%%%%%%%%%%%%%%%%%%%%%%%%%%%%%%%%%%
%\begin{remark}\label{remark:DoA-ambiguity}
\begin{corollary}\label{cor:DoA-ambiguity}
% it cannot be corollary as it is negative 
If the spacing between the antennas of the original array is $\Delta>\frac{\lambda}{2M_{\mathrm{RF}}}$, 
the effective spacing between the $L$ elements of the virtual antenna formed by the $L$ subarrays causes the DoA estimates generated by the SPC-MPM approach to be non-unique. In particular, each DoA will result in $M_{\mathrm{RF}}$ estimates, of which $M_{\mathrm{RF}}-1$ are spurious.
% The output signals $\{q_{k}[\ell]\}_{\ell=1}^L$  from $L$ RF chains in the considered PC-HAD architecture form an $L$-antenna digital virtual array with a spacing of $M_{\mathrm{RF}}\Delta$  between the virtual antennas, resulting in direction-finding ambiguity.
\end{corollary}
%\end{remark}
%%%%%%%%%%%%%%%%%%%%%%%%%%%%%%%%%%%%%%%%%%%%%%%%%%%%%%%%%%%%%%%%%%%%%%%%%%%
% \begin{IEEEproof}
% See Appendix \ref{sec: proof-corr1}.
% \end{IEEEproof}
%%%%%%%%%%%%%%%%%%%%%%%%%%%%%%%%%%%%%%%%%%%%%%%%%%%%%%%%%%%%%%%%%%%%%%%%%%
\begin{IEEEproof}
% From Theorem~\ref{Thr:SPC}, 
% %it can be verified that 
% %the output signals $\{q_{k}[\ell]\}_{\ell=1}^L$ in~\eqref{eq:q_n_ell_sec} form a digital virtual array, with each subarray acting as a 
% the spacing between the elements of the virtual antenna array is $M_{\mathrm{RF}}\Delta$, causing the phases in the virtual steering vector, $\psi_{r,\ell} = (\ell - 1) \mu_r M_{\mathrm{RF}}$ in~\eqref{eq:q_n_ell_sec} to be multiples of $M_{\mathrm{RF}}\mu_r$ instead of $\mu_r$, as in FD receivers. When $M_{\mathrm{RF}} \mu_r>2\pi$, phase ambiguity arises, because phases, $\psi_{r,\ell} = (\ell - 1) \mu_r M_{\mathrm{RF}}$ in~\eqref{~\eqref{eq:q_n_ell_sec}}, in the form of $M_RF \mu_r+2\pi i$ yield the same virtual steering vector. In other words, the phases $\mu_i=\mu_r+2\pi i/M_{RF}$, $\mu_i\in(-\pi,\pi]$ would yield the same virtual steering vector. 
From Theorem~\ref{Thr:SPC}, the spacing between the elements of the virtual antenna array is $M_{\mathrm{RF}}\Delta$, causing the phases in the virtual steering vector, $\psi_{r,\ell} = (\ell - 1) \mu_r M_{\mathrm{RF}}$ in~\eqref{eq:q_n_ell_sec}, to be multiples of $M_{\mathrm{RF}}\mu_r$ instead of $\mu_r$, as in FD receivers. 
When $M_{\mathrm{RF}} \mu_r > 2\pi$, phase ambiguity arises because phases of the form $M_{\mathrm{RF}} \mu_r + 2\pi i$, $i\in \mathbb{Z}$, yield the same virtual steering vector. In other words,  $\tilde{\mu}_i = \mu_r + \tfrac{2\pi i}{M_{\mathrm{RF}}}$,   $\tilde{\mu}_i \in (-\pi, \pi]$, represents the phase of the $i$-th grating lobe,  and  yields the same virtual steering vector.
Hence, there are $M_{\mathrm{RF}}$ distinct phases associated with $\mu_r$, and  the set of $M_{\mathrm{RF}} R$ phases for the $R$ sources can be expressed as:
 \begin{align}\label{eq:G_r}
\!\mathcal{G} &= \bigl\{ %\mu_r + \tfrac{2\pi i}{M_{\mathrm{RF}}} 
  \tilde{\mu}_i \;\big|\; %i \in \mathbb{Z},
  \bigl\lceil \tfrac{M_{\mathrm{RF}}}{2} \bigl(-1 - \tfrac{\mu_r }{\pi}\bigr) \bigr\rceil \leq i \leq \bigl\lfloor \tfrac{M_{\mathrm{RF}}}{2} \bigl(1 - \tfrac{\mu_r}{\pi}\bigr) \bigr\rfloor, i \in \mathbb{Z}, &\nonumber \\
  &\; \qquad \qquad \qquad \qquad  \qquad \qquad \qquad r = 1, \ldots, R \bigr\}.
  %  &\qquad \qquad \qquad \qquad \qquad \qquad \qquad \quad r=1,\ldots, R.
\end{align}
\end{IEEEproof}

We now present a method for resolving the DoA ambiguity elucidated in Corollary~\ref{cor:DoA-ambiguity}.

%---------------------------Power optimization--------------------------
% mention why snr maximize the true DoA not spurious, 
% Q, q is used before, change notation
% mention the analog combiner similar to DFT using the phases
% dont mention the K1 and K2
% add remark for total number of snapshots indicate percentage 10 or 1 for resolve ....
\subsubsection{Resolving DoAs Ambiguity} 
\label{sec:Optimizing-Power}
%For $R$ sources, after using the SPC-MPM approach, we have $M_{\mathrm{RF}} R$ angles, only $R$ are desired, and the rest correspond to the spurious angles. 
% refer to cor 1
%%%%%%%%%%%%%%%%%%%%%%%%%%%%%%%%%%%%%%%%%%%%%%%
%When $\Delta > \frac{\lambda}{2M_{\mathrm{RF}}}$, the SPC-MPM approach leads to $M_{\mathrm{RF}} R$ estimates for $R$  sources, as outlined in Corollary~\ref{cor:DoA-ambiguity}. 
%%%%%%%%%%%%%%%%%%%%%%%%%%%%%%%%%%%%%%%%%%%%%%%%%
%Among these, only $R$ estimates correspond to the true DoAs, while the remaining estimates are spurious.
%To resolve DoA ambiguity and obtain the $R$ true DoAs, our main philosophy is that to cycle over a set of analog combiners and construct these combiners based on the estimates obtained in~\eqref{eq:G_r}, then true $R$ DoAs are then identified as those associated with the phases that maximize the instantaneous SNR across the RF chains for all combiners.  
%To address DoA ambiguity, the proposed method is to cycle over 
%------------------------------------------------
% >> add The true DoAs correspond to phases that maximize the SNR of the received signals, while spurious angles result in lower SNR values due to phase misalignment.
% >> Why G = M_{\mathrm{RF}}R/L
%------------------------------------------------
To address DoA ambiguity, we note that only one of the $M_{\mathrm{RF}}$ estimated phases for each source corresponds to the true DoA.  %The true DoA corresponds to phase that maximize the instantaneous SNR of the received signals, while spurious angles result in lower SNR values. % due to phase misalignment.
Such phase 
%The true DoA 
can be identified to be the one that maximizes the SNR; a property that is not shared by  the remaining $M_{\mathrm{RF}}-1$  spurious phases. % yield lower SNR values.
To this end,  we use a set of $G$ analog combiners, $\{\boldsymbol{W}_{\!\!\mathrm{A},g}\}_{g=1}^G$, along with $G$ segments of snapshots of length $K$ each.
%with allocating $KG$ snapshots for resolving DoA ambiguity. 
Each element in this set, $\boldsymbol{W}_{\mathrm{\!\! A},g}\in\mathbb{C}^{M\times L}$, is constructed as %takes the form 
 %follows the form 
 in~\eqref{eq:W_A_PC}, but with the phases connected to its $L$ RF chains, i.e., the phases of individual columns,  %of  $\boldsymbol{W}_{\mathrm{\!\! A},g}$
%selected using
configured using $L$ distinct phases from  $\mathcal{G}$ in~\eqref{eq:G_r}. Hence, $G=\tfrac{M_{\mathrm{RF}}R}{L}=\tfrac{M R}{L^2}$, which, for ease of exposition, will be  assumed to be an integer.
%For $L \leq M_{\mathrm{RF}}R$, the number of combiners is $G=\tfrac{M_{\mathrm{RF}}R}{L}=\tfrac{M R}{L^2} \in \mathbb{Z}$, and 
The $g$-th analog combiner, $\boldsymbol{W}_{\!\!\mathrm{A},g}$, can be expressed as:
\begin{equation}
\label{eq:W_A_nu}
 \boldsymbol{W}_{\!\!\mathrm{A},g}= 
    % = \boldsymbol{d}_{(g-1)L+1} \oplus  \cdots   \oplus \boldsymbol{d}_{g L},
    % %----------------------------------------------
    %\bigoplus_{\ell =1}^{L}\boldsymbol{d}(\tilde{\mu}_{(g-1)L+\ell}),
    \bigoplus_{\ell =1}^{L}\boldsymbol{w}_{\mathrm{\!A},\ell}(\tilde{\mu}_{(g-1)L+\ell}), 
        %----------------------------------------------
    %\bigoplus_{\ell =1}^{L}\boldsymbol{d}_{j}(\tilde{\mu}_{j}),
\end{equation} 
where, for any $\beta$, the $m$-th entry of $\boldsymbol{w}_{\mathrm{\!A},\ell}(\beta)
%\tilde{\mu}_j)
\in\mathbb{C}^{M_{\mathrm{RF}}\times 1}$ is $e^{\jmath (m-1)\beta}$, $m=1,\ldots,M_{\mathrm{RF}}$, and $\tilde{\mu}_{(g-1)L+\ell}$ is the $((g-1)L+\ell)$-th element of $\mathcal{G}$ in~\eqref{eq:G_r}, for $g=1,\ldots,G$.

By cycling over the set of analog combiners, 
$\{\boldsymbol{W}_{\!\!\mathrm{A},g}\}_{g=1}^G$, 
we identify the true $R$ DoAs as the ones that maximize the SNR at the outputs of the RF chains. %In particular, $\mu_r$, for $r=1, \ldots, R$, corresponding to the true DoAs, is given by $\mu_r =  \tilde{\mu}_{j_r^\star}$, 
In particular, the estimate of true DoA, $\mu_r$, %for $r=1, \ldots, R$,
corresponds to the phase $\tilde{\mu}_{j_r^\star} \in \mathcal{G}$, 
where $j_r^\star$ is the index of the phase %for the $r$-th source
% index of optimal phase
that maximizes the SNR of the $r$-th source. 
In other words, %This index can  be obtained by solving the following optimization problem:
\begin{equation}
 j_r^\star \!\!= \!\!\!\! \argmax_{j_r \in \{(r-1)M_{\mathrm{RF}}+1, \ldots, rM_{\mathrm{RF}}\}} \!\!\!\!\!\!\!\!\!\!\!\!\!\!\!
\frac{\|\boldsymbol{w}^{H}_{\mathrm{\!A},\ell}(\tilde{\mu}_{j_r})  \boldsymbol{A}_{\ell}\boldsymbol{S}_{g} + \boldsymbol{w}^{ H}_{\mathrm{\!A},\ell}(\tilde{\mu}_{j_r})\boldsymbol{Z}_{g}\|^2}{ K\|\boldsymbol{w}_{\mathrm{\!A},\ell}(\tilde{\mu}_{j_r})\|^2 } - 1,
\label{eq:SNR_gamma_nu_ell}
\end{equation}
where  $\boldsymbol{A}_{\ell} \in \mathbb{C}^{M_{\mathrm{RF}} \times R}$ is the $\ell$-th partition of $\boldsymbol{A}$, i.e., $\boldsymbol{A}= \begin{bmatrix} \boldsymbol{A}^T_{1} & \cdots & \boldsymbol{A}^T_{L} \end{bmatrix}^T \in \mathbb{C}^{M \times R}$.
The indices $g$ and $\ell$ corresponding to  a given $j_r$ are given by $g = \lceil \frac{j_r }{L}  \rceil$  and $\ell=j_r-(g-1)L$.
%\blu{and $\alpha = 1$ because SPC-MPM is only applicable to PC-HAD.} %where each partition  %corresponds to $M_{\mathrm{RF}}$-sized elements of  is the $M_{\mathrm{RF}}$-row sub-matrix corresponding to $\boldsymbol{w}^{H}_{\mathrm{\!A},\ell}$. %, 
%with $g = \lceil \frac{j_r }{L}  \rceil$  and $\ell=j_r-(g-1)L$.}
%--------------------------------------------------------------
%the $\ell$-th column of $\boldsymbol{W}_{\!\!\mathrm{A},g}$ is  $\boldsymbol{w}_{\mathrm{\!A},\ell}^{(j_r)}=[\boldsymbol{0}_{(\ell-1)M_{\mathrm{RF}}}^T\alpha\boldsymbol{d}(\tilde{\mu}_{j_r})^T\boldsymbol{0}_{M-\ell M_{\mathrm{RF}}}^T]^T$, 
%--------------------------------------------------------------
%with $g = \lceil \frac{j_r }{L}  \rceil$ denoting the index of $g$-th analog combiner and $\ell=j_r-(g-1)L$  denoting the index of the RF chain within the analog combiner.  
The matrices $\boldsymbol{S}_{g}\in\mathbb{C}^{R\times K}$ and $\boldsymbol{Z}_{g}\in\mathbb{C}^{M\times K}$ are, respectively, the signal and noise components of the $g$-th length-$K$ segment of snapshots. %allocated for resolving DoA ambiguity, corresponding to $\boldsymbol{W}_{\mathrm{\!\!A},g}$.

\begin{remark}\label{remark:SPC-snapshot}
% In the SPC-MPM approach, the total number of snapshots $\tilde{K}=\tilde{K}_1+\tilde{K}_2$ is divided between DoA estimation, $\tilde{K}_1$, and resolving DoA ambiguity, $\tilde{K}_2$. 
% %The snapshots for resolving DoA ambiguity 
% The latter is further divided into $G$ partitions,  %each containing 
% with $K$ snapshots per analog combiner. %Only a small partition of snapshots
% Resolving DoA ambiguity requires only a small fraction, e.g., $10\%$,  of $\tilde{K}$ to identify the phases that correspond to the true DoAs with maximum SNR.\hfill~$\Box$
%--------------------------------------------------------------
%  our numerical simulations indicate that  allocating small number of snapshots to resolve the DoA ambiguity.....
Resolving  ambiguity requires a small fraction of the total  snapshots, e.g., $KG=\frac{\tilde{K}}{10}$, cf. Section~\ref{sec:simulation} below.   
%$10\%$,  of the total snapshots, $\tilde{K}$, 
%to identify the phases that correspond to the true DoAs, %with maximum SNR, 
%as demonstrated by the numerical simulations in 
%Section~\ref{} below.
\hfill~$\Box$
%Resolving DoA ambiguity uses only a small fraction, e.g., $10\%$ of $\tilde{K}$, as the goal is to identify DoAs with maximum SNR rather than estimate them. \hfill~$\Box$
%as the goal is to select DoAs with maximum instantaneous SNR, not to estimate them.
\end{remark}

\begin{remark}\label{remark:SPC-MPM and PCAP-MPM}
In SPC-MPM, the vectors 
$\{\boldsymbol{q}_k\}_{k=1}^{\tilde{K}}$, used to construct the Hankel matrices are  $L$-dimensional, 
whereas in  FD and PMPM, these vectors 
%\blu{they} 
are $M$-dimensional.
%construct the Hankel matrices, which are then used to form the matrix pencil, unlike the length $M$ used in the FD and PMPM approaches. 
The lower dimensionality of SPC-MPM results in 
%\blu{leads to} 
reduced 
%\blu{reduces}
resolution, cf. Section~\ref{sec:simulation} below.\hfill~$\Box$
\end{remark}
\begin{remark}\label{remark:number_of_sources}
% \blu{In FD and PMPM, the maximum number of sources estimable by MPM is bounded by $R\in [\xi,M-\xi]$. In contrast, SPC-MPM and PCAP-MPM bound $R$ by $R\in [\xi,L-\xi]$. Thus, the rank of each Hankel matrix in the MPM must be $R$, as defined in~\eqref{eq:x_n_first}, \eqref{eq:G_PMPM}, \eqref{eq:H_SPC}, and~\eqref{eq:G_PCAP}.}
In FD and PMPM, the number of sources that can be estimated via the MPM is  $R\in [\xi,M-\xi]$, whereas the corresponding range in SPC-MPM  is $R\in [\xi,L-\xi]$. These ranges ensure that the rank of the respective Hankel matrices in~\eqref{eq:x_n_first}, \eqref{eq:G_PMPM}, and~\eqref{eq:H_SPC} is $R$.\hfill~$\Box$
%, as defined in~\eqref{eq:x_n_first}, \eqref{eq:G_PMPM}, \eqref{eq:H_SPC}, and~\eqref{eq:G_PCAP}.
\end{remark}

The PMPM and SPC-MPM approaches address the two key challenges in MPM-based DoA estimation: the attenuation imposed by individual analog combiners and the tangling of received signals caused by the HAD architecture. 
For the first challenge,  both approaches mitigate attenuation by cycling over an exhaustive set of analog combiners. For the second challenge, PMPM depends on periodicity to disentangle the received signals. Extracting such periodicity, if exists, requires additional hardware, which may not be available for some receivers.  
In contrast, SPC-MPM disentangles the output signals of the analog combiner by using single-phase analog combiners. In the next section, numerical results   show that both techniques significantly outperform their existing counterparts, especially when the number of snapshots is limited. 
\section{Simulation}\label{sec:simulation}
In this section, we %conduct numerical simulations 
compare the RMSE between the actual and the DoA estimates generated by the proposed approaches and the approaches in:
% To evaluate the performance of the proposed approaches, we compare them 
%To do so, these approaches are compared 
%with: %the techniques based on
1) root Hybrid Digital Analog Phase Alignment~(root-HDAPA)~\cite{shu2018low}, 2) covariance matrix reconstruction~(CMR)~\cite{li2020covariance},
%(with 10 angles and outputs averaged) 
%and
3) full spatial coverage and successive refinement
%~(FSC-SR)~\cite{Mona2023doa}.
(FSCR)~\cite{Mona2023doa}, and 4) the MPM in FD receivers~(FD-MPM)~\cite{yilmazer2008multiple}.
As a benchmark, the CRLB of the FD~\cite{stoica1990performance, stoica1989music} and HAD receivers~\cite{Mona2023doa} are also depicted. 
The RMSE is given by %, where 
%$\mathrm{RMSE} =
$\sqrt{\frac{1}{RD} \sum_{r=1}^{R}\sum_{d=1}^{D}(\theta_{r}^{(d)}-\Hat{\theta}_{r}^{(d)})^2}$, where 
 $D= 2000$ is the number of Monte Carlo trials.
For the  FD receiver, the  $R \times R$ CRLB matrix  is~\cite{stoica1990performance, stoica1989music}:
% \begin{equation*}
% %\label{eq:CRLB_FD}
% \!\operatorname{CRLB_{FD}}(\boldsymbol{\theta})=\\\frac{\sigma_z^2}{2\tilde{K}}\Bigl(\Re \Bigl\{\boldsymbol{F}^H \boldsymbol{P}^{\perp}_{\!\!\boldsymbol{A}}\boldsymbol{F} \odot \left(\boldsymbol{\Phi}  \boldsymbol{A}^H \boldsymbol{\Sigma}^{-1}  \boldsymbol{A} \boldsymbol{\Phi}\right)^T  \Bigr\} \Bigr)^{-1}, 
% \end{equation*} 
%%%%%%%%%%%%%%%%%%%%%%%%%%%%%%%%%%%%%%%%%
\begin{multline}
\label{eq:CRLB_FD}
\operatorname{CRLB_{FD}}(\boldsymbol{\theta})=\\\frac{\sigma_z^2}{2\tilde{K}}\Bigl(\Re \Bigl\{\boldsymbol{F}^H \boldsymbol{P}^{\perp}_{\!\!\boldsymbol{A}}\boldsymbol{F} \odot \left(\boldsymbol{\Phi}  \boldsymbol{A}^H \boldsymbol{\Sigma}^{-1}  \boldsymbol{A} \boldsymbol{\Phi}\right)^T  \Bigr\} \Bigr)^{-1}, 
\end{multline} 
where %$\tilde{K}$ is the total number of snapshots, 
the $r$-th column of $\boldsymbol{F}\in\mathbb{C}^{M\times R}$ is  
$\boldsymbol{f}_{\!r}=\nabla_{\!\theta_r}{\boldsymbol{a}(\theta_r})$, $\boldsymbol{\Phi}$ is defined in~\eqref{eq:P_r}, and $ \boldsymbol{\Sigma}=\mathbb{E}\{\boldsymbol{X}_n\boldsymbol{X}_n^H\}= \boldsymbol{A}\boldsymbol{\Phi} \boldsymbol{A}^H +  \boldsymbol{I}_{\!M}$.
% \begin{equation}
%     \label{eq:sigma_fd}
%  \boldsymbol{\Sigma}=\blu{\mathbb{E}\{\boldsymbol{X}\boldsymbol{X}^H\}}= \boldsymbol{A}\boldsymbol{\Phi} \boldsymbol{A}^H +  \boldsymbol{I}_{\!M}.  
% \end{equation}
%%%%%%%%%%%%%%%%%%%%%%%%%%%%%%%%%%%%%%%%%%%%%%%%%%%%%%%%%%%

% %%%%%%%%%%%%%%%%%%%%%%%%%%%%%%%%%%%%%%%%%%%%%%%%%%%%%%%%%%%%%%%%%%%%%%
For SPC-MPM, the analog combiners, $\{\boldsymbol{W}_{\mathrm{\!\! A},n}\}_{n=1}^{N}$, satisfy $\boldsymbol{W}_{\mathrm{\!\! A},n}^H\boldsymbol{W}_{\mathrm{\!\! A},n}=\tfrac{ M}{L} \boldsymbol{I}_{\!L}$. Hence, the $R \times R$ CRLB matrix is~\cite{Mona2023doa}: % derived in~\cite{Mona2023doa} as:
\begin{align}
\label{eq:CRLB}
%&\mathbb{E}\{(\boldsymbol{\theta}-\boldsymbol{\hat{\theta}})(\boldsymbol{\theta}-\boldsymbol{\hat{\theta}})^T \} \succeq 
&\operatorname{CRLB_{SPC}}(\boldsymbol{\theta}) = \frac{\sigma_z^2  M}{2KL} \Bigl(\Re \Bigl\{ \sum_{n=1}^{N} \Bigl( \Bigl( \boldsymbol{F}^H \boldsymbol{W}_{\mathrm{\!\! A},n} \boldsymbol{P}^{\perp}_{\!\!\boldsymbol{E}_n} \boldsymbol{W}_{\mathrm{\!\! A},n}^H \boldsymbol{F}\Bigr)  
\nonumber \\&\qquad\qquad\qquad\quad  \quad 
\odot \Bigl(\boldsymbol{\Phi} \boldsymbol{E}^H_n \boldsymbol{\Upsilon}_{\!n}^{-1} \boldsymbol{E}_n \boldsymbol{\Phi}\Bigr)^T  \Bigr) \Bigr\}\Bigr) ^{-1},
\end{align}
where 
%the actual and estimated are denoted by $\boldsymbol{\theta}=[{\theta}_1\cdots{\theta}_R]^T$ and $ \hat{\boldsymbol{\theta}}=[{\hat{\theta}}_1\cdots {\hat{\theta}}_R]^T$, respectively,
%$\eta=\frac{\sigma_z^2  M}{2KL}$, 
%$K$ is the number of snapshots processed by each $\boldsymbol{W}_{\mathrm{\!\! A},n}$, 
%and 
$\boldsymbol{E}_n=\boldsymbol{W}_{\mathrm{\!\! A},n}^H\boldsymbol{A}$, and the $L\times L$ matrix 
%$\boldsymbol{\Upsilon}_{\!n}= \in\mathbb{C}^{L\times L}$ 
$ \boldsymbol{\Upsilon}_{\!n} 
  = \boldsymbol{W}^H_{\mathrm{\!\! A},n}\boldsymbol{A}\boldsymbol{\Phi} \boldsymbol{A}^H \boldsymbol{W}_{\mathrm{\!\! A},n}+ 
  \tfrac{M}{L}
  \boldsymbol{I}_{\!L}$, $n={1,\ldots, N}$. Using an approach analogous to the one in~\cite{Mona2023doa}, the CRLB for PMPM  can be shown to be given by the FD expression in~\eqref{eq:CRLB_FD}, but 
  %with FD receiver 
  with $K$ snapshots, cf. Theorem~\ref{Thr:PMPM} in Section~\ref{sec:PMP-set-analog-combiner-alg}.

For fair comparison, the total number of snapshots is set to $\tilde{K}$ for all receivers. For PMPM and SPC-MPM, the number of analog combiners, $N=\frac{M}{L}$.
For  PMPM, the $\tilde{K}$ snapshots  are %used for DoA estimation is set to $\tilde{K}$, which is 
used to construct the $K_{\mathrm{A}}=K=\frac{\tilde{K}}{N}$ Hankel matrices in~\eqref{eq:Y_E}.
%in~\eqref{eq:Y_E}, is set to \blu{$\tilde{K}$}. % $K_{\mathrm{A}}=\frac{\tilde{K}}{N}$ 
In contrast,  for SPC-MPM,  $\tilde{K}$ is partitioned into %$\tilde{K}_1$ and $\tilde{K}_2$, i.e., $\tilde{K}=
$\tilde{K}_1+\tilde{K}_2$, where $\tilde{K}_1=K_{\mathrm{A}}$ is used 
% DoA estimation, 
to construct the Hankel matrices in~\eqref{eq:Y_E_SPC},
and $\tilde{K}_2= 
\frac{\tilde{K}}{8}$ is used to resolve DoA ambiguity, cf.  Section~\ref{sec:Optimizing-Power}, and  $K =\frac{\tilde{K}_1}{N}$.
 %In contrast,  
%for PMPM, $\tilde{K}_1=\tilde{K}=K_{\mathrm{A}}$ is used only for DoA estimation. 
%For PMPM and SPC-MPM, $N=\frac{M}{L}$, and $K =\frac{\tilde{K}_1}{N}$. 
The matrix pencil parameter is set to  $\xi=\frac{M}{2}$ for PMPM and %\blu{MPM in the FD receiver}, 
FD-MPM,
and $\xi=\frac{L}{2}$ for SPC-MPM.
The matrix $\boldsymbol{S}_n$ in~\eqref{eq:X} 
comprising the signals of the $R$ sources is assumed to be zero-mean Gaussian with independent entries satisfying~\eqref{eq:P_r}  and $\boldsymbol{S}_n=\boldsymbol{S}$,  
for   $n=1,\ldots,N$. Since $\sigma_z^2=1$, the receive SNR of the $r$-th source is 
%$P_r/\sigma_z^2$. 
%The receive SNR of the $r$-th source is 
$P_r$. 
%We also set the spacing between
The antenna spacing $\Delta$ in~\eqref{eq:mu} is set to $\frac{\lambda}{2}$.
\begin{example}[Comparison with CRLB]\label{ex:CRLB}
In this example, we compare the 
RMSE versus the DoA $\theta$ for  FD-MPM~\cite{yilmazer2008multiple} and the approaches proposed herein for HAD receivers when the number of antennas is   $M = 32$, the number of RF chains is $L = 8$, the number of snapshots  $\tilde{K} = 128$, and the $\mathrm{SNR}=20$~dB.
%performance of 
This comparison is depicted in Figure~\ref{fig:crlb_vs_theta_v2}. As a benchmark,  we  plot the square root of the  corresponding root CRLBs. 
The RMSE of  PMPM  with FC-HAD and PC-HAD is compared with the root CRLB in~\eqref{eq:CRLB_FD} with $K$ snapshots,
%~\cite{stoica1990performance, stoica1989music}, cf. Theorem~\ref{Thr:PMPM} in Section~\ref{sec:PMP-set-analog-combiner-alg},
and that of %compare the RMSE of 
SPC-MPM is compared with the root CRLB in~\eqref{eq:CRLB}. 
From Figure~\ref{fig:crlb_vs_theta_v2}, it can be seen that PMPM with FC-HAD and PC-HAD exhibit indistinguishable performance, with a relatively small gap to the root CRLB in~\eqref{eq:CRLB_FD} with $K=32$ snapshots.
% This gap occurs because the CRLB is obtained using the FD receiver with $NK=128$ snapshots, whereas the PMPM with $NK$ snapshots resembles the FD receiver with $K$ snapshots, cf. Theorem~\ref{Thr:PMPM}. 
%\blu{For instance, at \blu{$\theta = 0^\circ$}, the RMSEs for PMPM with FC-HAD and PC-HAD are \blu{$0.0052^\circ$}, whereas the corresponding root CRLB is \blu{$0.0046^\circ$}.}
%for the FD receiver.
From Figure~\ref{fig:crlb_vs_theta_v2}, it can also be seen that that SPC-MPM exhibits a gap to the root CRLB in~\eqref{eq:CRLB} that is slightly larger than that exhibited by PMPM and its corresponding root CRLB. This gap is mainly due to the snapshots consumed to resolve DoA ambiguity. 
% \blu{For instance, at $\theta = 0^\circ$,  the RMSEs for PMPM with FC-HAD and PC-HAD are $0.005^\circ$, whereas the corresponding root CRLB is $0.004^\circ$
% the RMSE for SPC-MPM is $0.006^\circ$, compared to the corresponding CRLB of  $0.004^\circ$ for the PC-HAD receiver.} 
For instance, at $\theta = 0^\circ$, the RMSEs for PMPM with both FC-HAD and PC-HAD are $0.005^\circ$, and the corresponding root CRLB is $0.004^\circ$, whereas the RMSE for SPC-MPM is $0.006^\circ$ and its corresponding  root CRLB is $0.004^\circ$. % .
%Moreover, the performance of PMPM and SPC-MPM  is  only a \blu{$3.5$~dB} gap from the CRLB of the FD receiver with $NK=128$ snapshots. 
Finally, it can be seen from Figure~\ref{fig:crlb_vs_theta_v2} that FD-MPM~\cite{yilmazer2008multiple} with $\tilde{K}=128$ snapshots exhibits a gap to its root CRLB comparable to the corresponding gap between PMPM and SPC-MPM, and their respective root CRLBs.
% For instance, at $\theta = 14.5^\circ$, the RMSEs for PMPM with FC-HAD and PC-HAD are $0.0052^\circ$, while for SPC-MPM, it is $0.0057^\circ$, compared to the corresponding CRLB of $0.0023^\circ$ for FD and $0.0046^\circ$ for PC-HAD.
%and  the proposed approaches are lower-bounded by the CRLB of the FD receiver, with a gap of only $3.5$~dB.}
%This indicates that the performance of PMPM with FC-HAD and PC-HAD receivers is indistinguishable, as both exhibit behavior similar to the FD receiver
% This confirms the indistinguishable performance of PMPM with FC-HAD and PC-HAD,  cf. Theorem~\ref{Thr:PMPM} in Section~\ref{sec:PMP-set-analog-combiner-alg}. 
%Furthermore,  FC-HAD and PC-HAD receivers  in  PMPM outperform the PC-HAD receiver in the SPC-MPM approach  by using the periodicity in the received signals, whereas PC-HAD in SPC-MPM  handles non-periodic Gaussian signals.
% Figure~\ref{fig:crlb_vs_theta_v2} also shows the MPM in the FD receivers~\cite{yilmazer2008multiple} achieves the correspond CRLB and 
% that all the the proposed techniques are lower-bounded by the CRLB for the FD receiver~\cite{stoica1990performance} by only $3.7$~dB between the proposed techniques and the FD receiver.
% Figure~\ref{fig:crlb_vs_theta_v2} also shows that MPM in FD receivers achieves the corresponding CRLB and  the proposed approaches are lower-bounded by the CRLB of the FD receiver, with a gap of only $3.5$~dB.} 
%, with PCAP-HAD  performing slightly better than PC-HAD in SPC-MPM.
\hfill$\Box$
     \begin{figure}
        \centering
     \psfrag{XXX}[][][2]{DoA, $\theta^\circ$}
    \psfrag{YYY}[][][2]{RMSE (degrees)}
    \psfrag{PMPM-FCXXXXXXXXXXXX}[][][1.7]{PMPM, FC-HAD}
    \psfrag{PMPM-PCXXXXXXXXXXXX}[][][1.7]{PMPM, PC-HAD}
    \psfrag{SPC-MPMXXXXXXXXXXXX}[][][1.7]{\hspace*{-1.1cm} SPC-MPM}
     \psfrag{FD-MPMXXX}[][][1.7]{\qquad \ \ FD-MPM~\cite{yilmazer2008multiple}}
     \psfrag{CRLB-FCXXXXXXXXXX}[][][1.7]{\qquad \quad \ \ \ \ CRLB-FC-HAD}
     \psfrag{CRLB-PCXXXXXXXXXXXXXXX}[][][1.7]{\hspace*{-0.6cm} root CRLB-SPC}
        \psfrag{CRLB-FD-KXXXXXXXXXX}[][][1.7]{\qquad \qquad \ \ \ root CRLB-FD, $K$-snapshot}
               \psfrag{CRLB-FD-NKXXXXXXXXXX}[][][1.7]{\qquad \qquad \ \ \ \  root CRLB-FD, $NK$-snapshot}
    \resizebox{.5\textwidth}{0.3\textwidth} 
    {\includegraphics{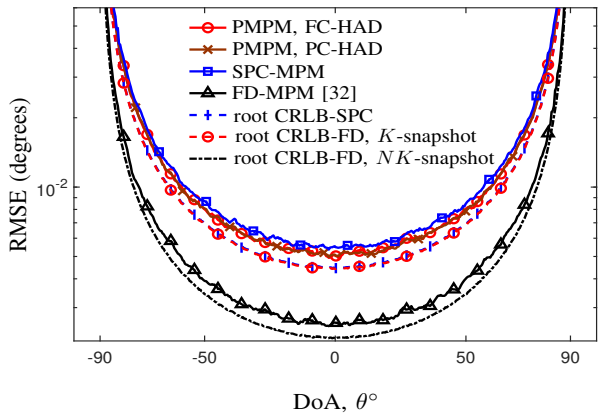}}
        \caption{Root CRLBs and RMSEs  of proposed approaches.} 
        \label{fig:crlb_vs_theta_v2}
        \end{figure}
\end{example}
%%%%%%%%%%%%%%%%%%%%%%%%%%%%%%%%%%%%%%%%%%%%%%%%%%%%%%%%%%%%%%%%%%%%

%-------------------------------------------------------------------------
\begin{example}[Verifying Theorems~\ref{Thr:PMPM} and~\ref{Thr:SPC}]\label{ex:example_1} 
In this example, we consider one source with a DoA of $\theta= 30^\circ$, 
 $M=64$,  $L=8$, $N=8$ and $\tilde{K} %=\tilde{K}_1+ \tilde{K}_2
= 256$.  The RMSE yielded by PMPM, SPC-MPM and that of root-HDAPA~\cite{shu2018low} and CMR~\cite{li2020covariance} are depicted in Figure~\ref{fig:single_source_vs_conv_v2}. 
For comparison, this figure also shows the RMSE yielded by the FD-MPM~\cite{yilmazer2008multiple} in two cases: the first uses $M$ antennas with  $\Delta=\frac{\lambda}{2}$ and  $K=32$ snapshots, whereas the second uses $L$ antennas with $M_{\mathrm{RF}}\Delta$ %= 8\Delta
spacing and $K=28$ snapshots.
The respective snapshots for PMPM, CMR, and root-HDAPA  
are $\tilde{K}_1=NK=256$. 
For SPC-MPM, $\tilde{K}$ is divided into $\tilde{K}_1= NK=224$ and $\tilde{K}_2 
= 32$ snapshots.

From Figure~\ref{fig:single_source_vs_conv_v2},  it can be seen that the proposed PMPM and SPC-MPM approaches significantly outperform root-HDAPA~\cite{shu2018low} and~CMR~\cite{li2020covariance}. 
%From this figure onward, we focus on FC-HAD with PMPM, as its performance is close to that of PC-HAD with PMPM, to improve illustration.
%From this figure onward, we focus on FC-HAD with PMPM, as its performance is close to  that of PC-HAD with PMPM, for better illustration.
% this  SPC-MPM and PCAP have the same performance mainly because of the estimation process itself and using the concatenation and the MP however if we are using anther method the difference will appear and PCAP will be better.
For instance, at an $\mathrm{SNR}=10$~dB, the RMSEs yielded by PMPM, and SPC-MPM are $0.0068^\circ$, and $0.0073^\circ$, respectively. In contrast, the RMSEs for root-HDAPA and CMR are 
$13.825^\circ$ and $0.23^\circ$, respectively.  
The performance of root-HDAPA can be attributed to the fixed analog combiner used therein~\cite{shu2018low}, which heavily attenuates signals arriving from particular DoAs. 
%As for CMR, the performance shown in Figure~\ref{fig:single_source_crlb} aligns with the authors' observation in~\cite[Section~IV]{li2020covariance}. 
The performance of CMR aligns with the observation made in~\cite[Section~IV]{li2020covariance} and can be attributed to the, potentially sub-optimal, diagonal loading used in reconstructing the covariance matrix. 
% From Figure~\ref{fig:single_source_vs_conv_v2}, we can note also that SPC-MPM and MPM in the FD with $K=28$ snapshots and $\theta=30^\circ$, which correspond to one of the beams of the $M_{\mathrm{RF}}$-DFT matrix, has the same RMSE performance over all SNRs, which coincides with Theorem~\ref{Thr:SPC}, cf. Section~\ref{sec:SPC}. In addition , MPM in the FD with $\tilde{K}$ snapshots outperforms PMPM, cf. Theorem~\ref{Thr:PMPM} in Section~\ref{sec:PMP}. For instance, at an RMSE of $0.007^\circ$, MPM in FD  offers a 10~dB advantage over PMPM.
Figure~\ref{fig:single_source_vs_conv_v2} also shows that PMPM and FD-MPM exhibit identical RMSE performance across all SNRs, coinciding with Theorem~\ref{Thr:PMPM}. 
%To verify the statement of Theorem~\ref{Thr:SPC} 
For SPC-MPM, we note that, since the DoA of $\theta=30^\circ$ coincides with the angle of one of the beams of the $M_{\mathrm{RF}}$-DFT matrix, Theorem~\ref{Thr:SPC} asserts that the $M$-antenna SPC-MPM outperforms the $L$-antenna FD-MPM by $10\log M_{\mathrm{RF}}$. Figure~\ref{fig:single_source_vs_conv_v2} verifies this theoretical finding. % is agrees with the num as proved in Theorem~\ref{Thr:SPC}. 
For instance, at an RMSE of $0.004^\circ$, SPC-MPM achieves a $9$~dB performance gain  % $10\log 8$
over FD-MPM. 
%Finally, it is also noted that PMPM and SPC-MPM exhibit almost the same performance as PMPM.}
% Notably, $\theta=30^\circ$ corresponds to one of the beams of the $M_{\mathrm{RF}}$-DFT matrix. Moreover, MPM in the FD receiver with $\tilde{K}$ snapshots outperforms PMPM, as shown in Theorem~\ref{Thr:PMPM}. For example, at an RMSE of $0.007^\circ$, MPM in FD  achieves a $10$~dB % it should be 10log64 >> 18.06 
% performance gain over PMPM.
%In contrast, our techniques, avoid diagonal loading and covariance matrix reconstruction.
% The figure also shows that the proposed techniques achieve the corresponding CRLB in~\eqref{eq:CRLB} for both FC and PC receivers in PMP, with only a slight deviation observed between SPC-MPM and PCAP-MPM and their respective CRLBs. This slight difference arises because these techniques allocate portions of the snapshots to resolve the DoA ambiguity. In contrast, the approach in~\cite{shu2018low} does not reach the corresponding CRLB, which is higher than those achieved by the proposed techniques.
%This is because the analog combiner used in~\cite{shu2018low} always steers in the broadside direction, resulting in heavy attenuation or nullification of certain DoAs. 
\hfill$\Box$
\begin{figure}
        \psfrag{XXX}[][][2]{SNR [dB]}
        \psfrag{YYY}[][][2]{RMSE (degrees)}
        \psfrag{PMPM-PC}[][][1.6]{ \ \ PMPM-PC}
        \psfrag{PMPM-FC}[][][1.6]{PMPM}
        \psfrag{SPC-MPM}[][][1.6]{\quad \  \  SPC-MPM}
        \psfrag{Root-HDAPA}[][][1.6]{\qquad \quad \ \ root-HDAPA~\cite{shu2018low}}
        \psfrag{CMR-Ref32}[][][1.6]{\quad    CMR~\cite{li2020covariance}}
        \psfrag{FD-MPM-K-M-antennas}[][][1.6]{ \qquad  \qquad \   FD-MPM~\cite{yilmazer2008multiple}, {$M$-antenna}}
         \psfrag{FD-MPM-NK-M-antennas}[][][1.6]{ \qquad \  \ \ FD-MPM, $256$ snapshots}
         \psfrag{FD-MPM-K-L-antennas-M-rfDelta}[][][1.6]{ \quad \ FD-MPM~\cite{yilmazer2008multiple}, {$L$-antenna}}
       \resizebox{.5\textwidth}{0.3\textwidth} 
        {\includegraphics{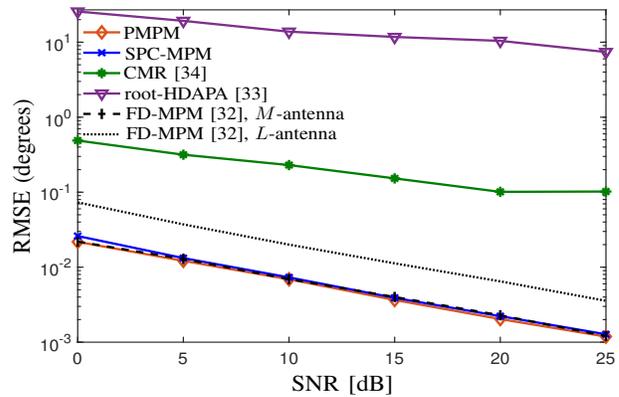}}
    \caption{RMSE comparison of proposed approaches.} 
    %compared with~root-HDAPA~\cite{shu2018low}, CMR~\cite{li2020covariance} and MPM in FD~\cite{yilmazer2008multiple}.}
    \label{fig:single_source_vs_conv_v2}
\end{figure}

\end{example}

%----------------------------------------------------------------------------
\begin{example}[Multiple DoA Estimation]\label{ex:example_2}
%%%%%%%%%%%%%%%%%%%%%%%%%%%%%%%%%%%%%%%%%%%%%%%%%%%
In this example, we compare the RMSE of the proposed approaches with that of FC-FSCR~\cite{Mona2023doa}\footnote{FC-FSCR outperforms PC-FSCR when the number of snapshots is small.}
 and FD-MPM~\cite{yilmazer2008multiple}
%in a scenario 
with four DoAs: $\theta_1=-60^\circ$, $\theta_2=-15^\circ$, $\theta_3=35^\circ$, and  $\theta_4=75^\circ$, when $M=64$, $L=16$,  and $\tilde{K}=128$. The RMSE of all techniques is shown in Figure~\ref{fig:Figure_4DoAs}.
%, \blu{$\xi=32$ for PMPM, and $\xi=8$ for SPC-MPM}.  
For FSCR, $\tilde{K}$, is divided equally between the two underlying algorithms. For the first algorithm, the resolution is set to $\delta_1=10^{-1}$. For the second algorithm, the number of inner iterations is set to $5$, % iterations,
each with $\nu=1550$ search steps.

% From Figure~\ref{fig:Figure_4DoAs}, it is evident that PMPM outperforms SPC-MPM.
% For instance, at an $\mathrm{SNR}=15$~dB, the RMSEs of PMPM, and SPC-MPM are $0.008^\circ$ and $0.009^\circ$, respectively, while MPM with the FD receiver yields an RMSE of $0.004^\circ$. This corresponds to a difference of only $3$~dB between the proposed techniques and the FD receiver.

% In this example, we consider the FSCR technique with the FC-HAD receiver since the estimation accuracy of the algorithms in~\cite{Mona2023doa} depends on the statistical averaging of the SNR received, and the SNR received from the $r$-th source for FSCR is upper-bounded by $MP_r$ in FC-HAD, whereas in PC-HAD, it is upper-bounded by $MP_r/L$, leading to lower estimation accuracy and poor estimation performance, particularly when $\tilde{K} = 128$. 
% In this example, we consider the FSCR technique with FC-HAD because the estimation accuracy of the algorithms in~\cite{Mona2023doa} depends on the statistical averaging of the received SNR. 
% In particular, the SNR of the received signals after the digital combiner in FC-HAD is upper-bounded by $MP_r$, compared to to $MP_r/L$ in PC-HAD, resulting in better estimation performance in FSCR, particularly when $\tilde{K}$ is limited, e.g.,  $\tilde{K} = 128$.
From Figure~\ref{fig:Figure_4DoAs}, it is evident that PMPM outperforms SPC-MPM and FC-FSCR. 
For instance, at an $\mathrm{SNR}=15$~dB, the RMSEs of PMPM, SPC-MPM and FC-FSCR are $0.007^\circ$, $0.009^\circ$ and $0.07^\circ$, respectively, while FD-MPM yields an RMSE of $0.004^\circ$, which corresponds to a $3$~dB gap between the proposed approaches and the FD receiver. The small gap between PMPM and SPC-MPM aligns with the results reported in Example~\ref{ex:CRLB}. The performance of FC-FSCR is due to the fact that the  underling algorithms rely on statistical averaging~\cite{Mona2023doa},  which requires a large number of snapshots for accurate estimation. 
%In Figure~\ref{fig:rmse_vs_snapshots_2} in Example~\ref{ex:example_4} below
The effect of %increasing 
the number of snapshots on performance will be discussed in  Example~\ref{ex:example_4} below.
%it can be seen that  FSCR %in~\cite{Mona2023doa} 
% is significantly superseded by the proposed
% approaches when $\tilde{K}=32$. For instance, at an   $\mathrm{SNR}=15$~dB, the RMSEs yielded by PMPM, SPC-MPM, and PCAP-MPM  are  $0.022^\circ$, $0.025^\circ$ and $0.027^\circ$, respectively, whereas that yielded by FSCR is $15.5^\circ$. The reason for that is that the estimation of the algorithms in~\cite{Mona2023doa} rely on calculating the received SNR, which requires a large number of snapshots for accurate estimation. Therefore,  when  $\tilde{K}$ increased to $128$ in Figure~\ref{fig:Figure_4DoAs}, the RMSE performance of all the techniques improved particularly FSCR. % in~\cite{Mona2023doa}.
% For instance, at an   $\mathrm{SNR}=15$~dB, the RMSE yielded by PMPM, SPC-MPM, and PCAP-MPM  are  $0.009^\circ$, $0.01^\circ$ and $0.01^\circ$, respectively, whereas that yielded by FSCR is $0.017^\circ$.
\hfill$\Box$
\begin{figure}
    \centering
    \psfrag{XXX}[][][2]{SNR [dB]}
    \psfrag{YYY}[][][2]{RMSE (degrees)}
    % \psfrag{DMP-SNR-5dB}[][][1.8]{\quad \  $\mathrm{SNR}=5$~dB}
    %  \psfrag{GMP-SNR-5dB}[][][1.8]{\quad \ $\mathrm{SNR}=5$~dB}
    % \psfrag{PCAP-MP-SNR-5dB}[][][1.8]{\hspace*{-0.1cm} $\mathrm{SNR}=5$~dB}
    %------------------------------------------------------------
    \psfrag{PMPM-FC}[][][1.8]{\quad \!\!\!\!\!\! PMPM}
     \psfrag{SPC-MPM}[][][1.8]{ \quad \ \  SPC-MPM}
    \psfrag{PCAP-MPM}[][][1.8]{\qquad \ PCAP-MPM}
    \psfrag{FSCR-FC}[][][1.8]{\qquad \quad \ \ FC-FSCR~\cite{Mona2023doa}}
    \psfrag{FD-MP}[][][1.8]{\qquad \qquad  \ FD-MPM~\cite{yilmazer2008multiple}}
 %---------------------------------------------------------------------
  % \psfrag{DMPXX}[][][1.8]{\!PMPM}
  %    \psfrag{GMPXXXXXX}[][][1.8]{FPC-MPM}
  %   \psfrag{PCAP-MPXX}[][][1.8]{\ \ PCAP-MPM}
    \resizebox{.5\textwidth}{0.3\textwidth} 
    {\includegraphics{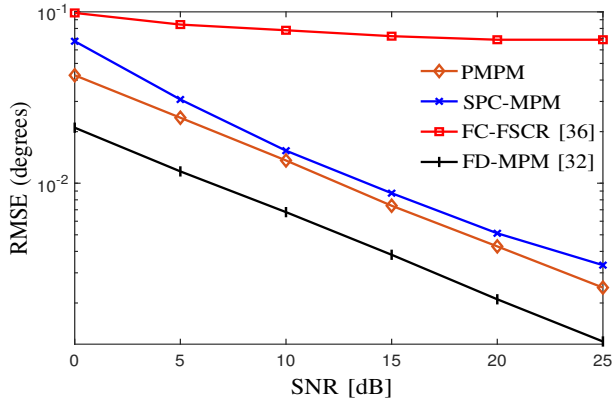}}
    \caption{%RMSE of proposed approaches versus snapshots.
    RMSE of proposed approaches  for
four DoAs.}
    \label{fig:Figure_4DoAs}
\end{figure}
\end{example}

%------------------------------------------------------------------------------
\begin{example}[Resolution]\label{ex:example_3} 
To assess the resolution of the proposed approaches, we consider $R=2$ sources. The DoA of the first source is $\theta_1=-15^\circ$, whereas that of the second source is  $\theta_2=\theta_1-\Delta\theta$, where $\Delta\theta \in \{0.3^\circ, 0.5^\circ, 2^\circ\}$.  We set $M=128$, $L=16$, $\tilde{K}=64$, $\tilde{K}_1=56$, and $\tilde{K}_2=8$.
From Figure~\ref{fig:res_0.3_0.5_2}, it can be seen that the RMSE of all proposed approaches improves monotonically as the angular separation, $\Delta\theta$, between the two DoAs increases from $0.3^\circ$ to $2^\circ$. For instance, at an $\mathrm{SNR}=10$~dB,  the RMSE values for PMPM improve from $0.93^\circ$ to $0.006^\circ$, while those for SPC-MPM 
%and PCAP-MPM 
improve from $0.36^\circ$ to $0.007^\circ$ under the same conditions.
% For $\Delta\theta = 0.5^\circ$ and $2^\circ$, PMPM outperforms SPC-MPM and PCAP-MPM due to the advantage of $M$-dimensional signals in moderate to high SNR regimes, rather than the $L$-dimensional signals used in SPC-MPM and PCAP-MPM,
% %, cf. Sections~\ref{sec:PMP}--\ref{sec:PCAP}. 
% cf.~Theorems~\ref{Thr:PMPM}--\ref{Thr:PCAP}.
% At $\Delta\theta = 0.3^\circ$and SNR levels between $0$~dB and $12$~dB, low SNR dominates, reducing PMPM performance despite the use of $M$-dimensional signals.
For $\Delta\theta = 0.5^\circ$ and $2^\circ$, PMPM outperforms SPC-MPM. This can be attributed to the fact that the signal used to construct the Hankel matrix in these techniques is $M$-dimensional, whereas that used to construct the   Hankel matrix in SPC-MPM is $L$-dimensional, cf.~Theorems~\ref{Thr:PMPM} and~\ref{Thr:SPC}. 
%which provide a significant advantage in moderate to high SNR levels compared to the $L$-dimensional signals used in SPC-MPM and PCAP-MPM, cf.~Theorems~\ref{Thr:PMPM}--\ref{Thr:PCAP}. 
It can also be seen from  Figure~\ref{fig:res_0.3_0.5_2}  that, when $\Delta\theta$ is reduced to $0.3^\circ$, the performance of PMPM is constrained by the SNR until  $12$~dB, favoring SPC-MPM. For instance, at %$\Delta\theta = 0.3^\circ$ and
an RMSE of $0.4^\circ$, SPC-MPM  offers a 4~dB advantage over PMPM.  Beyond that SNR, PMPM outperforms SPC-MPM.  For instance, at %$\Delta\theta = 0.3^\circ$ and
an RMSE of $0.02^\circ$, PMPM  offers a 5.7~dB advantage over SPC-MPM.
%, cf.~Theorems~\ref{Thr:PMPM}--\ref{Thr:PCAP}.
%For instance, at $\Delta\theta = 0.3^\circ$ and $\mathrm{SNR}=10$~dB, PMPM yields an RMSE of $0.93^\circ$, whereas SPC-MPM and PCAP-MPM yield $0.36^\circ$. However, at $\mathrm{SNR}=15$~dB, PMPM improves significantly, achieving an RMSE of $0.02^\circ$, compared to $0.07^\circ$ for SPC-MPM and PCAP-MPM.
%%%%%%%%%%%%%%%%%%% Note to Mona %%%%%%%%%%%%%%%%%%%%%%%%%%%%%%%
% when RMSE vs SNR>> it better to fix RMSE and compare with dB advantage
%%%%%%%%%%%%%%%%%%%%%%%%%%%%%%%%%%%%%%%%%%%%%%%%%%
%For instance, at $\Delta\theta = 0.3^\circ$ and an RMSE of $0.4^\circ$, SPC-MPM and PCAP-MPM offer an 4~dB advantage over PMPM. 
%However, at $\mathrm{SNR}=15$~dB, PMPM improves significantly, achieving an RMSE of $0.02^\circ$, compared to $0.07^\circ$ for SPC-MPM and PCAP-MPM.
% \blu{These results highlight the resolution superiority of PMPM, which can be attributed to using $M$-dimensions received signal, similar to the FD received signals, compared to  $L$-dimensions received signals in SPC-MPM and PCAP-MPM, cf. Sections~\ref{sec:PMP},~\ref{sec:SPC} and~\ref{sec:PCAP}.}
\hfill$\Box$
\begin{figure}
    \centering
    \psfrag{XXX}[][][2]{SNR [dB]}
    \psfrag{YYY}[][][2]{RMSE (degrees)}
    % %%%%%%%%%%%%%%%%%%%%%%%%%%%%%%%%%%%%%%%%%%%%%%%%%%%%%%%%%%%%%%%%%%%
    % \psfrag{PMPM-0.3XXX}[][][1.8]{\ \ \  $\Delta\theta=0.3^\circ$}
    %  \psfrag{SPC-MPM-0.3XXX}[][][1.8]{\ \ $\Delta\theta=0.3^\circ$}
    % \psfrag{PCAP-MPM-0.3XXX}[][][1.8] {\ $\Delta\theta=0.3^\circ$}
    % %------------------------------------------------------------
    %  \psfrag{PMPM-0.5XXX}[][][1.8]{\ \ \  $\Delta\theta=0.5^\circ$}
    %  \psfrag{SPC-MPM-0.5XXX}[][][1.8]{\    $\Delta\theta=0.5^\circ$}
    % \psfrag{PCAP-MPM-0.5XXX}[][][1.8]{\   $\Delta\theta=0.5^\circ$}
    % %----------------------------------------------------------------
    % \psfrag{PMPM-2XX}[][][1.8]{\ \ \  $\Delta\theta=2^\circ$}
    %  \psfrag{SPC-MPM-2XX}[][][1.8]{\ $\Delta\theta=2^\circ$}
    % \psfrag{PCAP-MPM-2XX}[][][1.8]{ $\Delta\theta=2^\circ$}
  %%%%%%%%%%%%%%%%%%%%%%%%%%%%%%%%%%%%%%%%%%%%%%%%%%%%%%%%%%%%%%%%%%%
      \psfrag{theta-0-3}[][][1.7]{\quad \ \ $\Delta\theta=0.3^\circ$}
     \psfrag{theta-0-5}[][][1.7]{\quad \ \ $\Delta\theta=0.5^\circ$}
    \psfrag{theta-2}[][][1.7]{\quad \ \ $\Delta\theta=2^\circ$}
 %---------------------------------------------------------------------
 \psfrag{PMPMXX}[][][1.7]{\!\!\!\!\! PMPM}
     \psfrag{SPCXXXXX}[][][1.7]{\ SPC-MPM}
    \psfrag{PCAP-MPXXX}[][][1.7]{\ PCAP-MPM}
    \resizebox{.5\textwidth}{0.3\textwidth} 
    {\includegraphics{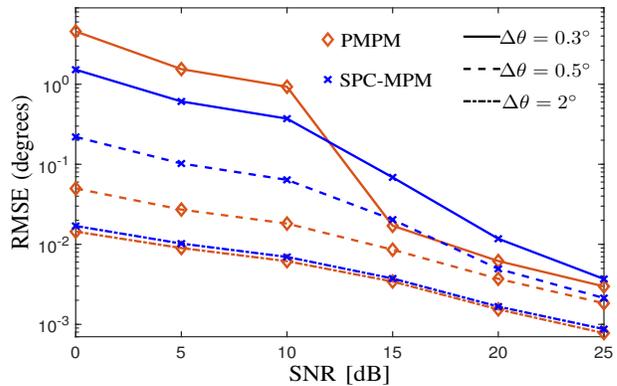}}
    \caption{Resolution investigation at $\Delta\theta = 0.3^\circ, 0.5^\circ, 2^\circ$.}
    \label{fig:res_0.3_0.5_2}
\end{figure}
\end{example}

%----------------------------------------------------------------------------
\begin{example}[Effect of Snapshots on Estimation Accuracy]\label{ex:example_4} 
%\blu{[Impact of snapshots%on RMSE performance on estimation]}
In this example, we  investigate 
the impact of the number of snapshots on the performance of the proposed approaches and the FC-FSCR technique in~\cite{Mona2023doa}. We consider a HAD receiver at  an $\mathrm{SNR}$ of $10$~dB with $M=32$,  $L=8$ and a random DoA, $\theta$ uniformly distributed over  
%\blu{$\theta \in [-80^\circ, 80^\circ]$,} 
%\blu{$\theta \sim \mathcal{U}[-80^\circ, 80^\circ]$,}
%\blu{$\theta \sim \mathcal{U}[-\frac{4\pi}{9}, \frac{4\pi}{9}]$,}
$[-\frac{\pi}{2} +a , \frac{\pi}{2} -a ]$, where $a=\frac{\pi}{100}$ is chosen to avoid the high estimation error at $\pm \frac{\pi}{2}$, cf. Figure~\ref{fig:crlb_vs_theta_v2}.
% 4\pi/9
For FC-FSCR, we use the same settings as in  Example~\ref{ex:example_2}. 
%uniformly distributed 
%with.
%For PMPM, $\xi=16$, while for SPC-MPM, $\xi=4$. 
We set $\tilde{K}_2=1$ for $\tilde{K}<8$, and  $\tilde{K}_2= \frac{\tilde{K}}{8}$ for $\tilde{K} \ge 8$. 
%The parameter $\tilde{K}_2$ is set to $1$ for $\tilde{K}=2$ and $4$, and to $\frac{\tilde{K}}{8}$ for $\tilde{K}=8$ to $128$.
% \blu{For FSCR, $\tilde{K}$, is divided equally between the two underlying algorithms. The resolution of the first algorithm is set to $\delta_1=10^{-1}$, the number of inner iterations of the second algorithm is set to $5$ iterations, each with $\nu=620$ search steps.} The RMSE of all approaches is shown in Figure~\ref{fig:rmse_vs_snapshots_2}. 

From Figure~\ref{fig:rmse_vs_snapshots_2}, it is evident that the RMSE %of all approaches decreases the total number of snapshots increases, while 
of FC-FSCR is significantly superseded by the proposed approaches. 
For instance, for $\tilde{K} = 8$, the RMSEs for PMPM, SPC-MPM are $2.6^\circ$, and $1^\circ$, respectively, compared to $32.1^\circ$ for FC-FSCR. When $\tilde{K}$ increases to $512$, the RMSE of the proposed approaches drops to $0.016^\circ$, while FSCR remains at $0.38^\circ$.
%\blu{The performance of FC-FSCR is due to the fact that the estimation of the algorithms in~\cite{Mona2023doa} rely on statistical averaging of the received SNR, which requires a large number of snapshots for accurate estimation.} 
%As for the proposed techniques, 
The poor performance of proposed techniques when $\tilde{K}\leq 3$ is due to the fact that both PMPM and SPC-MPM require a minimum number of snapshots, equal to the number of analog combiners, i.e., $\tilde{K}=4$.
%Therefore,  when  $\tilde{K}$ increased to $128$, the RMSE performance of all the techniques improved particularly FSCR. 
%PCAP-MPM provides superior spatial resolution and estimation accuracy  
% The proposed techniques exhibit poor performance when number of snapshots is small, e.g., $2$ to $8$ This is due to the fact that PMPM and SPC-MPM require a minimum of snapshots, equal to the number of analog combiners, i.e., $4$ snapshots.
%The superior performance of PCAP-MPM is attributed to allocating most of the snapshots, $\tilde{K}_1$, to one analog combiner. %, unlike PMPM and SPC-MPM, which distribute snapshots among a set of analog combiners. 
%As $\tilde{K}$ increases from $32$ to $128$, PMPM achieves slightly better RMSE performance compared to SPC-MPM and PCAP-MPM.
As $\tilde{K}$ increases, the performance of both  PMPM and SPC-MPM continue to improve.
%, \blu{whereas that of FC-FSCR saturates beyond $\tilde{K}=64$}.   %from $32$ to $128$, the RMSE performance of PMPM, and SPC-MPM becomes indistinguishable. % comparable.
\hfill$\Box$
% All approaches show improved RMSE performance as the SNR increases from $5$~dB to $10$~dB. For instance, at $\tilde{K} = 16$ and $\mathrm{SNR} = 5$~dB, the RMSEs for PMPM, SPC-MPM, and PCAP-MPM are $0.1^\circ$, $0.12^\circ$, and $0.113^\circ$, respectively. When the $\mathrm{SNR}$ increases to $10$~dB, these values reduce to $0.054^\circ$, $0.064^\circ$, and $0.059^\circ$, respectively.
%\input{Figures/Figure_snapshots}
\begin{figure}
    \centering
    \psfrag{XXX}[][][2]{$\Tilde{K}$}
    \psfrag{YYY}[][][2]{RMSE (degrees)}
    % \psfrag{DMP-SNR-5dB}[][][1.8]{\quad \  $\mathrm{SNR}=5$~dB}
    %  \psfrag{GMP-SNR-5dB}[][][1.8]{\quad \ $\mathrm{SNR}=5$~dB}
    % \psfrag{PCAP-MP-SNR-5dB}[][][1.8]{\hspace*{-0.1cm} $\mathrm{SNR}=5$~dB}
    %------------------------------------------------------------
    \psfrag{PMPM-FC-SNR-10}[][][1.7]{\!\!\!\!\!\!\!\!\!\!PMPM}
     \psfrag{SPC-PC-SNR-10}[][][1.7]{ SPC-MPM}
    \psfrag{PCAP-MP-SNR-10}[][][1.7]{\ \ PCAP-MPM}
    \psfrag{FSCR-SNR-10}[][][1.7]{\qquad  \ \  FC-FSCR~\cite{Mona2023doa}}
    % FSCR with FC
 %---------------------------------------------------------------------
  % \psfrag{DMPXX}[][][1.8]{\!PMPM}
  %    \psfrag{GMPXXXXXX}[][][1.8]{FPC-MPM}
  %   \psfrag{PCAP-MPXX}[][][1.8]{\ \ PCAP-MPM}
    \resizebox{.5\textwidth}{0.3\textwidth} 
    {\includegraphics{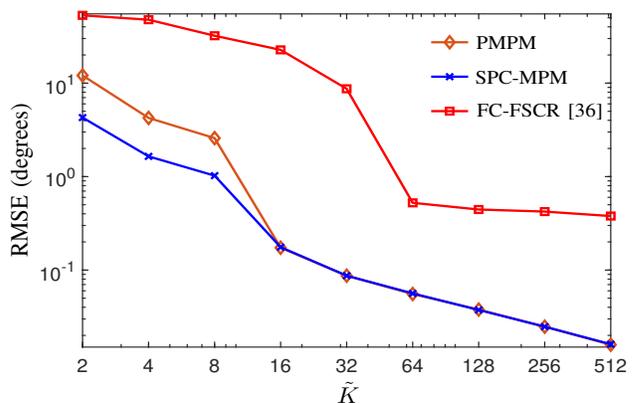}}
    \caption{%RMSE of proposed approaches versus snapshots.
    Impact of number of snapshots on RMSE.}
    \label{fig:rmse_vs_snapshots_2}
\end{figure}
\end{example}

% %----------------------------------------------------------------------------
% \begin{example}\label{ex:example_5} 
% \hfill$\Box$
% %\input{project/figures/Figure3_alg1_alg2_M_64_L_4}
% \end{example}
% %------------------------------------------------------------------------------

%%------------------------------------------------------------------------------
\section{Conclusion}\label{sec:conclusion}
We considered the problem of reliable DoA estimation in HAD receivers when the number of snapshots is too small for statistical averaging to be considered accurate. To overcome the difficulties incurred by the small number of snapshots, we invoked matrix pencil constructions. These constructions can be readily used to extract the DoA in FD receivers, but using them in HAD receivers requires several hurdles to be overcome. Towards that end, we developed two approaches. 
%Two novel DoA estimation approaches were developed for HAD receivers, addressing scenarios where the number of snapshots is too small to ensure reliable statistical averaging. 
%when the number of snapshots is too small for statistical averaging to be reliable.
% The first two approaches cycle over an exhaustive set of analog combiners that collectively span the entire space. 
% The first approach uses the periodicity of the received signals across these analog combiners in FC-HAD and PC-HAD receivers. The signals are then combined in such a way that the resulting received signals are analogous to those of the FD receiver, albeit with SNR and snapshot penalties.
% The second approach 
% eliminates the periodic signals of the first approach by
% by constructing each analog combiner in PC-HAD  using a single column from the DFT matrix.
%These approaches cycle over an exhaustive set of analog combiners that span the entire space. 
The first approach can be applied in FC-HAD and PC-HAD receivers and exploits the periodicity embedded in the received signals. In contrast, the second approach can be applied to PC-HAD receivers only, and does not rely on periodicity. Instead, it uses single-phase analog combiners, which results in DoA ambiguity, and requires further processing to be resolved.
% to render the output of the analog combiners mimic that o
%in FC-HAD and PC-HAD receivers. 
%Despite its efficacy, this approach requires accurate synchronization for extracting the periodicity embedded in the received signals. To avoid this cost, the second approach uses  single-phase analog combiners  in PC-HAD. In both approaches, the received signals resemble those of the FD receiver, albeit with a snapshot penalty.
%To eliminate the requirement 
%To void periodic signals 
% \blu{To eliminate the signal processing penalty associated with extracting periodicity and the set of analog combiners, the third approach introduces a novel class of analog combiners %for HAD receivers
% %, enabling the estimation of signals arriving from any DoA using one analog combiner.
% that estimates signals from any DoA using only one analog combiner.}
%In the second approach, DoA ambiguity is resolved by selecting the DoAs that maximize the received SNR.
% The construction of analog combiners in the latter two approaches introduces DoA ambiguity, which is resolved by selecting the DoAs that maximize the received SNR.
%\blu{Numerical simulations indicate that the three proposed approaches exhibit comparable RMSE performance. The first approach has slightly better performance, although it requires periodicity of the signals.
%%%%%%%%%%%%%%%%%%%%%%%%%%%%%%%%%%%%%%%%%%%%%%%%%%%%%%%%%%%%%%
% \org{Numerical simulations indicate that the two proposed approaches exhibit comparable RMSE performance, with the first approach showing slightly better results.}
Numerical simulations show that the proposed approaches exhibit comparable RMSEs, but the first achieves superior resolution at moderate-to-high SNRs.  Both approaches significantly outperform their existing counterparts.
%%%%%%%%%%%%%%%%%%%%%%%%%%%%%%%%%%%%%%%%%%%%%%%%%%%%%%%%%%%%%%%%%%%%%%
% The third approach offers better spatial resolution and estimation accuracy, particularly when the number of snapshots is small. 
% This comes at a low cost, and only additional shifters are required without the need for additional RF chains.
% The third approach offers better spatial resolution and accuracy when the number of snapshots is small. %, requiring only additional shifters and no extra RF chains.
% This comes at a low cost, and only additional shifters are required without the need for additional RF chains.
%Comparison of the RMSE of these approaches with the corresponding CRLB indicates that the performance of these approaches is within a relatively small gap to optimal.
Comparison with the corresponding CRLBs shows that the performance of the proposed approaches %is close to optimal
is within a small gap to optimal.

%%%%%%%%%%%%%%%%%%%%%%%%%%%%%%%%%%%%%%%%%%%%%%%%%%%%%%%%%%%%%%%%%%%%%%%%%%%%%%%%%%%%%%
\begin{appendices}
\section{Proof of Theorem~\ref{Thr:PMPM}}\label{sec: proof-threorm1}
We begin by stating the following preliminaries. 
\paragraph*{Preliminaries}
Let $\boldsymbol{B}\in\mathbb{C}^{M\times M}$ be an arbitrary full rank matrix partitioned as $\boldsymbol{B}=[\begin{matrix} \boldsymbol{B}_1 & \cdots & \boldsymbol{B}_N \end{matrix}]$. The matrix $\boldsymbol{B}_n$ is full column rank, and the projector on its column space is
%of this matrix is given by %an arbitrary matrix $\boldsymbol{B}$ by 
$\boldsymbol{P}_{\!B_{n}}= \boldsymbol{B}_{n}(\boldsymbol{B}^H_{n} \boldsymbol{B}_{n})^{-1}  \boldsymbol{B}^H_{n}$.
Furthermore, 
%Using the fact that 
$\sum_{n=1}^{N} \boldsymbol{P}_{\!\boldsymbol{B}_{n}}=\boldsymbol{I}_M$ and $\sum_{n\neq m}^{N} \boldsymbol{P}_{\!\boldsymbol{B}_{n}}=\boldsymbol{P}^{\perp}_{\!\boldsymbol{B}_{m}}$, where $\boldsymbol{P}_{\boldsymbol{\!B}_{m}}^{\perp}\in\mathbb{C}^{M\times M}$ is the projector on the  null space of $\boldsymbol{B}$.  
\paragraph*{Detailed Proof}
The output of the analog combiner, $\boldsymbol{W}_{\!\!{\mathrm{A}},n}$, is given in~\eqref{eq:Q_n}, where $\boldsymbol{W}_{\!\!{\mathrm{A}},n}$  is given in~\eqref{eq:W_A_FC_k}  for FC-HAD and in~\eqref{eq:W_A_PC_k} for PC-HAD. 
Setting $\boldsymbol{B}=\boldsymbol{W}_{\!\!{\mathrm{A}},n}$ yields $\boldsymbol{P}_{\boldsymbol{W}_{\!\!{\mathrm{A}},n}}=\frac{1}{\alpha^2 M_{\mathrm{RF}}}\boldsymbol{W}_{\!\!{\mathrm{A}},n} \boldsymbol{W}_{\!\!{\mathrm{A}},n}^H$, where, we have used that $\boldsymbol{W}_{{\mathrm{\!\! A}},n}^H\boldsymbol{W}_{{\mathrm{\!\! A}},n}=\alpha^2 M_{{\mathrm{RF}}}\boldsymbol{I}_{\!L}$.
In the baseband,  $\boldsymbol{Q}_n$ in~\eqref{eq:Q_n} is fed to the $n$-th baseband digital combiner, $\boldsymbol{W}_{\!{\mathrm{D}},n}$.  Our goal is to select $\boldsymbol{W}_{\!{\mathrm{D}},n}$ %in a way that
to facilitate synthesizing the projector  on the column space of 
$\boldsymbol{W}_{\!\!{\mathrm{A}},n}$. To do so, we set $\boldsymbol{W}_{\!{\mathrm{D}},n}=\frac{1}{\alpha^2M_{\mathrm{RF}}} \boldsymbol{W}^H_{\!\!{\mathrm{A}},n}$, $n=1,\ldots,N$. Applying this to~\eqref{eq:Q_n}
%$, is set as 
%$\boldsymbol{W}_{\!{\mathrm{D}},n}=\frac{1}{\alpha^2M_{\mathrm{RF}}} \boldsymbol{W}^H_{\!\!{\mathrm{A}},n}$, 
%$\frac{1}{\alpha^2M_{\mathrm{RF}}} \boldsymbol{W}^H_{\!\!{\mathrm{A}},n}$, 
yields:
\begin{equation}
\boldsymbol{T}_n = \boldsymbol{W}^H_{\!{\mathrm{D}},n}\boldsymbol{Q}_n
= \boldsymbol{P}_{\boldsymbol{W}_{\!\!{\mathrm{A}},n}} \boldsymbol{A}\boldsymbol{S}_{n} + \boldsymbol{P}_{\boldsymbol{W}_{\!\!{\mathrm{A}},n}} \boldsymbol{Z}_{n}.\label{eq:T_n} \end{equation}
%Doing so for $\boldsymbol{W}_{{\mathrm{\!\! A}},n}$, cf.~\eqref{eq:W_A_FC_k} and~\eqref{eq:W_A_PC_k}, and $\boldsymbol{W}_{{\mathrm{\!\! D}},n}$ for $n=1, \ldots, N$, and then 
Aggregating the outputs of the $N$ digital combiners yields $\boldsymbol{Y}_{\!\!\mathrm{PMPM}}=\sum_{n=1}^{N}\boldsymbol{T}_n$. This summation cannot be computed because the first term of the summand, $\boldsymbol{T}_n$, contains the unknown signal $\boldsymbol{S}_{n}$. However, when $\boldsymbol{S}_{n}$ is periodic, i.e., $\boldsymbol{S}_{n}=\boldsymbol{S}$, for $n=1,\ldots, N$,  we 
incorporate the identity
$\sum_{n=1}^{N} \boldsymbol{P}_{\boldsymbol{W}_{\!\!{\mathrm{A}},n}}= \boldsymbol{I}_{\!M}$, % requires that $\boldsymbol{S}_{n}=\boldsymbol{S}$, for $n=1,\ldots, N$. 
%This condition implies that the received signals, $\boldsymbol{S}_{n}$, must exhibit periodicity across the $N$ combiners. 
%Consequently, using this periodic property along with $\sum_{n=1}^{N} \boldsymbol{P}_{\boldsymbol{W}_{\!\!{\mathrm{A}},n}}= \boldsymbol{I}_{\!M}$, the aggregate received signal 
to obtain $\boldsymbol{Y}_{\!\!\mathrm{PMPM}}$ in~\eqref{eq:y_PMPM}.

Comparing $\boldsymbol{Y}_{\!\!\mathrm{PMPM}}$ in~\eqref{eq:y_PMPM} 
with the signal corresponding to the $n$-th length-$K$ segment in the FD receiver, i.e., $\boldsymbol{X}_n$ in~\eqref{eq:X}, it can be seen that
both $\boldsymbol{Y}_{\!\!\mathrm{PMPM}}$ and $\boldsymbol{X}_n$ are $M\times K$ matrices. However, $\boldsymbol{X}_n$ is obtained using $K$ snapshots, whereas $\boldsymbol{Y}_{\!\!\mathrm{PMPM}}$ is obtained using $\tilde{K}=NK$ snapshots. Hence, it can be concluded that 
%they are equivalent except that PMPM  processes $NK$ snapshots, while FD  processes $K$ snapshots, i.e.,
% $\boldsymbol{Y}_{\!\!\mathrm{PMPM}} \big|_{\substack{\tilde{K}=NK  
% }} = \boldsymbol{X} \big|_{\substack{\tilde{K}=K}}$.
$\boldsymbol{Y}_{\!\!\mathrm{PMPM}} \big|_{\substack{\!\!\!\!\!\!\!\! \tilde{K}=NK \\ \mathrm{SNR_{PMPM}}
}} 
\!\!\!\!\!= \boldsymbol{X} \big|_{\substack{\!\!\!\!\! \tilde{K}=K \\ \mathrm{SNR_{FD}}}}$.

To complete the proof, we note that: %it can be readily verified that 
\begin{equation} \label{eq:SNR_FD}
   \mathrm{SNR_{\mathrm{FD}}} = 
  \frac{ \Tr( \boldsymbol{A}\boldsymbol{\Phi}\boldsymbol{A}^H)}{M}.
\end{equation}
It remains to determine $\mathrm{SNR_{PMPM}}$. Towards that end, we have % and $\mathrm{SNR_{FD}}$. For $\mathrm{SNR_{PMPM}}$, we have
\begin{equation}
  \mathrm{SNR_{PMPM}}  = \frac{\mathbb{E}\{\bigl\|\boldsymbol{A}\boldsymbol{S}\bigr\|_F^2\}}{
   \mathbb{E} \Bigl\{ \Bigl\| \sum_{n=1}^{N} \boldsymbol{P}_{\boldsymbol{W}_{\!\!\mathrm{A},n}} \boldsymbol{Z}_n \Bigr\|_F^2 \Bigr\}}.
   %    \frac{ \Tr( \boldsymbol{A}\boldsymbol{\Phi}\boldsymbol{A}^H)}{ 
% M\sigma_z^2}, 
   %\frac{ \Tr( \boldsymbol{A}\boldsymbol{\Phi}\boldsymbol{A}^H)}{ }, 
    \label{eq:SNR_PMPM}
\end{equation}
Invoking~\eqref{eq:P_r}, it can be seen that $\mathbb{E}\{\bigl\|\boldsymbol{A}\boldsymbol{S}\bigr\|_F^2\}=K\Tr( \boldsymbol{A}\boldsymbol{\Phi}\boldsymbol{A}^H)$. Using the fact that $\boldsymbol{P}_{\boldsymbol{W}_{\!\!\mathrm{A},n}}$ is idempotent Hermitian, and
%$=\boldsymbol{P}^H_{\boldsymbol{W}_{\!\!\mathrm{A},n}}$, 
$\sum_{n=1}^{N} \boldsymbol{P}_{\boldsymbol{W}_{\!\!\mathrm{A},n}}=\boldsymbol{I}_M$, and the fact that: 
 % $\mathbb{E} \{ \boldsymbol{Z}_n \boldsymbol{Z}^\dag_m \}= K \boldsymbol{I}_M$, 
\begin{equation}
\mathbb{E} \{ \boldsymbol{Z}_n \boldsymbol{Z}^H_m \}= \!\begin{cases}
        K \boldsymbol{I}_M, \ \ n=m, \\
        0, \qquad \ \  n\neq m,
    \end{cases}
\end{equation}  
% Furthermore, the PMPM approach operates at an SNR that is $-10\log M$ lower than that of the FD receiver. In particular, 
% the output SNR of $\boldsymbol{Y}_{\!\!\mathrm{PMPM}}$ %in~\eqref{eq:y_PMPM} 
% is given by:
% \begin{equation}
%   \!\!\!\mathrm{SNR_{PMPM}}  \!\!= \!\! \frac{\mathbb{E}\{\bigl\|\boldsymbol{A}\boldsymbol{S}\bigr\|_F^2\}}{
%    \mathbb{E} \Bigl\{ \Bigl\| \sum_{n=1}^{N} \boldsymbol{P}_{\boldsymbol{W}_{\mathrm{A},n}} \boldsymbol{Z}_n \Bigr\|_F^2 \Bigr\}}
%    \!\!=\!\!
% %    \frac{ \Tr( \boldsymbol{A}\boldsymbol{\Phi}\boldsymbol{A}^H)}{ 
% % M\sigma_z^2}, 
%    \frac{ \Tr( \boldsymbol{A}\boldsymbol{\Phi}\boldsymbol{A}^H)}{ 
% M}, 
%     \label{eq:SNR_PMPM}
% \end{equation}
% using the fact that, $\boldsymbol{P}_{\boldsymbol{W}_{\!\!\mathrm{A},n}}=\boldsymbol{P}^H_{\boldsymbol{W}_{\!\!\mathrm{A},n}}$,$\sum_{n=1}^{N} \boldsymbol{P}_{\boldsymbol{W}_{\!\!\mathrm{A},n}}=\boldsymbol{I}_M$, and 
%  % $\mathbb{E} \{ \boldsymbol{Z}_n \boldsymbol{Z}^\dag_m \}= K \boldsymbol{I}_M$, 
% \begin{equation}
% \mathbb{E} \{ \boldsymbol{Z}_n \boldsymbol{Z}^H_m \}= \!\begin{cases}
%         K \boldsymbol{I}_M, \quad \ n=m, \\
%         0, \qquad \quad \  n\neq m,
%     \end{cases}
% \end{equation}  
we have %it can be verified that
 % \blu{$ \mathbb{E} \{ \| \boldsymbol{P}_{\boldsymbol{W}_{\!\!\mathrm{A},n}} \boldsymbol{Z}_n \|_F^2 \} = \sigma_z^2 K \Tr( \boldsymbol{P}_{\boldsymbol{W}_{\!\!\mathrm{A},n}})=KL\sigma_z^2$}, 
 %%%%%%%%%%%%%%%%%%%%%%%%%%%%%%%%%%%%%%%%%%%%%%%%%%%%%%%%%%%%%%%%%%%%%
% \begin{align}
% &\mathbb{E} \Bigl\{ \Bigl\| \sum_{n=1}^{N} \boldsymbol{P}_{\boldsymbol{W}_{\!\!\mathrm{A},n}} \boldsymbol{Z}_n \Bigr\|_F^2 \Bigr\} \nonumber  \\
%     &= \mathbb{E} \Bigl\{ \Tr ( \sum_{n=1}^{N} \boldsymbol{P}_{\boldsymbol{W}_{\!\!\mathrm{A},n}} \boldsymbol{Z}_n \sum_{m=1}^{N} (\boldsymbol{P}_{\boldsymbol{W}_{\!\!\mathrm{A},m}} \boldsymbol{Z}_m)^H ) \Bigr\} \nonumber  \\
%      &= \mathbb{E} \Bigl\{ \Tr \Bigl( \sum_{n=1}^{N} \sum_{m=1}^{N} \boldsymbol{P}_{\boldsymbol{W}_{\!\!\mathrm{A},n}} \boldsymbol{Z}_n \boldsymbol{Z}_m^H \boldsymbol{P}_{\boldsymbol{W}_{\!\!\mathrm{A},m}}\bigr) \Bigr\} \nonumber  \\
%      &=  \Tr \Bigl( \sum_{n=1}^{N} \sum_{m=1}^{N} \boldsymbol{P}_{\boldsymbol{W}_{\!\!\mathrm{A},n}} \mathbb{E} \Bigl\{\boldsymbol{Z}_n \boldsymbol{Z}_m^H\Bigr\} \boldsymbol{P}_{\boldsymbol{W}_{\!\!\mathrm{A},m}}\Bigr) \nonumber  \\
%      &=K \Tr \sum_{n=1}^{N}  \boldsymbol{P}_{\boldsymbol{W}_{\!\!\mathrm{A},n}}= KM.
% \end{align}
 %%%%%%%%%%%%%%%%%%%%%%%%%%%%%%%%%%%%%%%%%%%%%%%%%%%%%%%%%%%%%%%%%%%%%%
$\mathbb{E} \Bigl\{ \Bigl\| \sum_{n=1}^{N} \boldsymbol{P}_{\boldsymbol{W}_{\!\!\mathrm{A},n}} \boldsymbol{Z}_n \Bigr\|_F^2 \Bigr\}
    =  KM$, whence % and $\mathrm{SNR_{PMPM}}$ in~\eqref{eq:SNR_PMPM} can be given by:
    \begin{equation}
  \mathrm{SNR_{PMPM}}=  
   \frac{ \Tr( \boldsymbol{A}\boldsymbol{\Phi}\boldsymbol{A}^H)}{ 
M}.
    \label{eq:SNR_PMPM_2}
\end{equation}
% %the variance of the Gaussian noise in~\eqref{eq:X} is
% $\sigma_z^2=1$ in~\eqref{eq:X},
% and $\boldsymbol{\Phi}$ is defined in~\eqref{eq:P_r}.
% \begin{equation}
% \label{eq:Zn}
% \boldsymbol{P}_{\!\!z}
%     = \mathbb{E} \Bigl\{ \Bigl\| \sum_{n=1}^{N} \boldsymbol{P}_{\boldsymbol{W}_{\!\!\mathrm{A},n}} \boldsymbol{Z}_n \Bigr\|_F^2 \Bigr\}
%     = KLN \sigma_z^2=  KM \sigma_z^2,
% \end{equation}
%and  the output SNR 
% Finally, for $\mathrm{SNR_{FD}}$, it can be readily verified that: %we  have %FD receivers in~\eqref{eq:X}  can be given by:
% \begin{equation} \label{eq:SNR_FD}
%    \mathrm{SNR_{\mathrm{FD}}} = 
%   \frac{ \Tr( \boldsymbol{A}\boldsymbol{\Phi}\boldsymbol{A}^H)}{M}.
% \end{equation}
%Thus, using 
Combining~\eqref{eq:SNR_FD}  and~\eqref{eq:SNR_PMPM_2}, 
%yields~\eqref{eq:snr_PMPM_FD} and 
concludes the proof.
% $\boldsymbol{Y}_{\!\!\mathrm{PMPM}} \big|_{\substack{\mathrm{SNR}= \tfrac{1}{M}\Tr( \boldsymbol{A}\boldsymbol{\Phi}\boldsymbol{A}^H)  
% }} = \boldsymbol{X} \big|_{\substack{\mathrm{SNR}=\Tr( \boldsymbol{A}\boldsymbol{\Phi}\boldsymbol{A}^H)}}$.

%%%%%%%%%%%%%%%%%%%%%%%%%%%%%%%%%%%%%%%%%%%%%%%%%%%%%%%%%%%%%%%%%%%%%%%%%%%%%%%%%%%%%%
\section{Proof of Theorem~\ref{Thr:SPC}}\label{sec: proof-threorm2}
To show the equivalence between the input to the $L$-antenna FD receiver and the output of the $n$-th analog combiner,  $\boldsymbol{W}_{\mathrm{\!\! A_{PC}},n}$, we compare $\{x_{k}[m]\}_{m=1}^M$ in~\eqref{eq:x_k} with the signal component of $\{q_{k}[\ell]\}_{\ell=1}^L$  in~\eqref{eq:q_n_ell_sec}. 
%This comparison reveals that the source signal in the FD receiver, $\{s_r\}_{r=1}^R$, are replaced with the signal $\{s_r g(\mu_r - \phi_{n})\}_{r=1}^R$ 
This comparison reveals that they are equivalent but with: i. $M \leftarrow L$,  ii. $s_r\leftarrow s_r g(\mu_r - \phi_{n})$ and iii.
$e^{\jmath (m-1) \mu_r} \leftarrow e^{\jmath (\ell - 1) \mu_r M_{\mathrm{RF}}}$. The latter assignment and~\eqref{eq:mu} imply that $\Delta \leftarrow M_{\mathrm{RF}}\Delta$.

To see that the number of snapshots needed in the considered PC-HAD architecture is $NK$ as opposed to the $K$ snapshots needed for FD receivers, we note from~\eqref{eq:gamma_n} and~\eqref{eq:mu} with $\Delta=\frac{\lambda}{2}$, that the spatial sector covered by the $n$-th analog combiner, $\boldsymbol{W}_{\mathrm{\!\! A_{PC}},n}$, %constructed using $\phi_n$ in~\eqref{eq:gamma_n},
 %given by $\Theta_n$. In particular,  $\Theta_n$ is defined as:
%with width equals to $\frac{2\pi}{M_{{\mathrm{RF}}}}$ cf. Figure~\ref{fig:PC_2}, 
%Given that the width of the main lobe for PC-HAD is $\frac{2\pi}{M_{{\mathrm{RF}}}}$, the spatial sector covered by the $n$-th analog combiner, $\boldsymbol{W}_{\mathrm{\!\! A_{PC}},n}$, constructed using $\phi_n$ in~\eqref{eq:gamma_n},
%^$\phi_n$ in~\eqref{eq:gamma_n} corresponding to
%Using $\phi_n$ in~\eqref{eq:gamma_n} and 
%the fact that the width of the main lobe is $\frac{2\pi}{M_{{\mathrm{RF}}}}$ for PC-HAD, 
%%%%%%%%%%%%%%%%%%%%%%%%%%%%%%%%%%%%%%%%%%%%%%%%%%%%%%%%%%%%%%%%%
% %%  To get the width of sectors: right side
% $$\pi \sin \theta= \frac {2\pi(n-1)}{M_{\mathrm{RF}}} + \frac{\pi}{M_{\mathrm{RF}}}$$
% %%  To get the width of sectors: left side
% $$\pi \sin \theta= \frac {2\pi(n-1)}{M_{\mathrm{RF}}} - \frac{\pi}{M_{\mathrm{RF}}}$$
%%%%%%%%%%%%%%%%%%%%%%%%%%%%%%%%%%%%%%%%%%%%%%%%%%%%%%%%%%%%%%%
is 
$\Theta_n =\bigl(\arcsin\bigl(\frac{2n-3}{M_{\mathrm{RF}}}\bigr),\arcsin\bigl(\frac{2n-1}{M_{\mathrm{RF}}}\bigr)\bigr]$ for $n\in\{1,\ldots,\frac{N}{2}\}$, 
% \blu{$\Theta_n= \bigl(\arcsin\bigl(\frac{2n-1}{M_{\mathrm{RF}}}-2\bigr),\arcsin\bigl(\frac{2n-3}{M_{\mathrm{RF}}}\bigr)\bigr]$ }
$\Theta_n = \bigl(-\frac{\pi}{2}, \arcsin\bigl(\frac{2n-1}{M_{\mathrm{RF}}}-2\bigr)\bigr] \cup \bigl(\arcsin\bigl(\frac{2n-3}{M_{\mathrm{RF}}}\bigr), \frac{\pi}{2} \bigr)$
for $n=\frac{N}{2}+1$
and $\Theta_n= \bigl(\arcsin\bigl(\frac{2n-3}{M_{\mathrm{RF}}}-2\bigr),\arcsin\bigl(\frac{2n-1}{M_{\mathrm{RF}}}-2\bigr)\bigr]$ for $n\in\{\frac{N}{2}+2,\ldots,N\}$. As illustrated in Figure~\ref{fig:PC_2}, the width of each spatial sector is $\frac{2\pi}{M_{{\mathrm{RF}}}}$. %,  as illustrated in Figure~\ref{fig:PC_2}.
%Hence, this set of analog combiners spans the entire space.
% Hence, cycling over $\{\boldsymbol{W}_{\mathrm{\!\! A_{PC}},n}\}_{n=1}^{N}$ will enable all DoAs in $\bigl[\frac{-\pi}{2},\frac{\pi}{2}\bigr]$ to be estimated. In particular, each $\boldsymbol{W}_{\mathrm{\!\! A_{PC}},n}$ processes the $n$-th segment of $K$ snapshots, $n=1,\ldots,N$. 
% Hence, the total number of snapshots is $NK$.
Any DoA lies within the spatial sector defined by  $\boldsymbol{W}_{\mathrm{\!\! A_{PC}},n}$, i.e.,  $\theta_r \in \Theta_n$, can be estimated, otherwise it will be heavily attenuated or nullified. Thus,  using  $\{\boldsymbol{W}_{\mathrm{\!\! A_{PC}},n}\}_{n=1}^{N}$ ensures coverage of all DoAs within $\bigl[\frac{-\pi}{2},\frac{\pi}{2}\bigr]$. In particular, each $\boldsymbol{W}_{\mathrm{\!\! A_{PC}},n}$ processes the $n$-th segment containing $K$ snapshots, $n=1,\ldots,N$.  
%Hence, the total number of snapshots is $NK$. 
In other words, the projection underlying the PC-HAD receiver requires a total of  $NK$ snapshots to ensure that $K$ snapshots are received from each DoA. In contrast, such a projection is not part of the FD receiver, and hence, a total of  $K$ snapshots suffices to be received from each DoA.
%While the FD receiver covers the entire space using $K$ snapshots, the PC-HAD receiver uses the total number of snapshots  $NK$.
This completes the proof of the first two points of the theorem.

It remains to bound  %upper and lower bounds of 
$\mathrm{SNR}_{\mathrm{SPC},r}$. 
%and $\mathrm{SNR}_{\mathrm{FD},r}$ 
%for the $r$-th source. 
Let $\theta_r$ and $\boldsymbol{a}_r$ denote the DoA and the corresponding steering vector (cf.~\eqref{eq:a} and~\eqref{eq:mu}) of the $r$-th source, respectively, $r=1,\ldots,R$.
Let $\boldsymbol{a}_{r,\ell}\in\mathbb{C}^{M_{{\mathrm{RF}}}}$ denote the $\ell$-th partition of 
%the steering vector 
$\boldsymbol{a}_r$, i.e., $\boldsymbol{a}_r=\begin{bmatrix} \boldsymbol{a}_{r,1}^T&\ldots&\boldsymbol{a}_{r,L}^T  \end{bmatrix}^T$.
%%%%%%%%%%%%%%%%%%%%%%%%%%%%%%%%%%%%%%%%%%%%%%%%%%%%%%%%%%%%%%%
% %% ----------------------Detailed proof -------------------
% \begin{align}
%    \mathrm{SNR}_{{\mathrm{SPC}},r}
%     &=  \frac{\mathbb{E}\{\bigl\|\boldsymbol{W}_{\!\!\mathrm{A}}^H\boldsymbol{a}\boldsymbol{s}\bigr\|_F^2\}}{
%    \mathbb{E} \Bigl\{ \Bigl\| \boldsymbol{W}_{\!\!\mathrm{A}} \boldsymbol{Z} \Bigr\|_F^2 \Bigr\}}=  \frac{\mathbb{E}\{\Tr \bigr( \boldsymbol{W}_{\!\!\mathrm{A}}^H\boldsymbol{a}\boldsymbol{s}\boldsymbol{s}^H \boldsymbol{a}^H \boldsymbol{W}_{\!\!\mathrm{A}}   \bigl)}{
%    \mathbb{E} \Bigl\{ \Tr \Bigl(\boldsymbol{W}_{\!\!\mathrm{A}}^H \boldsymbol{Z} \boldsymbol{Z}^H \boldsymbol{W}_{\!\!\mathrm{A}}  \Bigr) \Bigr\}} \nonumber \\ 
%    &=  \frac{K P_r \Tr \bigr( \boldsymbol{W}_{\!\!\mathrm{A}}^H\boldsymbol{a} \boldsymbol{a}^H \boldsymbol{W}_{\!\!\mathrm{A}}   \bigl)}{
%  K  \Tr \Bigl(\boldsymbol{W}_{\!\!\mathrm{A}}^H  \boldsymbol{W}_{\!\!\mathrm{A}}  \Bigr) } = \frac{P_r}{M} \|\boldsymbol{W}_{\!\!\mathrm{A}}^H\boldsymbol{a}\|^2
%  \\
%     &= \frac{P_r|g(\mu_r - \phi_n)|^2 }{M_{\mathrm{RF}}}
% \end{align}
%%%%%%%%%%%%%%%%%%%%%%%%%%%%%%%%%%%%%%%%%%%%%%%%%%%%%%%%%%%%%%%
Let %hen $\theta_r$ lies within the spatial sector of  $\boldsymbol{W}_{\!\!\mathrm{A_{PC}},n}$, i.e., 
$\theta_r \in \Theta_n$, then,  using~\eqref{eq:q_n_ell} and~\eqref{eq:q_n_ell_sec}, the output SNR for the $r$-th source %using~\eqref{eq:q_n_ell} and~\eqref{eq:q_n_ell_sec} 
is given by:
\begin{align}
   \mathrm{SNR}_{{\mathrm{SPC}},r}
    &= P_r\Bigr(\frac{1}{M}\sum_{\ell=1}^L|\boldsymbol{w}_{{\mathrm{A_{PC}},\ell}}^H\boldsymbol{a}_{r,1}|^2\Bigl) \label{eq:SNR_SPC_max_1} \\
    &= \frac{P_r|g(\mu_r - \phi_n)|^2 }{M_{\mathrm{RF}}}
    \leq   M_{\mathrm{RF}}P_r,
    \label{eq:SNR_SPC_max}
\end{align}
% where the first part of~\eqref{eq:SNR_SPC_max} follows from the fact that $\boldsymbol{a}_{r,\ell}=\boldsymbol{a}_{r,1}e^{\jmath\psi_\ell}$ and  the bound %in~\eqref{eq:SNR_SPC_max} %equals that of the FD receiver and it 
% is achieved when 
% %$\boldsymbol{a}_{r,\ell}$ is multiple scaler of $\boldsymbol{v}_{n}$.
% $\boldsymbol{a}_{r,\ell}$ lies in the range space of $\{\boldsymbol{w}_{{\mathrm{A_{PC}},\ell}}\}_{\ell=1}^L$, i.e, multiple scaler of $\boldsymbol{v}_{n}$.
where~\eqref{eq:SNR_SPC_max_1} follows from the fact that $\boldsymbol{a}_{r,\ell}=\boldsymbol{a}_{r,1}e^{\jmath\psi_{r,\ell}}$. 
The first part of~\eqref{eq:SNR_SPC_max}  follows from setting $\boldsymbol{w}_{{\mathrm{A_{PC}},\ell}}=\boldsymbol{v}_n$, and $\phi_\ell=\phi_n$ in $g(\mu_r - \phi_\ell),\, \forall \ell$, cf.~\eqref{eq:f_ell_mu}. For the second part of~\eqref{eq:SNR_SPC_max}, we note that $\sup_{\mu_r}g(\mu_r - \phi_n)=M_{\mathrm{RF}}$, and is achieved when $\mu_r=\phi_n$, which corresponds to $\boldsymbol{a}_{r,\ell}=\boldsymbol{v}_{n}$. 
%\blu{Using~\eqref{eq:SNR_FD} in Appendix~\ref{sec: proof-threorm1}, it can be readily verified that the upper bound in~\eqref{eq:SNR_SPC_max} coincides with the  SNR for the $r$-th source in the FD receiver with $L$-antennas.} %, $\mathrm{SNR_{\mathrm{FD},r}}$
% we note from~\eqref{eq:SNR_FD} in Appendix~\ref{sec: proof-threorm1} that the SNR for the $r$-th source  in the FD receiver is $\mathrm{SNR_{\mathrm{FD},r}}=MP_r$. Hence
% The upper bound in~\eqref{eq:SNR_SPC_max} matches that of the FD receiver and is achieved when 
% $\boldsymbol{a}_{r,\ell}=\boldsymbol{v}_{n}$. 
% It can be verified that the SNR for the $r$-th source  in the FD receiver is $\mathrm{SNR_{\mathrm{FD},r}} %=\Tr(\boldsymbol{a}_r\boldsymbol{\Phi}_{r,r}\boldsymbol{a}_r^H)
% =MP_r$.
%is multiple scaler of $\boldsymbol{v}_{n}$.
%$\boldsymbol{a}_{r,\ell}$ 
%lies in the range space of $\{\boldsymbol{w}_{{\mathrm{A_{PC}},\ell}}\}_{\ell=1}^L$, i.e, 
% multiple scaler of $\boldsymbol{v}_{n}$.
%%-----------------------------------------------------------------
%Let 
%$\theta_r$ falls at the edges of the spatial sectors of 
%$\boldsymbol{W}_{\!\!\mathrm{A_{PC}},n}$ and $\boldsymbol{W}_{\!\!\mathrm{A_{PC}},n+1}$, i.e.,
%$\theta_r \in \Theta_n \cap \Theta_{n+1}$, %by considering the output of one analog combiner,
The lower bound on $\mathrm{SNR}_{\mathrm{SPC}, r}$ in~\eqref{eq:snr_SPC_FD} corresponds to the case in which  $\theta_r$ lies at the edge of $\Theta_n$,
%In particular, %the DoA of the $r$-th source, 
i.e.,  $\mu_r=\phi_n+\frac{\pi}{M_{\mathrm{RF}}}$,
%$\mu_r\in(\phi_n,\phi_{n+1})$, 
%information about $\mu_r$ will be dominantly contained in 
%the output of the RF chains corresponding to $\boldsymbol{v}_n$, 
%and $\boldsymbol{v}_{n+1}$
as illustrated in Figure~\ref{fig:PC_2} for $\mu_r=\phi_1+\frac{\pi}{4}$.
%Considering the output of one analog combiner, 
Substituting $\mu_r=\phi_n+\frac{\pi}{M_{\mathrm{RF}}}$  into $g(\mu_r - \phi_n)$ in~\eqref{eq:f_ell_mu} and~\eqref{eq:SNR_SPC_max} yields: %the output SNR for the $r$-th source
%considering one analog combiner, %is given by:
\begin{equation}
\!\!\!\!\mathrm{SNR}_{\mathrm{SPC}, r}
%\!\!
%=\!\!\frac{P_r\sum_{\ell=1}^L|g(\mu_r - \phi_\ell)|^2 }{M_{\mathrm{RF}} } 
%\!\!= \!\!
=\frac{P_r}{M_{\mathrm{RF}}\sin^2({\frac{\pi}{2M_{\mathrm{RF}}}})}.
    \label{eq:SNR_SPC_min}
\end{equation}
Using~\eqref{eq:SNR_FD} in Appendix~\ref{sec: proof-threorm1},
it can be verified that $ \mathrm{SNR}_{\mathrm{FD},r} = P_r$.
% \begin{equation} \label{eq:SNR_FD_spc}
%    \mathrm{SNR_{\mathrm{FD},r}} = 
%   \frac{ \Tr( \boldsymbol{A}\boldsymbol{\Phi}\boldsymbol{A}^H)}{L}= .
% \end{equation}}
Combining~\eqref{eq:SNR_SPC_max} and~\eqref{eq:SNR_SPC_min}, 
yields~\eqref{eq:snr_SPC_FD}. % and concludes the proof.

\end{appendices}

\bibliographystyle{ieeetr}\bibliography{References}

\end{document}